%% file: paper_v2.tex
\documentclass{article}
\pdfoutput=1
\usepackage{jheppub}
\usepackage{bm}
\usepackage{braket}
\usepackage[table]{xcolor}
\usepackage{mathrsfs}
\usepackage[english]{babel}
\usepackage{graphicx,multirow}
\graphicspath{{figures_stochastic_inflation/},{Old_files/figuresLockdown/}}
\usepackage{caption}
\usepackage{subcaption}
\usepackage{amssymb}
\usepackage{cleveref}
\usepackage{makecell}
\usepackage{physics}
\usepackage{placeins}
\usepackage{hhline}
\usepackage{tcolorbox}
\usepackage{upgreek}
\usepackage[makeroom]{cancel}
\usepackage{dsfont}
\usepackage{tabularx}
\usepackage{arydshln}
\usepackage[normalem]{ulem}
\usepackage{tikz}
\usetikzlibrary{arrows.meta,calc,fit,positioning}

\makeatletter
\gdef\@fpheader{}
\g@addto@macro\bfseries{\boldmath}
\makeatother

\def\nn{\nonumber}

\def\cF{{\cal F}}

\def\exd{{\hbox{d}}}
\def\bea{\begin{eqnarray}}
	\def\eea{\end{eqnarray}}
\def\be{\begin{equation}}
	\def\ee{\end{equation}}

\newcommand{\roughly}[1]{\mathrel{\raise.3ex\hbox{$#1$\kern-0.85em\lower1ex\hbox{$\sim$}}}}

\newcommand{\openone}{\hat{\mathds{1}}}
\def\Tr{\mathrm{Tr}}

\newcommand{\VF}{\mathtt{vol}_{\rm com}}
\newcommand{\Vphys}{\mathtt{vol}_{\rm phys}}
\newcommand{\hvarphi}{\hat{\Phi}}
\newcommand{\hpi}{\hat{\Pi}}

\numberwithin{equation}{section}
\title{Quantum Stochastic Inflation}

\author[a]{Robson Christie,}
\author[a]{Jaewoo Joo,}
\author[b,c]{Greg Kaplanek,}
\author[d]{Vincent Vennin,}
\author[e]{David Wands}

\affiliation[a]{School of Mathematics and Physics, University of Portsmouth, PO1 3FX, United Kingdom}
\affiliation[b]{Department of Electrical Engineering and Computer Science, Syracuse University, NY 13210, USA}
\affiliation[c]{Institute for Quantum \& Information Sciences, Syracuse University, NY 13210, USA}
\affiliation[d]{Laboratoire de Physique de l'Ecole Normale Sup\'{e}rieure, ENS, CNRS, Universit\'{e} PSL, Sorbonne Universit\'{e}, Universit\'{e} Paris Cit\'{e}, 75005 Paris, France}
\affiliation[e]{Institute of Cosmology \& Gravitation, University of Portsmouth, Dennis Sciama Building, Burnaby Road, Portsmouth, PO1 3FX, United Kingdom}

\emailAdd{robson.christie1995@gmail.com, jaewoo.joo@port.ac.uk, gkaplane@syr.edu, vincent.vennin@phys.ens.fr, david.wands@port.ac.uk}

\hypersetup{
  pdftitle={Quantum Stochastic Inflation},
  pdfauthor={Robson Christie, Jaewoo Joo, Greg Kaplanek, Vincent Vennin, David Wands},
  pdfsubject={Canonical open-system derivation of stochastic inflation for a free scalar field in de Sitter space},
  pdfkeywords={stochastic inflation, quantum stochastic inflation, stochastic cosmology, inflationary cosmology, quantum fields in curved spacetime, open quantum systems, de Sitter space, scalar fields in de Sitter space, free scalar field, coarse graining, canonical coarse graining, horizon crossing, super-Hubble modes, sub-Hubble modes, quantum-to-classical transition, decoherence, environmental decoherence, Lindblad dynamics, GKLS equation, Gorini-Kossakowski-Lindblad-Sudarshan equation, master equation, reduced density matrix, phase-space dynamics, Wigner function, Fokker-Planck equation, Langevin equation, stochastic noise, quantum noise, inflationary perturbations, curvature perturbations, cosmological perturbation theory, slow-roll inflation, Starobinsky stochastic inflation, stochastic-delta-N formalism, separate universe approach, infrared modes, ultraviolet modes, squeezing, squeezed states, canonical quantisation, Hamiltonian formalism, quantum Brownian motion, non-equilibrium quantum field theory, primordial fluctuations, early universe cosmology, open-system effective theory}
}

\date{\today}

\abstract{
We formulate stochastic inflation in an open quantum system framework. The field coarse-grained in a patch of fixed physical size, and the total momentum of that patch, form a canonical pair and act on a one-mode Fock space which we identify as the ``bulk''. At each time step, new comoving modes join the coarse-grained patch and the bulk has to be redefined. This redefinition produces an entangled mode that is traced over, yielding a non-unitary evolution equation for the bulk's density matrix. For a free test field in de Sitter, one obtains GKLS dynamics, generated by an effective Hamiltonian and a single non-Hermitian Lindblad operator, hence diffusion and Hubble friction originate from the same quantum channel. The Wigner-Weyl transform of the GKLS equation leads to a Fokker-Planck equation for the Wigner function, which matches the one that applies to the classical phase-space distribution of stochastic inflation. We also provide several schemes under which one can unravel the GKLS dynamics into stochastic Schr\"odinger equations when continuous measurements of the decoupled mode are performed, making contact with Langevin formulations of stochastic inflation. In the light-field regime, an additional overdamped reduction can be performed by integrating out the momentum variable in the Wigner distribution, leading to Starobinsky's slow-roll Fokker-Planck equation. In that regime, the purity of the patch is strongly suppressed. In contrast, for heavy fields, field diffusion is suppressed and the coarse-grained patch remains close to a pure underdamped oscillator, which prevents a classical stochastic treatment. }

\begin{document}

	\sloppy
	\maketitle
	\flushbottom

	\section{Introduction}
	\label{sec:intro}

		Stochastic inflation, pioneered by Starobinsky \cite{Starobinsky1986, NambuSasaki1988, StarobinskyYokoyama1994}, provides an effective theory for the super-Hubble dynamics of scalar fields in the early universe. The standard construction partitions the field degrees of freedom into a long-wavelength infrared (IR) system, $k<k_\sigma$, and a short-wavelength ultraviolet (UV) environment sector, $k>k_\sigma$. Here,
	\begin{equation} \label{ksigma_intro}
		k_\sigma(N)=\sigma aH
	\end{equation}
    is a fixed physical momentum scale (hence time-dependent comoving scale), where $a$ is the universe's scale factor and $H$ is the Hubble rate. The dimensionless parameter $\sigma$ measures the cutoff in Hubble units.

	Since the physical wavelength of a fixed comoving mode grows during inflation, modes continuously cross the coarse-graining scale and source the IR dynamics. The long-wavelength field then follows the classical drift inherited from the Klein-Gordon equation, while differentiating the moving cutoff produces a boundary source supported by the modes crossing \(k_\sigma\). Since different Fourier shells cross this boundary at different times and are uncorrelated in the free Bunch-Davies state, this source is represented as white noise, and the Langevin equations for a coarse-grained scalar field $\phi$ and the coarse-grained momentum $\pi$ read~\cite{grain2017stochastic, vennin2025quantum}
    \begin{align}
    \label{eq:Lang:Staro:phi}
        \frac{\dd\phi}{\dd N}=& \dfrac{\pi}{Ha^3} +\xi_\phi\, ,\\
        \frac{\dd\pi}{\dd N}= & -\dfrac{a^3}{H}m^2\phi+\xi_\pi\, .
    \label{eq:Lang:Staro:pi}
\end{align}
    Time is labelled by the number of e-folds $N=\ln(a)$, $m$ denotes the mass of the field $\phi$ which we assume to be a test free field throughout this work, and $\xi_\phi$ and $\xi_\pi$ are white Gaussian noises satisfying
    \begin{equation}
        \left\langle \xi_f (N) \xi_g(N')\right\rangle=\frac{\dd \ln k_\sigma(N)}{\dd N} \mathcal{P}_{f,g}\left[k_\sigma(N),N\right]\delta(N-N')\, ,
    \end{equation}
Here, $f$ and $g$ are either $\phi$ or $\pi$ and \begin{equation}
\mathcal{P}_{f,g}=\frac{k^3}{2\pi^2} \mathrm{Re}\left(f_k g_k^*\right)
\end{equation}
denotes the reduced power spectrum of $f$ and $g$,  $f_k$ and $g_k$ being their Fourier mode functions. Equivalently, the probability density $P(\phi,\pi,N)$ obeys the Starobinsky Fokker-Planck equation
\begin{equation}
\label{eq:Staro:FP:2d}
\frac{\partial P}{\partial N} =
-\frac{\pi}{Ha^3}\frac{\partial P}{\partial \phi}
+\frac{a^3}{H}m^2\phi\,\frac{\partial P}{\partial \pi}
+\frac12 D_{\phi\phi}\frac{\partial^2 P}{\partial \phi^2}
+D_{\phi\pi}\frac{\partial^2 P}{\partial \phi\,\partial \pi}
+\frac12 D_{\pi\pi}\frac{\partial^2 P}{\partial \pi^2}
\end{equation}
with the matrix $D_{fg}(N)=(\dd \ln k_\sigma(N)/\dd N)\, \mathcal{P}_{f,g}[k_\sigma(N),N]$.

If the field is light $(m<3H/2)$, once the system reaches the overdamped slow-roll attractor~\cite{grain2017stochastic}, phase-space collapses to one dimension (say $\phi$) and \cref{eq:Lang:Staro:phi,eq:Lang:Staro:pi} reduce to
	\begin{equation}
		\frac{\dd\phi}{\dd N} = -\frac{ m^2 \phi}{3H^2} +
        \sqrt{D_{\phi\phi}}
        \,\xi(N), \qquad \langle \xi(N)\xi(N')\rangle=\delta(N-N') .	\label{eq:StarobinskyLangevinIntro}
	\end{equation}
    The associated Fokker-Planck equation reads
	\begin{equation}
		\frac{\partial }{\partial N}P(\phi,N) = \frac{\partial}{\partial\phi}\!\left[\frac{{m^2}}{3H^2}\phi { P}(\phi,N) \right] +
        \frac{1}{2}
        \frac{\partial^2}{\partial\phi^2 }\!\left[D_{\phi\phi}P(\phi,N)\right].
		\label{eq:StarobinskyFPIntro}
	\end{equation}
	The Langevin form displays the stochastic force generated by the crossing shells, while the Fokker-Planck form gives the corresponding diffusion of the probability density. In the limit where the field is light ($m\ll H$) and coarse-grained at super-Hubble scales ($\sigma\ll 1$), $D_{\phi\phi}=H^2/(4\pi^2)$ and \cref{eq:StarobinskyLangevinIntro} reduces to its most common form.

	To justify the emergence of classical stochastic dynamics out of a quantum-field theoretic framework, a form of ``quantum-to-classical transition'' of primordial fluctuations is usually invoked. Usual arguments include: quantum squeezing and obstructions to access the so-called cosmological ``decaying mode'', associated with the momentum conjugate to the curvature perturbation~\cite{Polarski:1995jg, Kiefer:1998qe, Martin:2017zxs}; decoherence induced by other hidden fields, space-time regions outside the observer’s cosmological horizon or inaccessible scales~\cite{Brandenberger:1990bx, Lombardo:1995fg, Calzetta:1995ys, Lombardo:2005iz, Burgess:2006jn, Sharman:2007gi, Martin:2018zbe, Martin:2018lin, Martin:2021znx, Burgess:2022nwu}; the suppression of quantum entanglement between distinct coarse-grained patches in de Sitter~\cite{Martin:2021xml, Martin:2021qkg, Espinosa-Portales:2022yok, Agullo:2024cln, Ribes-Metidieri:2025nfw, Agullo:2025dxp}. Even though these mechanisms likely contribute to the emergence of classical structures out of gravitationally-enhanced quantum fluctuations, their detailed role in the reduction of quantum field theory on an inflating background to a classical, stochastic setup has never been made explicit. The goal of this paper is to fill this gap, by formulating stochastic inflation in a fully quantum manner and identifying the physical quantities (if any) that obey Langevin equations of the form \eqref{eq:StarobinskyLangevinIntro}.

	Inflationary expansion squeezes the vacuum state of perturbations, suppressing phase information and producing correlations between linear observables that mostly admit an effectively classical description~\cite{Martin:2015qta}. However, at the non-linear level, quantum phase-coherence may become more prominent~\cite{Green:2020whw, Launay:2024trh, Ireland:2026txt}, and at any rate a consistent stochastic theory must preserve not only classical-looking correlations, but also the uncertainty relation between the variables retained after coarse-graining. In many coarse-grained formulations, the retained variables are taken to be the averaged field and a correspondingly averaged momentum density, effectively treating these quantities as a canonical pair. We show that this identification misses an important time-dependent normalisation: if \(\hat Q\) is defined as the field averaged over a physical region, then its canonical partner $\hat{P}$ is not the averaged momentum density, but the total momentum contained in that same region. This distinction is essential for preserving the commutator of the retained \textit{bulk} observables, \([\hat Q,\hat P]=i\), after coarse-graining.

	The Fock space on which $\hat{Q}$ and $\hat{P}$ act defines a Hilbert space in which the density matrix of the coarse-grained patch lives. As time proceeds, new comoving scales join the coarse-grained patch and the homogeneous mode is continuously redefined. This redefinition produces an entangled decoupled mode, whose trace gives the effective non-unitary evolution equation for the density matrix, characterised by an effective Hamiltonian and a single non-Hermitian Lindblad operator. The latter acts as the annihilation operator for the component of the incoming boundary shell that couples to the bulk. Because this shell enters in a pure state, the induced noise and dissipation are tied together rather than added independently. Equivalently, the \(2\times2\) Kossakowski matrix for \(\hat Q\) and \(\hat P\) is positive and rank one, so the corresponding Gorini-Kossakowski-Lindblad-Sudarshan (GKLS) equation \cite{gorini1976completely,lindblad1976generators,schlosshauer2007decoherence,wiseman2009quantum,petruccione} contains only one dissipative channel. This rank-one structure is the central result of the paper: Hubble friction and stochastic-inflation diffusion arise from the same completely positive boundary map, while the canonical uncertainty relation is preserved. 

	The GKLS for the density matrix leads to a Fokker-Planck equation for the Wigner function (a phase-space function that provides a complete representation of the quantum state). We show that it coincides with the Starobinsky Fokker-Planck equation~\eqref{eq:Staro:FP:2d}. The crucial difference is that, in the quantum formulation, it applies to the Wigner function, which contrary to the distribution function $P$ is not necessarily positive (although it is positive -- since Gaussian -- for a free test field). In the overdamped limit, we also recover the slow-roll limit of the Starobinsky Fokker-Planck equation~\eqref{eq:StarobinskyFPIntro}. 

    The GKLS equation can also be stochastically ``unravelled'', i.e.~it can be cast into a stochastic Schr\"odinger equation (SSE) for the wavefunction. In this picture, the density matrix obtained by evaluating the ensemble average of the realisations of the SSE satisfies the GKLS equation. This is similar to saying that a classical Fokker-Planck equation can be ``unravelled'' by a stochastic Langevin equation. However, in the quantum case, the noise term in the SSE models an external measurement performed on the wavefunction, and different measurement protocols lead to different SSEs. We discuss a few possibilities, and show in which cases the equations of motion obtained for the quantum expectation values $\langle \hat{Q} \rangle$ and $\langle \hat{P} \rangle$ from the SSE match the Langevin equation~\eqref{eq:StarobinskyLangevinIntro} of stochastic inflation. 
    
	We also compute the purity of the coarse-grained patch, which need not be conserved because the reduced density matrix evolves under decohering, non-unitary dynamics. We show that the purity either asymptotes to a finite late-time value or converges to zero at late times. We discuss how this behaviour is related to the emergence of a classical limit. This article is part of wider efforts that include several other works. In particular, recent articles have developed open-system descriptions of inflationary perturbations by tracing over ultraviolet degrees of freedom \cite{Burgess:2014eoa, Burgess:2022nwu, Colas:2022hlq, Colas:2022kfu, goldman2024lindbladian,Colas:2024ysu, Burgess:2024eng, Beneke:2026nlo,Cespedes:2026nrg, Lopez2025, Salcedo2026}. Other approaches emphasise the role of gravitational or hidden-variable environments \cite{Akama:2026ent}, or derive stochastic-inflation equations using Schwinger-Keldysh methods \cite{Tokuda:2017fdh,Tokuda:2018eqs,Gorbenko:2019rza,Cespedes:2023aal,Green:2025hmo}. Related questions concerning how decoherence modifies stochastic-inflationary probability distributions were considered in \cite{Boddy:2016zkn}, while earlier attempts to model decoherence effects through additional noise sources in stochastic-inflationary Langevin equations include \cite{Haba:2018ork,Haba:2018akk}. Connections between decoherence of cosmological perturbations and stochastic inflation have also been discussed in \cite{Weenink:2011dd}.

	Closest to the present construction, Li \cite{Li2026Stochastic,Li2026lwl} derived a Lindblad equation for the inflaton Wigner function using the Schwinger-Keldysh formalism, together with the influence functional. This derivation finds a single Hermitian jump operator linear in the inflaton field and momentum densities, recovers the Fokker-Planck description and organises slow-roll corrections to the diffusion coefficient. Our construction takes a different route: the canonical bulk algebra fixes a different channel structure, where the boundary-adapted operator is the Bunch-Davies annihilation operator on the bulk mode and the non-Hermitian dissipator contributes to both Starobinsky diffusion and Hubble friction. As we argue below, the momentum density is not the canonical partner of the averaged field, which is one difference with respect to \cite{Li2026Stochastic,Li2026lwl}; those variables may be natural on the slow-roll attractor but are not canonical variables for a coarse-grained patch.

	We organise the paper by first constructing the canonical bulk algebra in \cref{sec:coarsegraining_hamiltonian}. We then derive its one-channel GKLS dynamics and write the equivalent Wigner/Fokker-Planck equation in \cref{sec:boundary_fluctuations} and isolate the massless limit in \cref{sec:light_mass_limit}, where the Starobinsky diffusion coefficient follows after the overdamped reduction. We finally study the fixed light-field range \(0<m<3H/2\), the critical point $(m=3H/2)$, and the heavy regime \(3H/2<m\) in the remainder of \cref{sec:quantum_to_classical}. In these massive regimes the restoring drift admits a stationary Gaussian covariance, and the stationary purity distinguishes a finite-purity quantum damped oscillator from a classical field random walk. We summarise our results and mention several prospects in \cref{sec:Summary}, before ending the paper with three appendices where technical details of the calculations laid out in the main text are deferred.

	\section{Canonical Coarse Graining}
	\label{sec:coarsegraining_hamiltonian}
	We consider a free massive real spectator scalar field $\Phi$ on an assumed spatially-flat de Sitter background,
	\begin{equation}\label{eq:frw-metric}
		\dd s^{2}=-\dd t^{2}+a^{2}(t)\dd \bm{x}^{2}, \qquad a(t)=e^{Ht},
	\end{equation}
	with constant Hubble rate $H$. We label time using the number of \(e\)-folds \(N = H t\). The action is
	\begin{equation}\label{eq:action}
		S=\int \exd^{4}x\; \sqrt{-g}\, \mathcal{L}_{\Phi}, \qquad \mathcal L_\Phi = -\frac12 g^{\mu\nu}\partial_\mu\Phi\,\partial_\nu\Phi -\frac{m^2}{2}\Phi^2 .
	\end{equation}
    We consider a spectator field \(\Phi\) whose energy density remains small compared to
	\( 3M_{\rm Pl}^2H^2\), so that	backreaction onto the geometry is negligible. The canonical conjugate of the field $\Phi$ is
	\begin{equation}\label{eq:pivarphi}
		\Pi = \frac{\partial(\sqrt{-g}\,\mathcal L_\Phi)} {\partial \dot\Phi} = a^3\dot\Phi .
	\end{equation}
	Canonical quantisation in comoving coordinates imposes
	\begin{equation}\label{eq:ETCR}
		[\hvarphi(\bm x, N),\hpi(\bm y,N)] = i\delta^{(3)}(\bm x-\bm y), \qquad [\hvarphi,\hvarphi]=[\hpi,\hpi]=0 ;
	\end{equation}
	in comoving $k$-space this becomes
	\begin{equation}\label{eq:ETCRk}
		[\hvarphi_{\bm k}(N),\hpi_{\bm k'}(N)] = i(2\pi)^3\delta^{(3)}(\bm k+\bm k') .
	\end{equation}
	We coarse grain the field and its conjugate momentum by integrating over a window function $W_N(\bm{y}-\bm{x})$. We call the degree of freedom selected by	this smearing the ``coarse-grained patch'' or ``bulk'', and define its intensive field amplitude and extensive bulk momentum as
	\begin{align} \label{eq:cg:fields:def}
		\hat Q_N(\mathbf x)
		&\equiv \frac{1}{\VF(N)} \int d^3y\,W_N(\mathbf y-\mathbf x)\hat\Phi(\mathbf y, N),
		\\
		\hat P_N(\mathbf x)
		&\equiv \int d^3y\,W_N(\mathbf y-\mathbf x)\hat\Pi(\mathbf y, N).
	\end{align}
	The prefactor $\VF(N)$ is the effective comoving volume of the patch, set such that $\hat Q_N$ and $\hat P_N$ satisfy canonical commutation relations, i.e., $[\hat Q_N,\hat P_N]=i$. Together with \cref{eq:ETCR}, this imposes that~\cite{Martin:2021xml} for the real even smearing kernels considered here
	\begin{equation}
		\label{eq:vol:gen}
		\VF(N) = \frac{1}{\int\dd^3\bm{y}\, W_N(\bm{y}-\bm{x})^{2}} \, .
	\end{equation}
	We now consider the case where the window function is sharp in Fourier space, i.e.,
	\begin{equation} \label{eq:heaviside}
		\int\dd^3\bm{x}\, W_N(\bm{x}) e^{-i \bm{k}\cdot\bm{x}}=\Theta_\sigma(N,\bm{k})
	\end{equation}
	where $\Theta_\sigma$ is the step function such that $\Theta_\sigma(N,\bm{k})=1$ if $k<k_\sigma(N)=\sigma a H$ and $0$ otherwise.
	In this case, the Fourier representation of the bulk variables in \cref{eq:cg:fields:def} reads:
	\begin{align}\label{eq:PQ-def-sharp}
		\hat Q_N(\mathbf x) &= \int\frac{d^3k}{(2\pi)^3}\,\Theta_\sigma(N,k)\hat\Phi_{\mathbf k}(N)e^{i\mathbf k\cdot\mathbf x}, \\
		\hat P_N(\mathbf x) &= \VF(N) \int\frac{d^3k}{(2\pi)^3}\,\Theta_\sigma(N,k)\hat\Pi_{\mathbf k}(N)e^{i\mathbf k\cdot\mathbf x}.
	\end{align}
	The label \(\bm{x}\) only specifies the centre of the coarse-grained patch. Since the de Sitter background is spatially homogeneous and the cutoff \(\Theta_\sigma(N,k)\) depends only on \(k=|\bm{k}|\), no physical result depends on this choice, and below we set \(\bm{x}=0\) for a single representative patch. Since the sharp window satisfies \(\Theta_\sigma^2=\Theta_\sigma\), Parseval's identity gives
\begin{equation}
\int \dd^3\bm{x}\,W_N(\bm{x})^2=\VF^2(N)\int\frac{\dd^3\bm{k}}{(2\pi)^3}\Theta_\sigma(N,k)=\VF^2(N)\frac{k_\sigma^3(N)}{6\pi^2}.
\end{equation}
Combining this with \cref{eq:vol:gen} gives
\begin{equation}\label{eq:VF-integral}
\VF(N)=\frac{6\pi^2}{k_\sigma^3(N)} .
\end{equation}
    see \cref{appA:bulkvars}. The prefactor $\VF(N)$ is thus the effective comoving bulk volume, corresponding to a fixed effective physical volume
	\begin{equation}\label{eq:VphysF-sharp}
		\Vphys = e^{3N}\VF(N) = \frac{6\pi^2}{\sigma^3H^3}\, .
	\end{equation}
	The \(N\)-dependence of \(\VF(N)\) simply reflects the fact that a fixed effective physical coarse-graining volume occupies a shrinking comoving volume during inflation.

	Other time-dependent canonical parametrisations of the same coarse-grained degree of freedom are possible. We use the intensive field amplitude and extensive bulk momentum~\eqref{eq:PQ-def-sharp} because this choice connects directly to the standard stochastic-inflation variables. One may instead define, for example, 
\begin{equation}
\tilde Q_N = \sqrt{\VF(N)} \int \frac{\dd ^3\bm{k}}{(2\pi)^3} \, \Theta_\sigma(N,\bm k)\hvarphi_{\bm k}(N), \qquad \tilde P_N = \sqrt{\VF(N)} \int \frac{\dd^3\bm{k}}{(2\pi)^3} \, \Theta_\sigma(N,\bm k)\hpi_{\bm k}(N)\, . 
\end{equation}
This is a time-dependent symplectic transformation of the bulk variables, hence it does not affect the calculation of the purity below (purity is a symplectic-invariant quantity), and such descriptions are physically equivalent so long as the Hamiltonian and Lindblad operator are transformed consistently.

	The moving cutoff defines a family of instantaneous one-mode canonical algebras, generated by the two averaged bulk variables $\hat Q_N$ and $\hat P_N$, and labelled by \(N\). The algebra at each time is associated with the infrared sector selected by the cutoff \(k_\sigma(N)\), and that sector changes as new modes cross the boundary. The dynamics of this time-dependent bulk algebra, including the effect of the moving boundary, are obtained next.

	\section{Boundary Channel and Reduced Open-System Dynamics}
	\label{sec:boundary_fluctuations}

	\begin{figure}[ht]
		\begin{center}
			\input{figures/GKLS_tikz.tex}
			\caption{\small Steps undertaken to evolve the homogeneous, coarse-grained mode of the field. We use the following terminology: \textit{incoming bulk} denotes the homogeneous mode at \(N\), \textit{incoming boundary shell} denotes the homogeneous shell mode added between \(N\) and \(N+\dd N\), \textit{redefined bulk} denotes the updated homogeneous mode at \(N+\dd N\), and \textit{decoupled mode} denotes the collective mode traced over after the redefinition.
            }
		\label{fig:bulk_boundary}
		\end{center}
	\end{figure}
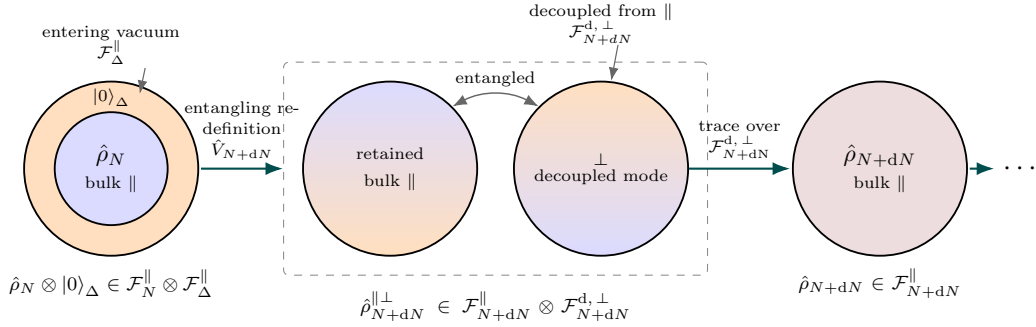

		Let $\mathcal{F}$ denote the Hilbert space in which the quantum state of the field lives,
		\begin{equation}
			\mathcal{F}=\bigotimes_{\bm{k}\in \mathbb{R}^{3}}  \mathcal{F}_{\bm{k}}\, ,
		\end{equation}
		where $\mathcal{F}_{\bm{k}}$ is the single-mode Fock space generated by $\hat{a}_k^{\dagger}$. The mode labels range over all momenta
        \footnote{Hermiticity of the real field implies the operator relation $\hat\Phi_{\mathbf{k}}^{\dagger}=\hat\Phi_{-\mathbf{k}}$. Thus independent field modes may equivalently be labelled by a half-space $\mathbb{R}^{3+}$, or one may work in the real/imaginary basis \cite{Lesgourgues:1996jc,Kiefer:1998jk,Martin:2018zbe}. In the present convention, however, the Fock spaces $\mathcal F_{\mathbf k}$ are associated with the independent annihilation operators $\hat a_{\mathbf k}$, which are labelled by all $\mathbf k\in\mathbb R^3$. This distinction does not affect the following derivations.}; reality is encoded in $\hat{\Phi}_{-\bm{k}}=\hat{\Phi}_{\bm{k}}^\dagger$ and $\hat{\Pi}_{-\bm{k}}=\hat{\Pi}_{\bm{k}}^\dagger$.

        For a free test field, in the Bunch-Davies vacuum different Fourier modes decouple hence the quantum state of the field can be written as the product
		\begin{equation}
			\hat{\rho} = \bigotimes _{\bm{k}\in \mathbb{R}^{3}} \hat{\rho}_{\bm{k}}\, .
		\end{equation}
		Let us now consider the infrared Fock space \(\mathcal{F}_N\) containing all modes with \(k < k_\sigma(N)\),
		\begin{equation}
			\label{eq:FN:def}
			\mathcal{F}_N = \bigotimes_{|\bm{k}| < k_\sigma(N)} \mathcal{F}_{\bm{k}}\, .
		\end{equation}
		The operators~\eqref{eq:PQ-def-sharp} act on a one-mode Fock space denoted $\mathcal{F}_N^\parallel$ and made of a single, collective degree of freedom pertaining to $\mathcal{F}_N$. Denoting by $\mathcal{F}_N^\perp$ its tensor complement, we have
		\begin{equation}
			\mathcal{F}_N = \mathcal{F}_N^\parallel \otimes \mathcal{F}_N^\perp\, .
		\end{equation}

	Before accounting for the moving boundary, we can isolate the unitary evolution of the Fourier modes already inside the cutoff, which live in the infrared Fock space $\mathcal{F}_N$. In $e$-fold time, the full free-field Hamiltonian acting on the entire field Fock space $\mathcal{F}$ is
	\begin{equation}\label{eq:KFullModesMain}
		\hat K_{\rm full}(N)
		= \int \frac{\dd^3\bm{k}}{(2\pi)^3}
		\left[
		\frac{\hpi_{\bm{k}}\hpi_{-\bm{k}}}{2H e^{3N}}
		+\frac{e^{3N}}{2H}\left(k^2e^{-2N}+m^2\right)
		\hvarphi_{\bm{k}}\hvarphi_{-\bm{k}}
		\right] .
	\end{equation}
	Restricting this Hamiltonian to the infrared sector $\mathcal{F}_N$ selected by $k_\sigma(N)=\sigma e^NH$, the gradient contribution is suppressed by $O(\sigma^2)$. At leading order in this super-Hubble expansion, we therefore use the cutoff-restricted Hamiltonian acting on $\mathcal{F}_N$:
	\begin{equation}\label{eq:KIRModesMain}
		\hat K_{{\rm IR},N}
		= \int \frac{\dd^3\bm{k}}{(2\pi)^3}\,\Theta_\sigma(N,\bm{k})
		\left(
		\frac{\hpi_{\bm{k}}\hpi_{-\bm{k}}}{2H e^{3N}}
		+\frac{e^{3N}}{2H}m^2
		\hvarphi_{\bm{k}}\hvarphi_{-\bm{k}}
		\right) .
	\end{equation}
	While this operator is defined on the full infrared field Fock space $\mathcal{F}_N$, its action on the coarse-grained homogeneous sector $\mathcal{F}_N^\parallel$ generated by $(\hat Q_N,\hat P_N)$ is identical to the action of the one-mode bulk Hamiltonian
	\begin{equation}\label{eq:KbulkMain}
		\hat K_{\rm IR, N}^{\parallel}
		= \frac{\hat P_N^2}{2H\Vphys}
		+ \frac{m^2\Vphys}{2H}\hat Q_N^2 .
	\end{equation}
	Indeed, since the complement $\mathcal{F}_N^\perp$ is traced out, we only require the evolution of the homogeneous observables, for which we have
	\begin{equation}
		i[\hat K_{{\rm IR},N},\hat Q_N]
		=i[\hat K_{\rm IR, N}^{\parallel},\hat Q_N]
		= \frac{\hat P_N}{H\Vphys},
		\qquad
		i[\hat K_{{\rm IR},N},\hat P_N]
		=i[\hat K_{\rm IR, N}^{\parallel},\hat P_N]
		= -\frac{m^2\Vphys}{H}\hat Q_N .
	\end{equation}
	Thus, while $\hat K_{{\rm IR},N}$ and $\hat K_{\rm IR, N}^{\parallel}$ act on different Fock spaces ($\mathcal{F}_N$ and $\mathcal{F}_N^\parallel$ respectively), they generate the same unitary evolution on the retained coarse-grained canonical variables. This represents the free unitary evolution of the pre-existing bulk modes. The change of the cutoff and the associated redefinition of the instantaneous homogeneous algebra are treated by the boundary-channel map below; they are not included in $\hat K_{\rm IR, N}^{\parallel}$.

	The quantum state of the scalar field when reduced to the Fock space $\mathcal{F}_N^\parallel$ (or, equivalently, when restricted to the subalgebra generated by \(\hat Q_N\) and \(\hat P_N\)~\cite{Haag1996LocalQuantumPhysics, Fewster:2019ixc}), is denoted $\hat\rho_N$. At two different times $N$ and $N'$, strictly speaking, the two Fock spaces $\mathcal{F}_N^\parallel$ and $\mathcal{F}_{N'}^\parallel$ are not the same, since they comprise different comoving Fourier modes. However, both describe a continuous one-mode system, hence they are isomorphic and a dynamical map can be derived that relates the density matrix at these two different times. This map is derived in detail in \cref{app:bulk_boundary_derivation}; here we only outline the main ingredients, sketched in \cref{fig:bulk_boundary}. The incoming bulk first evolves, the incoming boundary shell is then included, and the homogeneous variables are redefined according to
    \begin{equation}
        \mathcal F_N^\parallel\otimes\mathcal F_\Delta^\parallel
        \xrightarrow{\hat V_{N+\dd N}}
        \mathcal F_{N+\dd N}^\parallel\otimes\mathcal F^{\rm d,\,\perp}_{N+\dd N}.
    \end{equation}
    The Hilbert space \(\mathcal F^{\rm d,\,\perp}_{N+\dd N}\) describes the decoupled mode produced by this redefinition. It is contained in the tensor complement of the updated bulk,
    \begin{equation}
        \mathcal F^{\rm d,\,\perp}_{N+\dd N}
        \subset
        \mathcal F^\perp_{N+\dd N},
    \end{equation}
    and this decoupled mode is then traced out. The key ingredients of this process are itemised below:
		\begin{enumerate}

			\item \textit{Incoming boundary shell.} Consider the set of comoving modes entering the coarse-grained patch between $N$ and $N+\mathrm{d}N$. They live in the Hilbert space $\mathcal{F}_\Delta = \mathcal{F}_{k_\sigma(N)<|\bm{k}| <  k_\sigma(N+\mathrm{d}N)}$, which can again be decomposed into an isotropic, collective degree of freedom that lives in $\mathcal{F}_\Delta^\parallel$, and its tensor complement  $\mathcal{F}_\Delta^\perp$, so  $\mathcal{F}_\Delta =\mathcal{F}_\Delta^\parallel \otimes \mathcal{F}_\Delta^\perp$. The quantum state of the field when reduced to the incoming boundary shell $\mathcal{F}_\Delta^\parallel$ is denoted $\hat{\rho}_\Delta$. Since different Fourier modes decouple in the Bunch-Davies vacuum, and given that $\mathcal{F}_N^\parallel$ and $\mathcal{F}_\Delta^\parallel$ are made of disjoint sets of Fourier modes, the joint state for the bulk-and-shell is the product state $\hat{\rho}_N\otimes\hat{\rho}_\Delta$, which lives in $\mathcal{F}_N^\parallel \otimes \mathcal{F}_\Delta^\parallel$. In other words, the incoming bulk and the incoming boundary shell are \textit{disentangled}.

			\item \textit{Incoming-bulk dynamics.} Starting from $\hat{\rho}_N$ at time $N$, consider the unitary evolution of all Fourier  modes $\bm{k}$ with $\vert\bm{k}\vert<k_\sigma(N)$, generated by \cref{eq:KIRModesMain}. In the $\sigma\ll 1$ limit, gradient contributions can be neglected (this is the ``separate-universe'' approximation~\cite{Salopek:1990, Sasaki:1995, Wands:2000, Lyth:2003, Rigopoulos:2003, Lyth:2005}), hence they all evolve with the same Hamiltonian, represented on the homogeneous algebra by \cref{eq:KbulkMain}.
			The collective degree of freedom contained in  $\hat{\rho}_N$ thus evolves likewise, and between the times $N$ and $N+\mathrm{d}N$, this leads to $\hat{\rho}_N\to e^{-i\hat K_{\rm IR, N}^{\parallel}\mathrm{d}N}\hat{\rho}_N e^{i\hat K_{\rm IR, N}^{\parallel}\mathrm{d}N}$. This map is internal to the bulk $\mathcal{F}_N^\parallel$.

			\item \textit{Mode redefinition.} The  third step is to redefine the homogeneous degree of freedom after the incoming boundary shell has joined the infrared sector. The incoming bulk together with the incoming boundary shell can be rewritten as the redefined bulk at time \(N+\mathrm{d}N\), plus one decoupled mode. In \cref{appA:tracing}, the \textit{complete} (Hamiltonian evolutions$+$redefinition) one-step map is represented by the unitary \(\hat V_{N+\dd N}\):
			\[
			\mathcal F_N^\parallel\otimes\mathcal F_\Delta^\parallel
			\;\xrightarrow{\;\hat V_{N+\dd N}\;}\;
			\mathcal F_{N+\dd N}^\parallel
			\otimes
			\mathcal F^{\rm d,\, \perp}_{N+\dd N}.
			\]
			The unitary \(\hat V_{N+\dd N}\) implements this redefinition of the bulk. In general it entangles the redefined bulk state living in \(\mathcal F_{N+\dd N}^\parallel\) with the decoupled mode living in \(\mathcal F^{\rm d,\, \perp}_{N+\dd N}\).

			\item \textit{Tracing out the decoupled mode.} Finally, since we keep only the redefined bulk variables, we trace over this decoupled mode. Non-unitarity enters only after the decoupled-mode information is discarded. The result is the reduced state
			\[
			\hat\rho_{N+\dd N}
			=
			\Tr_{\mathcal F^{\rm d,\, \perp}_{N+\dd N}}
			\!\left[
			\hat V_{N+\dd N}
			\left(\hat\rho_N\otimes\hat\rho_\Delta\right)
			\hat V_{N+\dd N}^\dagger
			\right],
			\]
			which lives on \(\mathcal F_{N+\dd N}^\parallel\). This partial tracing is the origin of the non-unitary part of the effective evolution.
		\end{enumerate}
		This defines a (non-unitary) map $\mathcal{F}_{N}^\parallel\to \mathcal{F}_{N+\mathrm{d}N}^\parallel$, derived at leading order in $\mathrm{d}N$ in \cref{app:bulk_boundary_derivation}, and by repeating the procedure iteratively one can map $\mathcal{F}_{N}^\parallel$ to any $\mathcal{F}_{N'}^\parallel$ with $N'>N$. At each time, the microscopic embedding of the homogeneous one-mode algebra in the full field Fock space changes, while the reduced description remains a one-mode open quantum system.

	\subsection{Lindblad equation for the density matrix}

	For the free quadratic theory with a sharp moving cutoff, the above construction gives a closed one-step GKLS equation on the projected bulk algebra. In practice, we use the Bunch-Davies mode expansion
	\begin{equation}\label{eq:modeexpansion}
		\hvarphi_{\bm k}(N) = \phi_k(N)\hat a_{\bm k} + \phi_k^*(N)\hat a_{-\bm k}^{\dagger}, \qquad \hpi_{\bm k}(N) = \pi_k(N)\hat a_{\bm k} + \pi_k^*(N)\hat a_{-\bm k}^{\dagger},
	\end{equation}
	where the mode functions are given by
	\begin{align}
		\label{eq:HankelBoundaryModesMain:phi}
		\phi_{k}(N) &= \frac{\sqrt{\pi}}{2\sqrt H}\, e^{-3N/2} e^{\tfrac{i\pi}{4}(2\nu+1)} H_\nu^{(1)}\left(\frac{k}{aH}\right), \\ \pi_{k}(N) &= -\frac{\sqrt{\pi H}}{2}\, e^{3N/2} e^{\tfrac{i\pi}{4}(2\nu+1)} \left[ \frac{k}{e^{N}H} H_{\nu-1}^{(1)}\left(\frac{k}{e^{N}H}\right) + \left(\frac32-\nu\right)H_\nu^{(1)}\left(\frac{k}{aH}\right) \right] ,
		\label{eq:HankelBoundaryModesMain:pi}
	\end{align}
	with
	\begin{equation}
		\nu = \sqrt{\frac94-\frac{m^2}{H^2}}
	\end{equation}
	and Wronskian normalisation
	\begin{equation}\label{eq:Wronskian}
		\phi_k\pi_k^*-\phi_k^*\pi_k=i .
	\end{equation}
	As mentioned above, the full incoming boundary shell $\mathcal{F}_\Delta=\mathcal{F}_\Delta^\parallel \otimes \mathcal{F}_\Delta^\perp$ contains many angular modes, but the homogeneous bulk observables \(\hat Q_N\) and \(\hat P_N\) couple only to one collective combination. Appendix~\ref{appA:boundary} constructs this collective boundary mode explicitly. The appendix does the shell book-keeping; here we give the result that the projected boundary supplies a single Markovian vacuum input channel for the homogeneous bulk algebra, described by the boundary-adapted annihilation operator
	\begin{equation}\label{eq:Adef}
		\hat A_N = i\left[ \frac{\phi_{k_\sigma}^*(N)}{\sqrt{\VF(N)}}\hat P_N - \sqrt{\VF(N)}\,\pi_{k_\sigma}^*(N)\hat Q_N \right], \qquad [\hat A_N,\hat A_N^\dagger]=1 .
	\end{equation}
	Here \(\phi_{k_\sigma}(N)\equiv\phi_{k_\sigma(N)}(N)\) and \(\pi_{k_\sigma}(N)\equiv\pi_{k_\sigma(N)}(N)\). Eq.~\eqref{eq:Adef} is \cref{eq:AdefApp} in the appendix, and the inverse relations between \((\hat Q_N,\hat P_N)\) and \((\hat A_N,\hat A_N^\dagger)\) are given in \cref{eq:QPfromA}. The defining property of \(\hat A_N\) is that its boundary contribution contains only the entering annihilation increment and no independent creation increment, as shown in \cref{eq:dAboundary}. This is why the homogeneous shell projection gives one vacuum input channel rather than two independent noises.

	The corresponding Lindblad operator on the instantaneous bulk algebra is
	\begin{equation}\label{eq:Lsqrt3A}
		\hat L_N = \sqrt3\,e^{i\chi}\hat A_N, \qquad \chi\in\mathbb R .
	\end{equation}
	The factor \(\sqrt3\) comes from the phase-space volume entering the bulk per $e$-fold and appears in the Kraus derivation in \cref{eq:LdefTracingApp}. The phase \(e^{i\chi}\) is conventional: multiplying \(\hat L_N\) by a global phase leaves the unconditional GKLS generator, and hence the Wigner/Fokker-Planck dynamics, unchanged. It only rotates the measured quadratures in the stochastic unravellings discussed in \cref{sec:SSE}.

	The Hamiltonian generator that accompanies this boundary channel is fixed by the same one-step moving-cutoff map. Appendix~\ref{appA:Bulk}, first obtains the bulk drift equations~\eqref{eq:BulkDrift1} and~\eqref{eq:BulkDrift2}. The unitary matching in Appendix~\ref{appA:Unitary}, especially \cref{eq:KeffGeneralUnitaryApp,eq:KeffFinal}, then gives
	\begin{equation}\label{eq:Keff}
		\hat K_{\rm eff}(N) = \frac{\hat P_N^2}{2H\Vphys} + \frac{m^2\Vphys}{2H}\hat Q_N^2 + \frac34\{\hat Q_N,\hat P_N\}.
	\end{equation}
	The first two terms are the homogeneous bulk Hamiltonian  \(\hat K_{\rm IR, N}^{\parallel}\) introduced in \cref{eq:KbulkMain}, while the anticommutator term is induced by the continuous redefinition of the bulk.

	Tracing over the decoupled mode gives the one-channel GKLS equation for the density matrix on the instantaneous homogeneous bulk algebra,
	\begin{equation}\label{eq:Master1DN}
		\frac{\dd \hat\rho_N}{\dd N} = -i[\hat K_{\rm eff}(N),\hat\rho_N] + \hat L_N\hat\rho_N\hat L_N^\dagger - \frac12 \bigl\{ \hat L_N^\dagger\hat L_N,\hat\rho_N \bigr\}.
	\end{equation}
	This equation is also derived in Appendix~\ref{app:bulk_boundary_derivation}, \cref{eq:GKLSFinalApp}. In the free quadratic theory, different shell increments are built from disjoint annihilation operators and are independent in the Bunch-Davies vacuum. The Markovianity is thus a direct consequence of using the sharp cutoff in Fourier space defined in \cref{eq:heaviside}. 

	Let us stress again that the microscopic infrared Fock space changes with \(N\), since \(k_\sigma(N)\) selects a different set of field modes at each $e$-fold. The bulk algebra \(\mathcal A_N={\rm Alg}\{\hat Q_N,\hat P_N\}\) is therefore embedded differently in the field algebra at different times. For bulk observables, we represent these instantaneous one-mode algebras on a fixed canonical algebra \(\mathcal A={\rm Alg}\{\hat Q,\hat P\}\) by the replacement
	\begin{equation}
		\hat Q_N\to\hat Q, \qquad \hat P_N\to\hat P, \qquad [\hat Q,\hat P]=i, \qquad \hat\rho_N\to\hat\rho(N).
	\end{equation}
	Note that the coefficients in \(\hat K_{\rm eff}(N)\) are time independent because \(\Vphys=e^{3N}\VF(N)\) is constant, and in exact de Sitter space, the coefficients in \(\hat L_N\) also become time independent after evaluating the mode functions at \(k=k_\sigma(N)\). We call this fixed-operator description in which we drop the time subscripts, the \textit{stationary representation}; the instantaneous canonical pairs are represented on a single fixed algebra. The corresponding GKLS equation is a stationary master equation.

	The effective Hamiltonian, Lindblad operator and stationary master equation are
	\begin{tcolorbox}[colback=white,colframe=black,arc=0pt,outer arc=0pt]
		\vspace{-10pt}
		\begin{align}
			\hat K_{\rm eff} &= \frac{\hat P^{2}}{2H\Vphys} + \frac{m^2\Vphys}{2H}\hat Q^2 + \frac34\{\hat Q,\hat P\},
			\label{eq:KeffStationaryMain}
			\\[4pt] \hat L_{\chi} &= i\sqrt3\,e^{i\chi}\left[ \frac{\phi_{k_\sigma}^*(N)}{\sqrt{\VF(N)}}\hat P - \sqrt{\VF(N)}\pi_{k_\sigma}^*(N)\hat Q \right],
			\label{eq:LindbladTimeIndep}
			\\[4pt] \frac{\dd\hat\rho}{\dd N} &= -i[\hat K_{\rm eff},\hat\rho] + \hat L_\chi\hat\rho\hat L_\chi^\dagger - \frac12 \bigl\{ \hat L_\chi^\dagger\hat L_\chi,\hat\rho \bigr\} .
			\label{eq:Master1D}
		\end{align}
	\end{tcolorbox}
	Equivalently, the dissipator may be written in the operator basis \((\hat Q,\hat P)\), where the corresponding Kossakowski matrix is positive and rank one, see \cref{eq:RankOneQPKossakowski} in \cref{app:diagonal}. The form of \cref{eq:Master1D} mirrors the fluctuation-dissipation picture of Brownian motion, in which the friction is fixed by the system-bath coupling alone while the noise depends on the state of the bath~\cite{caldeira1983path,petruccione}. The same asymmetry appears here: the damping and the diffusion both originate in the single boundary map, yet the damping is mass-independent while the diffusion is not. The \(-3p\) damping is set by the geometry of the moving cutoff, namely the rate at which the coarse-graining volume grows, and is therefore common to every test field, whereas the mass enters the diffusion coefficients~\eqref{eq:DiffusionGeneralMain} through the covariance of the entering Bunch-Davies modes.

	\subsection{Fokker-Planck equation for the Wigner function}
	
		For a quantum state described by the density matrix $\hat{\rho}$ on phase space $(q,p)$, the Wigner function is defined as the Wigner-Weyl transform of the density matrix~\cite{Wigner:1932}
		\bea\label{eq:Wgendef}
		W(q, p) = \frac{1}{2\pi} \int_{-\infty}^{\infty} \dd u\, e^{- i p u} \left\langle q - \frac{u}{2} \bigg\vert \, \hat{\rho} \, \bigg\vert q + \frac{u}{2} \right\rangle \,.
		\eea
		Since the Wigner-Weyl transform is invertible, this provides a representation of the quantum state which is fully equivalent to the density matrix. Moreover, the expectation value of any operator $\hat{O}$ can be written as
		\bea
		\langle \hat{O}\rangle = \mathrm{Tr}(\hat{O}\hat{\rho}) = \int \dd q \int \dd p\, O(q,p) W(q,p)\, ,
		\eea
		where $O(q,p)=\int\dd u\,e^{-ipu}\langle q-u/2\vert \hat{O} \vert q+u/2\rangle$ is the Wigner-Weyl transform of $\hat{O}$. The Wigner function is real (given that $\hat{\rho}$ is Hermitian) and normalised to one ($\int \dd q \dd p W(q,p)=1$, since $\mathrm{Tr}(\hat{\rho})=1$), thus one may view $W$ as a quasi-distribution function against which observables can be computed. It is only a ``quasi'' distribution since the Wigner function is not necessarily everywhere positive. However, for the case of an initially Gaussian state of a free test field, the Wigner function remains Gaussian and always positive.

	Because \(\hat K_{\rm eff}\) is quadratic and \(\hat L_\chi\) is linear in \((\hat Q,\hat P)\), the Wigner transform of \cref{eq:Master1D} is an exact Fokker-Planck equation
	\begin{align}
		\partial_N W ={}& -\partial_q\!\left(\frac{p}{H\Vphys}W\right) -\partial_p\!\left[\left(-3p-\frac{m^2\Vphys}{H}q\right)W\right] \nonumber\\ &+ \frac12D_{QQ}W_{qq} + D_{QP}W_{qp} + \frac12D_{PP}W_{pp},
		\label{eq:FPGeneralMain}
	\end{align}
	with
	\begin{equation}
		D_{QQ}=\frac{3e^{3N}}{\Vphys}\phi_{k_\sigma}^*\phi_{k_\sigma}, \qquad D_{QP}=3\,{\rm Re}\!\left(\phi_{k_\sigma}^*\pi_{k_\sigma}\right), \qquad D_{PP}=3e^{-3N}\Vphys\,\pi_{k_\sigma}^*\pi_{k_\sigma} ,
		\label{eq:DiffusionGeneralMain}
	\end{equation}
	as can be shown by standard Wigner-Weyl transform methods (see e.g., \cite{wiseman2009quantum, petruccione}). The arbitrary phase \(\chi\) drops out, as expected since rephasing the single Lindblad operator leaves the GKLS generator invariant.

	For the free field in the Bunch-Davies vacuum, the Wigner function is Gaussian and positive, so it can be identified with the Starobinsky phase-space probability density after changing variables to \(q=\phi\) and \(p=\VF(N)\pi\), as follows from \cref{eq:cg:fields:def}. The corresponding Jacobian relation is
	\begin{equation}
	W(q,p,N)\to P\left[\phi=q,\pi=\frac{p}{\VF(N)},N\right]\frac{1}{\VF(N)} .
	\end{equation}
	With this identification, \cref{eq:FPGeneralMain,eq:DiffusionGeneralMain} reproduce the Starobinsky phase-space Fokker-Planck equation~\eqref{eq:Staro:FP:2d}. For more general states, however, the Wigner function need not be positive, so it should not be interpreted as an ordinary phase-space probability distribution.

	Even in that case, one can always define a positive field marginal probability distribution by integrating out the momentum variable,
	\begin{equation}
		{\cal P}_Q(q,N)=\int \dd p\,W(q,p;N)=\bra{q}\hat\rho(N)\ket{q}\geq0 .
	\end{equation}
	Together with the current \(J_Q(q,N)=(H\Vphys)^{-1}\int \dd p\,pW(q,p;N)\), integration of \cref{eq:FPGeneralMain} over \(p\) gives
	\begin{equation}
		\partial_N{\cal P}_Q=-\partial_qJ_Q+\frac12D_{QQ}\partial_q^2{\cal P}_Q ,
	\end{equation}
    provided the boundary terms vanish. This identity is exact within the free quadratic theory, but it is not closed: the current \(J_Q\) obeys its own evolution equation, involving higher momentum moments. In the overdamped regime, the momentum/current sector relaxes quickly compared with the field marginal, so one may use the quasistatic closure \(\partial_N J_Q\simeq0\), solve algebraically for \(J_Q\), and obtain an effective one-dimensional Fokker-Planck equation for \({\cal P}_Q\). As we show in \cref{sec:light_mass_limit}, for light fields ($m<3H/2$, for which an overdamped attractor exists) this Wigner marginal reproduces the Starobinsky Fokker-Planck equation at leading order in $\sigma$.

	\subsection{Stochastic Schr\"odinger equation for the wavefunction}
    \label{sec:SSE}

    	\begin{figure}[ht]
		\centering
		\input{figures/pure_state_povm_tikz.tex}
		\caption{\small Selective SSE branch. The same entangling bulk--boundary redefinition as in \cref{fig:bulk_boundary} is followed by a rank-one measurement of the decoupled mode, rather than a partial trace. Conditioning on outcome \(r\) leaves the bulk in the pure state \(\ket{\psi_{c,r}}\).}
		\label{fig:PureStatePOVM}
	\end{figure}
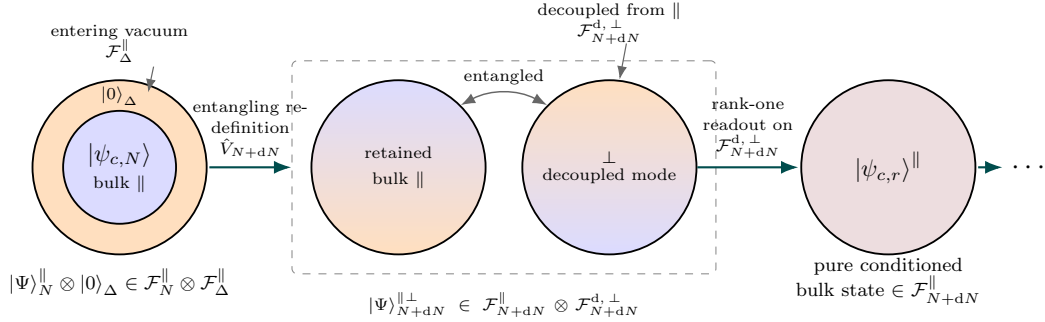

	It is useful at this point to separate two possible descriptions of the same bulk-boundary evolution. Before the decoupled mode is traced out, the system contains both the retained bulk variables and the decoupled mode. The one-step entangling unitary generally entangles them. The leading vacuum-one-particle block of this entangled state is displayed in \cref{eq:FullStateBlock01App} and \eqref{eq:FullStateBlock01ExpandedApp}. Since an entangled state cannot be written as a product of an independent pure bulk state and an independent pure decoupled-mode state, the bulk alone does not in general have a pure state.

	There are then two ways to proceed. If the decoupled mode is not measured, its information is removed from the reduced description by taking the partial trace. This gives a mixed reduced state of the bulk and leads to the GKLS equation derived above. In this sense the GKLS equation is the quantum master equation underlying the classical Fokker-Planck description of stochastic inflation.

A second route consists in measuring the decoupled mode, hence projecting it onto some of the eigenstates of the operator being measured. Since the outcome of such measurement is random, the resulting Schr\"odinger equation for the bulk is stochastic. The details of that stochastic Schr\"odinger equation (SSE) depend on the measurement being performed, which in practice is specified by a positive operator-valued measure (POVM). Different possibilities are explored in \cref{appA:povm_unravellings}, such as the explicit jump, homodyne and heterodyne measurements; here we only highlight that they are different efficient measurements of the same decoupled mode and average back to the same GKLS equation.
     
     To make contact with the Langevin formulation, one may choose a perfectly efficient rank-one measurement of the decoupled mode and condition on its outcome. For a pure pre-measurement joint trajectory, a definite measurement outcome collapses the decoupled mode to a pure state and leaves the bulk in the corresponding pure conditional state. The selective pure-state branch is shown in \cref{fig:PureStatePOVM}. Different choices of decoupled-mode measurement give different stochastic unravellings~\cite{wiseman2009quantum}, but averaging over the conditioned trajectories gives back the same GKLS evolution. As shown in Appendix~\ref{appA:povm_unravellings}, in the massless super-Hubble and overdamped limit, the field-adapted homodyne choice of phase in the Lindblad operator (see \cref{eq:LindbladTimeIndep}) \(\chi=\chi_Q\) gives the one-real-noise Starobinsky Langevin equation~\eqref{eq:StarobinskyLangevinIntro}. The heterodyne unravelling gives two real noises whose combined field-centre diffusion can be written as the same effective Starobinsky Langevin noise.

	\section{The (mass-dependent) Quantum to Classical Transition}
	\label{sec:quantum_to_classical}

	The route to classical stochastic behaviour depends on the mass. As we shall now show, for light fields ($m<3H/2$), the phase-space Wigner equation~\eqref{eq:FPGeneralMain} contains an overdamped field marginal whose positive distribution obeys the Starobinsky one-dimensional Fokker-Planck equation~\eqref{eq:StarobinskyFPIntro}. At the critical point \((m=3H/2)\) and for heavy fields \((m>3H/2)\), the \(D_{QQ}\) channel is suppressed; the remaining late-time diagnostic is the purity of the stationary Gaussian state approached at late times by the retained \((Q,P)\) system. The common structure behind these regimes is the closed low-moment sector; because the Hamiltonian is quadratic and the Lindblad operator is linear, the Moyal expansion terminates and the first moments and covariance close for any reduced state with finite second moments.

	Let \(W(q,p;N)\) be the Wigner function of the retained mode, with \(\hat{\mathbf{Z}}=(\hat Q,\hat P)^\top\) and phase-space coordinates \(\mathbf{z}=(q,p)^\top\). The first moments and covariance matrix elements are
	\begin{equation}
		\ev*{\hat{\mathbf{Z}}}=\int \dd q\,\dd p\,\mathbf{z} W(q,p;N),
		\qquad
		\Sigma_{ij}(N) = \int \dd q\,\dd p\, \left(z_i-\ev*{\hat{Z}_i}\right) \left(z_j-\ev*{\hat{Z}_j}\right)W(q,p;N).
	\end{equation}
	We write this covariance matrix as
	\[
	\mathbf{\Sigma}=
	\begin{pmatrix}
		\Delta_{QQ} & \Delta_{QP}\\ \Delta_{QP} & \Delta_{PP}
	\end{pmatrix}.
	\]
	For the quadratic Wigner generator in \cref{eq:FPGeneralMain}, the moment hierarchy closes at second order~\cite{MyPaper}. The first moments are deterministic and obey
	\begin{equation}\label{eq:massiveDriftBoth}
		\frac{\dd \ev*{\hat Q}}{\dd N}=\frac{\ev*{\hat P}}{H\Vphys}, \qquad \frac{\dd \ev*{\hat P}}{\dd N}=-3\ev*{\hat P}-\frac{m^2\Vphys}{H}\ev*{\hat Q} ,
	\end{equation}
	and combine into the damped-oscillator equation
\begin{equation}\label{eq:dampedOscillator}
		\frac{\dd ^2 \ev*{\hat Q}}{\dd N^2}+3\frac{\dd \ev*{\hat Q}}{\dd N}+\frac{m^2}{H^2}\ev*{\hat Q}=0.
	\end{equation}
For $m>0$, this describes a damped oscillator, which is overdamped for $m<3H/2$ and underdamped for $m>3H/2$.
The corresponding covariance equations~\cite{MyPaper} are
    \begin{align}
	    \dv{\Delta_{QQ}}{N}
		&=\frac{2}{H\Vphys}\Delta_{QP}+D_{QQ},
		\nonumber\\
		\dv{\Delta_{QP}}{N}
		&=\frac{1}{H\Vphys}\Delta_{PP}
		-\frac{m^2\Vphys}{H}\Delta_{QQ}
		-3\Delta_{QP}+D_{QP},
		\label{eq:CovComponentMain}
		\\
		\dv{\Delta_{PP}}{N}
		&=-\frac{2m^2\Vphys}{H}\Delta_{QP}
		-6\Delta_{PP}+D_{PP},
		\nonumber
	\end{align}
	with diffusion entries \(D_{QQ},D_{QP},D_{PP}\) given in \cref{eq:DiffusionGeneralMain}. The first-moment equations do not feed back into this covariance system. Thus the covariance may reach a stationary value while the expectation values continue to evolve or vice versa.

	Equivalently, \cref{eq:CovComponentMain} can be written as the closed Lyapunov equation
	\begin{align}
		\dv{\mathbf{\Sigma}}{N}
		&=\mathbf{A}\mathbf{\Sigma}+\mathbf{\Sigma} \mathbf{A}^\top+\mathbf{D},
		\label{eq:CovLyapunovMain}
		\\
		\mathbf{A}&=
		\begin{pmatrix}
			0 & \dfrac{1}{H\Vphys}\\[6pt]
			-\dfrac{m^2\Vphys}{H} & -3
		\end{pmatrix},
		\nonumber\\
		\mathbf{D}&=3
		\begin{pmatrix}
			\dfrac{e^{3N}}{\Vphys}\abs{\phi_{k_\sigma}}^2
			& \Re(\phi_{k_\sigma}^*\pi_{k_\sigma}) \\[8pt]
			\Re(\phi_{k_\sigma}^*\pi_{k_\sigma})
			& e^{-3N}\Vphys\abs{\pi_{k_\sigma}}^2
		\end{pmatrix}.
		\nonumber
	\end{align}
	For \(m>0\), the deterministic drift stabilises this covariance sector. The stationary covariance is determined by the algebraic Lyapunov equation
    \begin{equation}
    	\mathbf A\boldsymbol{\Sigma}_\infty
    	+
    	\boldsymbol{\Sigma}_\infty\mathbf A^\top
    	+
    	\mathbf D
    	=0.
    	\label{eq:StationaryLyapunovMain}
    \end{equation}
    this equation has the matrix solution
    \begin{equation}
    	\boldsymbol{\Sigma}_\infty
    	=
    	\int_0^\infty
    	e^{\mathbf A s}\mathbf D e^{\mathbf A^\top s}\,\dd s, \qquad 		\mathbf{\Sigma}_\infty=
		\begin{pmatrix}
			\Delta_{QQ}^{\infty} & \Delta_{QP}^{\infty}\\
			\Delta_{QP}^{\infty} & \Delta_{PP}^{\infty}
		\end{pmatrix},
    	\label{eq:StationaryCovMatrixMain}
    \end{equation}
    Writing out the components we have
	\begin{align}
		\Delta_{QQ}^{\infty}
		&=
		\frac{
			e^{-3N}\abs{\pi_{k_\sigma}}^2
			+6H\,\Re(\phi_{k_\sigma}^*\pi_{k_\sigma})
			+e^{3N}(9H^2+m^2)\abs{\phi_{k_\sigma}}^2
		}{2m^2\Vphys},\\
		\Delta_{QP}^{\infty}
		&=
		-\frac32 H e^{3N}\abs{\phi_{k_\sigma}}^2,\\
		\Delta_{PP}^{\infty}
		&=
		\frac{\Vphys}{2}
		\left(
			e^{-3N}\abs{\pi_{k_\sigma}}^2
			+m^2e^{3N}\abs{\phi_{k_\sigma}}^2
		\right).
	\end{align}
	Here the value of $N$ is irrelevant since the explicit dependence is inverse to the time dependence of the mode functions. The stable fixed point of this Wigner evolution is Gaussian. A normalised one-mode Gaussian Wigner function with covariance \(\mathbf{\Sigma}(N)\) has the form
	\begin{equation}
		W_G(z;N) = \frac{1}{2\pi\sqrt{\det\mathbf{\Sigma}(N)}} \exp\!\left[ -\frac12(\mathbf{z}-\ev*{\hat{\mathbf{Z}}})^\top\mathbf{\mathbf{\Sigma}}^{-1}(N)(\mathbf{z}-\ev*{\hat{\mathbf{Z}}}) \right].
	\end{equation}
	The purity \(\gamma(N)=\Tr[\rho^2(N)]\) measures how much of the retained one-mode state can still be represented by a single quantum state rather than by a statistical mixture. It equals one for a pure reduced bulk state and decreases as the degree of entanglement with the traced decoupled modes increases. As discussed in \cref{sec:coarsegraining_hamiltonian}, the one-mode purity is invariant under canonical reparametrisations of the retained bulk degree of freedom, so this diagnostic is independent of which canonically-related bulk variables are used. The entangled bulk-boundary block before the trace is displayed in \cref{eq:FullStateBlock01App,eq:FullStateBlock01ExpandedApp}. For a Gaussian Wigner state, the purity is entirely determined by the covariance determinant \(\det\mathbf{\Sigma}(N) = \Delta \hat{Q}^2 \Delta \hat{P}^2 - (\Delta \hat{Q}\hat{P})^2\):
	\begin{equation}
		\gamma(N) = \frac{1}{2\sqrt{\det\mathbf{\Sigma}(N)}} .
		\label{eq:GaussianPurityFromCovMain}
	\end{equation}
	The stationary purity \(\gamma_\infty\) follows by inserting \cref{eq:StationaryCovMatrixMain} into \cref{eq:GaussianPurityFromCovMain}. The same equation also determines how fast the covariance approaches the stationary Gaussian state. The exact covariance solution is
	\begin{equation}
		\mathbf{\Sigma}(N)=\mathbf{\Sigma}_\infty+e^{\mathbf{A}N}\left(\mathbf{\Sigma}_0-\mathbf{\Sigma}_\infty\right)e^{\mathbf{A}^\top N}.
		\label{eq:CovRelaxationMain}
	\end{equation}
	Thus the covariance gap is governed by pairwise sums of the drift eigenvalues
	\begin{equation}
		-\frac32\pm\frac12\sqrt{9-\frac{4m^2}{H^2}} .
		\label{eq:CovGapExponentsMain}
	\end{equation}
	For perturbations around the stationary Gaussian covariance, Taylor expanding \(\gamma(\mathbf{\Sigma})=1/(2\sqrt{\det \mathbf{\Sigma}})\) around \(\mathbf{\Sigma}_\infty\) shows that \(\gamma(N)-\gamma_\infty\) is linearly controlled by \(\mathbf{\Sigma}(N)-\mathbf{\Sigma}_\infty\). The slowest covariance mode therefore fixes the linearised purity relaxation rate,
	\begin{equation}
		\dv{\gamma}{N} \simeq
		\begin{cases}
			-\left[	\dfrac{2m^2}{3H^2}	+	O\!\left(\dfrac{m^4}{H^4}\right)	\right]
			\left(\gamma-\gamma_\infty\right),
			& \dfrac{m}{H}\ll1,
			\\[8pt]
			-\left(3-\sqrt{9-\dfrac{4m^2}{H^2}}\right)
			\left(\gamma-\gamma_\infty\right),
			& 0< \dfrac{m}{H}<\dfrac{3}{2},
			\\[8pt]
			-3\left(\gamma-\gamma_\infty\right),
			& \dfrac{m}{H}\geq\dfrac{3}{2}.
		\end{cases}
		\label{eq:PurityRelaxationRateMain}
	\end{equation}
For light fields, the coarse-grained bulk may thus take a large number of e-folds, of the order $H^2/m^2$, to decohere to a highly mixed state. In contrast, decoherence for heavy fields occurs over order-one e-folds, though to a much lesser final extent than for light fields.

	\begin{figure}[htpb]
	\centering
	\begin{minipage}{0.48\textwidth}
		\centering
		\includegraphics[width=\linewidth]{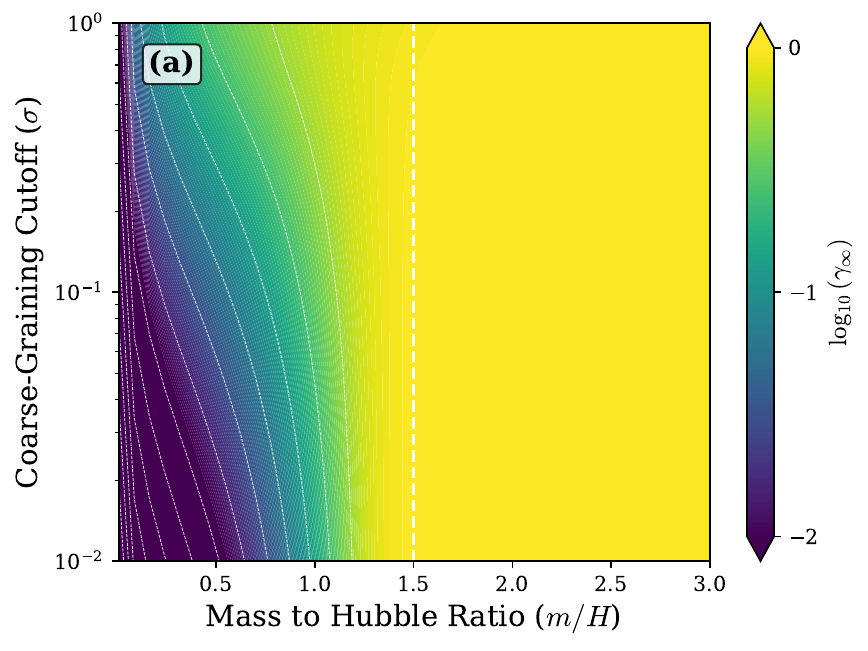}
	\end{minipage}\hfill
	\begin{minipage}{0.48\textwidth}
		\centering
		\includegraphics[width=\linewidth]{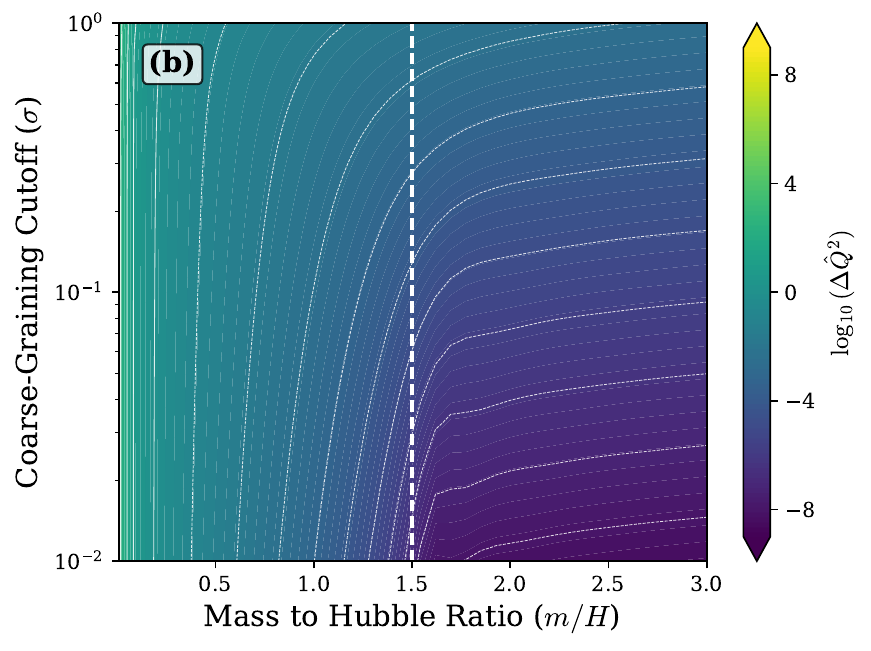}
	\end{minipage}
	\vspace{1em}
	
	\begin{minipage}{0.48\textwidth}
		\centering
		\includegraphics[width=\linewidth]{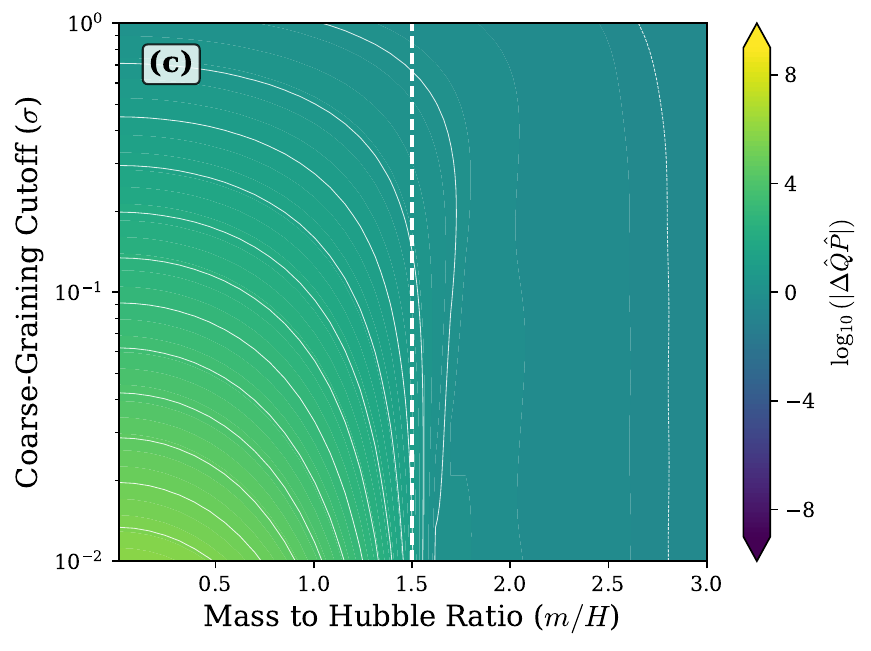}
	\end{minipage}\hfill
	\begin{minipage}{0.48\textwidth}
		\centering
		\includegraphics[width=\linewidth]{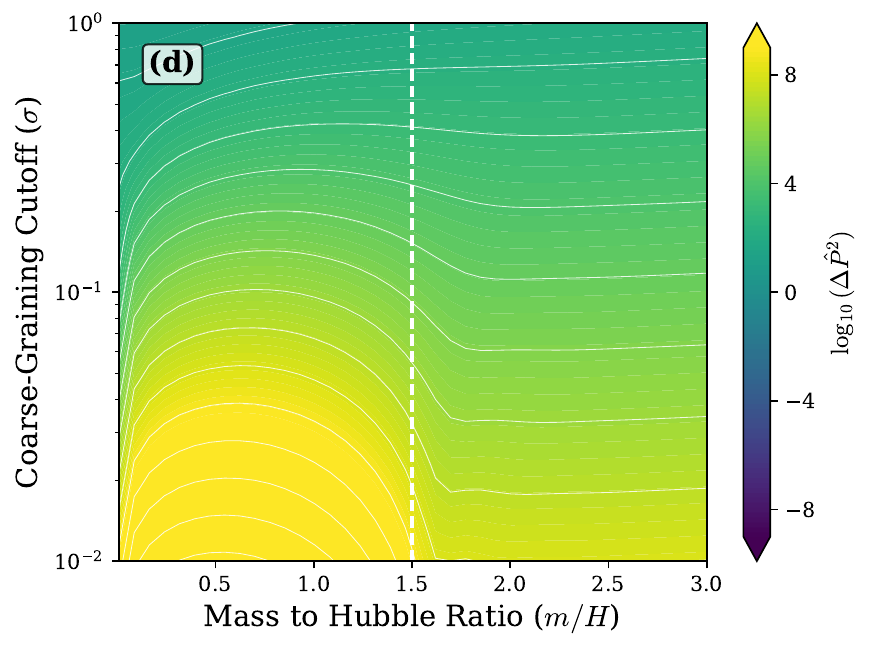}
	\end{minipage}
	\caption{Phase-space diagrams of the Gaussian stationary state derived explicitly in \cref{eq:StationaryCovMatrixMain}. (a) The stationary purity $\gamma_{\infty}$ of the coarse-grained state. (b) Field variance $\Delta \hat{Q}^2$. (c) Magnitude of the cross-covariance $|\Delta \hat{Q}\hat{P}|$. (d) Momentum variance $\Delta \hat{P}^2$. The solid colour contours (isolines) map lines of constant magnitude across the two-dimensional parameter space, while the dashed white line denotes the critical mass threshold $m = 1.5H$. All three covariance elements share the identical logarithmic colour axis to accurately compare their relative divergence magnitudes. In the light-field regime, the variances rapidly diverge as the cutoff $\sigma$ is tightened, resulting in an increasingly mixed state ($\gamma_{\infty} \to 0$).}
	\label{fig:heatmaps}
\end{figure}

Figure~\ref{fig:heatmaps} displays the stationary purity and covariance entries obtained from \cref{eq:StationaryCovMatrixMain}, and provides the map for the mass-regime analysis below. In the light-field regime, decreasing \(\sigma\) increases the field variance and can drive the stationary purity very small; this is the regime where the field marginal admits the Starobinsky stochastic limit. Critical and heavy fields approach finite-purity stationary states with equilibrated covariance and suppressed field diffusion.

\subsection{Light-mass fields}
\label{sec:light_mass_limit}

For a light free spectator field \(0\leq m<3H/2\), the order
\begin{equation}
    \nu=\sqrt{\frac94-\frac{m^2}{H^2}}
\end{equation}
is real and lies in \(0<\nu\leq 3/2\), with the strictly massless limit corresponding to \(\nu=3/2\). The diffusion coefficients of the Wigner generator in \cref{eq:FPGeneralMain} evaluate to
\begin{align}
    D_{QQ}^{\rm light}
    &=
    \frac{H^2\sigma^3}{8\pi}
    \left|H_\nu^{(2)}(\sigma)\right|^2 ,
    \label{eq:DQQLightMain}
    \\
    D_{QP}^{\rm light}
    &=
    \frac{H^2\sigma^3}{16\pi}
    \left[
    H_{\nu-1}^{(1)}(\sigma)H_\nu^{(2)}(\sigma)
    +
    H_\nu^{(1)}(\sigma)H_{\nu-1}^{(2)}(\sigma)
    -
    \frac{3}{\sigma}\left|H_\nu^{(2)}(\sigma)\right|^2
    \right],
    \label{eq:DQPLightMain}
    \\
    D_{PP}^{\rm light}
    &=
    \frac{H^2\sigma^3}{8\pi}
    \left|
    H_{\nu-1}^{(2)}(\sigma)
    -
    \frac{3}{2\sigma}H_\nu^{(2)}(\sigma)
    \right|^2 .
    \label{eq:DPPLightMain}
\end{align}
The corresponding Fokker--Planck equation for the Wigner function \(W(q,p;N)\) can be written in conservative form as
\begin{align}
    \partial_N W
    ={}&
    -\partial_q\!\left(
    \frac{\sigma^3H^2}{6\pi^2}pW
    \right)
    +
    \partial_p\!\left[
    \left(
    3p+\frac{6\pi^2m^2}{\sigma^3H^4}q
    \right)W
    \right]
    \nonumber
    \\
    &+
    \frac12D_{QQ}^{\rm light}\partial_q^2W
    +
    D_{QP}^{\rm light}\partial_q\partial_pW
    +
    \frac12D_{PP}^{\rm light}\partial_p^2W .
    \label{eq:FPLightConservativeMain}
\end{align}
The reduction to a field-only equation proceeds in three steps.
\begin{enumerate}
    \item Integrate the canonical Wigner equation over \(p\) to obtain an exact equation for the field marginal \({\cal P}_Q\), whose current \(J_Q\) is not yet closed.
    \item Use the overdamped hierarchy to solve \(J_Q\) quasistatically, estimating the needed second velocity moment from the fast conditional \(p\)-profile at fixed \(q\).
    \item After this field-only equation is obtained, take the small-mass slow-roll hierarchy that gives the standard Starobinsky limit.
\end{enumerate}
The canonical momentum \(p\) is extensive in the coarse-grained patch. The combination
\begin{equation}
    \frac{p}{H\Vphys}=\frac{\sigma^3H^2 p}{6\pi^2},
\end{equation}
is the intensive velocity that enters the field current. The overdamped reduction below uses the light-field, super-Hubble hierarchy, in which the momentum/current sector relaxes faster than the field marginal. The additional slow-roll expansion in \(m^2/H^2\) is only needed when the resulting field equation is compared with the standard Starobinsky form.

	We now reduce \cref{eq:FPLightConservativeMain} by integrating over the momentum variable following a similar treatment to that given in \cite{pavliotis2014stochastic}. We define the field marginal probability distribution
	\begin{equation}
		{\cal P}_Q(q,N)
		=
		\int_{-\infty}^{+\infty}\dd p\,W(q,p;N),
		\label{eq:FieldMarginalWignerMain}
	\end{equation}
	and the associated field current
	\begin{equation}
		J_Q(q,N)
		=
		\frac{\sigma^3H^2}{6\pi^2}
		\int_{-\infty}^{+\infty}\dd p\,pW(q,p;N).
		\label{eq:FieldCurrentMain}
	\end{equation}
	Integrating \cref{eq:FPLightConservativeMain} over \(p\) gives a closed expression for the evolution of the field marginal probability distribution, provided the boundary terms generated by integrations by parts in \(p\) vanish. This is the case for the Gaussian states considered here, and more generally whenever \(W\), \(\partial_p W\), and \(pW\) decay sufficiently fast as \(|p|\to\infty\). One obtains
	\begin{equation}
		\partial_N{\cal P}_Q
		=
		-\partial_qJ_Q
		+
		\frac12D_{QQ}^{\rm light}\partial_q^2{\cal P}_Q .
		\label{eq:FieldMarginalMain}
	\end{equation}
	The equation is not yet closed because it contains \(J_Q\). To obtain a closed equation for \({\cal P}_Q\), multiply \cref{eq:FPLightConservativeMain} by
	\[
	\frac{\sigma^3H^2}{6\pi^2}p
	\]
	and integrate over \(p\). A direct integration by parts yields
	\begin{align}
		\partial_NJ_Q
		=
		-3J_Q
		-
		\frac{m^2}{H^2}q{\cal P}_Q
		-
		\partial_qM_2
		-
		\frac{\sigma^3H^2}{6\pi^2}
		D_{QP}^{\rm light}\partial_q{\cal P}_Q
		+
		\frac12D_{QQ}^{\rm light}\partial_q^2J_Q \, ,
		\label{eq:FieldCurrentEquationMain}
	\end{align}
    where the second velocity moment
	\begin{equation}
		M_2(q,N)
		=
		\left(\frac{\sigma^3H^2}{6\pi^2}\right)^2
		\int_{-\infty}^{+\infty}\dd p\,p^2W(q,p;N)
		\label{eq:SecondVelocityMomentMain}
	\end{equation}
    has been introduced.
	Equation~\eqref{eq:FieldCurrentEquationMain} is then expanded in the light, super-Hubble regime. The term \(-3J_Q\) is leading, while the slow-roll drift source is \(O(m^2/H^2)\). The diffusion terms retained below scale as \(D_{QQ}^{\rm light}=O(H^2)\), \((\sigma^3H^2)D_{QP}^{\rm light}=O(H^2\sigma^2)\), and \((\sigma^6H^4)D_{PP}^{\rm light}=O(H^2\sigma^4)\), up to numerical factors. Terms with an additional \(N\)-derivative or extra \(q\)-gradients of \(J_Q\) are therefore beyond the leading overdamped closure. We set
	\begin{equation}
		\partial_NJ_Q\simeq0,
		\qquad
		\frac12D_{QQ}^{\rm light}\partial_q^2J_Q\simeq0
	\end{equation}
	at this order.
	It remains to approximate \(M_2\). At leading overdamped order, $q$ changes slowly compared with $p$. We define $G_q(p;N)$ as the solution of \cref{eq:FPLightConservativeMain} at fixed $q$, obtained by dropping all derivatives with respect to $q$. The resulting momentum-space dynamics can be written in the form
    \begin{equation}
    \partial_N G_q = -\partial_p J_P \, ,
    \label{eq:FastMomentumDynamicsMain}
    \end{equation}
 where
    \begin{equation}
    J_P = -\left[\left(3p+\frac{6\pi^2m^2}{\sigma^3H^4}q\right)G_q+\frac12D_{PP}^{\rm light}\partial_pG_q\right].
    \label{eq:MomentumCurrentMain}
    \end{equation}
    The stationary profile with no flux through $|p|=\infty$ satisfies $J_P=0$, from which we obtain the normalised Gaussian solution
	\begin{equation}
		G_q(p)
		=
		\sqrt{\frac{3}{\pi D_{PP}^{\rm light}}}
		\exp\!\left[
		-\frac{3}{D_{PP}^{\rm light}}
		\left(
		p+\frac{2\pi^2m^2}{\sigma^3H^4}q
		\right)^2
		\right],
		\qquad
		\int\dd p\,G_q(p)=1 .
		\label{eq:FastMomentumKernelMain}
	\end{equation}
	In the quasistatic approximation the momentum dependence rapidly approaches this Gaussian profile, \(W(q,p;N)\simeq{\cal P}_Q(q,N)G_q(p)\). Computing \(M_2\) then reduces to Gaussian moments,
	\begin{equation}
		\int \dd p\,pG_q(p)
		=
		-\frac{2\pi^2m^2}{\sigma^3H^4}q ,
		\label{eq:FastMomentumFirstMomentMain}
	\end{equation}
	and
	\begin{equation}
		\int \dd p\,p^2G_q(p)
		=
		\frac{4\pi^4m^4}{\sigma^6H^8}q^2
		+
		\frac{D_{PP}^{\rm light}}{6}\, .
		\label{eq:FastMomentumSecondMomentMain}
	\end{equation}
	The projected second velocity moment is therefore
	\begin{equation}
		M_2
		=
		\left(
		\frac{m^4}{9H^4}q^2
		+
		\frac{\sigma^6H^4}{216\pi^4}D_{PP}^{\rm light}
		\right)
		{\cal P}_Q .
		\label{eq:ProjectedSecondVelocityMomentMain}
	\end{equation}
	Keeping both terms in \cref{eq:ProjectedSecondVelocityMomentMain} gives
	\begin{equation}
		J_Q
		=
		-\frac{m^2}{3H^2}
		\left(
		1+\frac{2m^2}{9H^2}
		\right)
		q{\cal P}_Q
		-
		\left(
		\frac{m^4}{27H^4}q^2
		+
		\frac{\sigma^3H^2}{18\pi^2}D_{QP}^{\rm light}
		+
		\frac{\sigma^6H^4}{648\pi^4}D_{PP}^{\rm light}
		\right)
		\partial_q{\cal P}_Q
		+
		O\!\left(\text{higher gradients}\right).
		\label{eq:LightCurrentQuasistaticMain}
	\end{equation}
	Inserting this current into the exact marginal equation \cref{eq:FieldMarginalMain} yields
	\begin{align}
		\partial_N{\cal P}_Q(q,N)
		=
		\frac{m^2}{3H^2}
		\left(
		1+\frac{2m^2}{9H^2}
		\right)
		\partial_q\!\left(q{\cal P}_Q\right)
		+
		\partial_q\!\left[
		\left(
		\frac{m^4}{27H^4}q^2
		+
		D_Q^{\rm eff}
		\right)
		\partial_q{\cal P}_Q
		\right]
		+
		O\!\left(\text{higher gradients}\right),
		\label{eq:LightFieldMarginalGeneralMain}
	\end{align}
	with
	\begin{equation}
		D_Q^{\rm eff}
		=
		\frac12D_{QQ}^{\rm light}
		+
		\frac{\sigma^3H^2}{18\pi^2}D_{QP}^{\rm light}
		+
		\frac{\sigma^6H^4}{648\pi^4}D_{PP}^{\rm light}\, .
		\label{eq:DQEffectiveLightMain}
	\end{equation}
	For a light but non-zero mass, the index admits the expansion
    \begin{equation}
    \nu=\frac32-\frac{m^2}{3H^2}+O\!\left(\frac{m^4}{H^4}\right).
    \label{eq:LightNuExpansionMain}
    \end{equation}
    The small-\(\sigma\) expansion contains factors of the form \(\sigma^{3-2\nu}\). Expanding these factors in \(m^2/H^2\) generates logarithms of \(\sigma\):
    \begin{equation}
    \sigma^{3-2\nu}=\sigma^{\frac{2m^2}{3H^2}+O(m^4/H^4)}=1+\frac{2m^2}{3H^2}\ln\sigma+\cdots .
    \label{eq:LightSigmaNonUniformMain}
    \end{equation}
    Therefore the light-mass expansion of the effective diffusion coefficient is controlled by
	\begin{equation}
		\frac{m^2}{H^2}\ll1,
		\qquad
		\sigma\ll1,
		\qquad
		\frac{m^2}{H^2}|\ln\sigma|\ll1.
		\label{eq:LightMassExpansionConditionsMain}
	\end{equation}
	Thus, under the controlled light-field expansion hierarchy in \cref{eq:LightMassExpansionConditionsMain}, the drift is kept at order \(m^2/H^2\), while the \(m^4/H^4\) terms in \cref{eq:LightFieldMarginalGeneralMain} are dropped and the diffusion coefficient is evaluated at leading order in \(m^2/H^2\). At the same order,
	\begin{equation}
		M_2
		=
		\frac{\sigma^6H^4}{216\pi^4}D_{PP}^{\rm light}{\cal P}_Q
		+
		O\!\left(\frac{m^4}{H^4}\right).
		\label{eq:ProjectedSecondVelocityMomentLeadingMain}
	\end{equation}
	For light fields near the slow-roll regime, \(D_{PP}^{\rm light}\sim\sigma^{-2}\), so the velocity-width contribution scales as
	\begin{equation}
		\frac{\sigma^6H^4}{216\pi^4}D_{PP}^{\rm light}
		=
		O(\sigma^4).
		\label{eq:VelocityWidthScalingMain}
	\end{equation}
	The canonical momentum width need not vanish; the super-Hubble suppression applies to the intensive velocity entering the current.
	This gives
	\begin{equation}
		D_Q^{\rm eff}
		=
		\frac{H^2}{8\pi^2}
		+
		O\!\left(
		\sigma^2,
		\frac{m^2}{H^2}\ln\sigma,
		\frac{m^2}{H^2}
		\right).
		\label{eq:DQEffectiveLightExpansionMain}
	\end{equation}
	Combining \cref{eq:LightFieldMarginalGeneralMain,eq:DQEffectiveLightExpansionMain}, the leading light-mass, super-Hubble field equation is
	\begin{align}
		\partial_N{\cal P}_Q(q,N)
		=
		\frac{m^2}{3H^2}\partial_q\!\left(q{\cal P}_Q\right)
		+
		D_Q^{\rm eff}\partial_q^2{\cal P}_Q
		+
		O\!\left(
		\text{higher gradients},
		\frac{m^4}{H^4}
		\right),
		\label{eq:StarobinskyDiffusionFromReductionMain}
	\end{align}
	and therefore
	\begin{equation}
		\partial_N{\cal P}_Q(q,N)
		=
		\frac{m^2}{3H^2}
		\partial_q \left[q{\cal P}_Q(q,N)\right]
		+
		\frac{H^2}{8\pi^2}
		\partial_q^2{\cal P}_Q(q,N).
		\label{eq:StarobinskyLightMassFPMain}
	\end{equation}
    Identifying $q$ with the coarse-grained field variable, \cref{eq:StarobinskyLightMassFPMain} reproduces the Starobinsky Fokker--Planck one dimensional equation~\eqref{eq:StarobinskyFPIntro} on the slow-roll attractor.

	\subsection{Comparison with previous works}
	\label{sec:similar_works_comparison}

In the massless limit, the Lindblad operator \cref{eq:LindbladTimeIndep} reduces to
\begin{equation}
	\hat{L}_\chi^{m\to 0} = e^{i(\chi-\sigma)}\!\left(\frac{3\pi}{\sigma H}\hat{Q} - \frac{H}{2\pi}(1-i\sigma)\hat{P}\right) ,
	\label{eq:LMasslessLimit}
\end{equation}
which is in contrast with the single Lindblad operator ($\Pi_s$) found in Refs.~\cite{Li2026Stochastic,Li2026lwl}. The phase-space equation obtained in these works in the same massless limit,
	\begin{equation}
		\partial_N W(\phi,\pi,N) = \left( - \frac{\pi}{e^{3 N}H}\partial_\phi + \frac{H^2}{8\pi^2}\partial_\phi^2 \right)W ,
	\end{equation}
	reproduces the same field-only Starobinsky diffusion coefficient as \cref{eq:StarobinskyDiffusionFromReductionMain}. The distinction is that the infrared momentum density \(\Pi_s\) is not the canonical partner of the averaged field, since
	\begin{equation}
		P_N=\VF(N)\Pi_s, \qquad [Q_N,P_N]=i, \qquad [Q_N,\Pi_s]=\frac{i}{\VF(N)} .
	\end{equation}
	Thus the momentum-density variables are adequate for the leading field-only limit, but they are not canonical variables for a coarse-grained patch and are not a harmless convention when reconstructing the quantum Wigner or Lindblad equation; this is also the normalisation issue present in v1 of our own arXiv preprint \cite{Christie:2025knc}, corrected in v2 and the published version, as well as here, by using the extensive momentum \(P_N=\VF\Pi_s\).
	\subsection{Critical and Heavy Fields}
	\label{sec:massive_limits}

	The comparison with the massless/light stochastic-field limit is controlled by two diagnostics: whether the field marginal retains an unsuppressed random-walk channel, and whether the reduced phase-space state loses purity. In the critical and heavy regimes, as we shall now see, the \(D_{QQ}\) term and the velocity-weighted \(D_{QP}\) and \(D_{PP}\) feed-through terms vanish as \(\sigma\to0\), so the field random-walk channel is suppressed, while the full \((Q,P)\) state relaxes to a finite-purity stationary Gaussian.

	\subsubsection{Critical field}

	At \(m=3H/2\), one has \(\nu=0\). The Lindblad operator becomes
	\begin{equation}
		\hat L_{\rm crit}=e^{-i\pi/4}\left[\frac{3\pi^{3/2}}{\sqrt2\,\sigma^{3/2}H} \left(\sigma H_{-1}^{(2)}(\sigma)+\frac32H_0^{(2)}(\sigma)\right)\hat Q +\frac{H\sigma^{3/2}}{2\sqrt{2\pi}}H_0^{(2)}(\sigma)\hat P\right].
	\end{equation}
	The corresponding Fokker--Planck equation for the Wigner function is
	\begin{align}
		\partial_N W_{\rm crit}
		={}&
		-\partial_q\!\left(
		\frac{\sigma^3H^2}{6\pi^2}pW_{\rm crit}
		\right)
		+
		\partial_p\!\left[
		\left(
		3p+\frac{27\pi^2}{2\sigma^3H^2}q
		\right)W_{\rm crit}
		\right]
		\nonumber
		\\
		&+\frac12D^{\rm crit}_{QQ}\partial_q^2W_{\rm crit}
		+D^{\rm crit}_{QP}\partial_q\partial_pW_{\rm crit}
		+\frac12D^{\rm crit}_{PP}\partial_p^2W_{\rm crit}.
		\label{eq:FPCritMain}
	\end{align}
	Using \(H_{-1}^{(2)}=-H_1^{(2)}\), and for \(\sigma\ll1\),
	\begin{equation}
		H_0^{(2)}(\sigma)\simeq 1-i\frac{2}{\pi}\left[\ln\left(\frac{\sigma}{2}\right)+\upgamma_{\rm E}\right], \qquad H_1^{(2)}(\sigma)\simeq \frac{\sigma}{2}+i\frac{2}{\pi\sigma},
	\end{equation}
	and denoting the Euler--Mascheroni constant by \(\upgamma_{\rm E}\), the final super-Hubble limit gives
	\begin{align}
		D_{QQ}^{\rm crit}
		&=
		\frac{H^2\sigma^3}{8\pi}
		\left[
		1+\frac{4}{\pi^2}
		\left(\ln\frac{\sigma}{2}+\upgamma_{\rm E}\right)^2
		\right]
		\to 0\, ,
		\label{eq:DQQcrit}
		\\
		D_{QP}^{\rm crit}
		&=
		-\frac{3\pi}{4}
		\left[
		\frac32
		+\frac{2}{\pi^2}
		\left(\ln\frac{\sigma}{2}+\upgamma_{\rm E}\right)
		\left(2+3\ln\frac{\sigma}{2}+3\upgamma_{\rm E}\right)
		\right],
        \label{eq:DQPcrit}
		\\
		D_{PP}^{\rm crit}
		&=
		\frac{9\pi^3}{2H^2\sigma^3}
		\left[
		\frac94
		+\frac{1}{\pi^2}
		\left(2+3\ln\frac{\sigma}{2}+3\upgamma_{\rm E}\right)^2
		\right].
		\label{eq:CriticalDiffusionSuperHubbleMain}
	\end{align}
	The Wronskian normalisation $\mathrm{Im}(\phi^*_{k_\sigma}\pi_{k_\sigma})=-1/2$ gives the mass-independent identity
	\begin{equation}
		D_{QQ}^{\rm crit}D_{PP}^{\rm crit}
		-\left(D_{QP}^{\rm crit}\right)^2
		=
		\frac94\, ,
		\label{eq:CriticalDiffusionInvariantMain}
	\end{equation}
	Thus the dissipator remains active as a squeezed phase-space noise source; field diffusion vanishes, while the momentum diffusion and \(QP\) correlation compensate so that the one-channel quantum area is fixed. The connection between the Lindblad operator and the Wigner diffusion matrix is summarised in \cref{app:diagonal}. The deterministic part is given by the shared drift equations~\eqref{eq:massiveDriftBoth}; at \(m=3H/2\), \cref{eq:dampedOscillator} is critically damped. The boundary channel still injects vacuum fluctuations, but its \(D_{QQ}\) entry, \cref{eq:DQQcrit}, vanishes in the super-Hubble cutoff limit $\sigma\to 0$; the \(D_{QP}^{\rm crit}\) and \(D_{PP}^{\rm crit}\) feed-through terms carry the field-current factors \(\sigma^3H^2\) and \(\sigma^6H^4\), respectively, and are suppressed as well, so no unsuppressed field random-walk channel emerges. The explicit super-Hubble critical Fokker--Planck operator is \cref{eq:FPCritMain}. Its stationary covariance is
	\begin{equation}
		\mathbf{\Sigma}_\infty^{\rm crit}
		=
		\begin{pmatrix}
			\frac56D_{QQ}^{\rm crit}
			+\frac{4D_{QP}^{\rm crit}}{9H\Vphys}
			+\frac{2D_{PP}^{\rm crit}}{27H^2\Vphys^2}
			&
			-\frac12H\Vphys D_{QQ}^{\rm crit}
			\\[8pt]
			-\frac12H\Vphys D_{QQ}^{\rm crit}
			&
			\frac16D_{PP}^{\rm crit}
			+\frac38H^2\Vphys^2D_{QQ}^{\rm crit}
		\end{pmatrix},
		\label{eq:CriticalStationaryCovMain}
	\end{equation}
	and
	\begin{equation}
		\gamma_\infty^{\rm crit}
		=
		\frac{1}{2\sqrt{\det\mathbf{\Sigma}_\infty^{\rm crit}}}>0 .
		\label{eq:CriticalPurityMain}
	\end{equation}
	The purity in the deep super Hubble limit can be calculated explicitly as
	\begin{equation}
		\lim_{\sigma\to0^+}\gamma_\infty^{\rm crit}
		=
		\frac{3\pi}{\sqrt{4+9\pi^2}} = 0.978...
		\label{eq:CriticalPurityLimitMain}
	\end{equation}
	The first correction contains no term linear in \(\sigma\):
	\begin{equation}
		\gamma_\infty^{\rm crit}(\sigma)
		=
		\frac{3\pi}{\sqrt{4+9\pi^2}}
		+
		\frac{12\pi\sigma^2}{(4+9\pi^2)^{3/2}}
		\left(
		\ln\frac{\sigma}{2}
		+\upgamma_{\rm E}
		-\frac12
		\right)
		+O(\sigma^4\ln^2\sigma).
		\label{eq:CriticalPurityFirstCorrectionMain}
	\end{equation}
	Although individual covariance entries scale with \(\sigma\), the determinant remains finite and non-zero. The critical bulk therefore has a stable damped-oscillator covariance instead of the purity-losing field spreading of the massless limit.

	\subsubsection{Heavy fields}

	For heavy fields, \(\nu=i\mu\), with
	\begin{equation}
		\mu=\sqrt{\frac{m^2}{H^2}-\frac94}.
	\end{equation}
	The shared deterministic equation~\eqref{eq:dampedOscillator} is then underdamped, with
	\begin{equation}
		\ev*{\hat Q}(N)=e^{-3N/2} \left[ C_1\cos(\mu N)+C_2\sin(\mu N) \right].
	\end{equation}
	This underdamped phase-space rotation is the qualitative feature that is absent at the critical point. The dissipator still has the same one-channel form, but the heavy shell coefficients now oscillate with \(\ln\sigma\). The corresponding Fokker--Planck equation for the Wigner function is
	\begin{align}
		\partial_N W_{\rm heavy}
		={}&
		-\partial_q\!\left(
		\frac{\sigma^3H^2}{6\pi^2}pW_{\rm heavy}
		\right)
		+
		\partial_p\!\left\{
		\left[
		3p+\frac{6\pi^2}{\sigma^3H^2}
		\left(\frac94+\mu^2\right)q
		\right]W_{\rm heavy}
		\right\}
		\nonumber
		\\
		&+\frac12D^{\rm heavy}_{QQ}\partial_q^2W_{\rm heavy}
		+D^{\rm heavy}_{QP}\partial_q\partial_pW_{\rm heavy}
		+\frac12D^{\rm heavy}_{PP}\partial_p^2W_{\rm heavy}.
		\label{eq:FPHeavyMain}
	\end{align}
	For fixed \(\mu\), the small-\(\sigma\) expansion has the form
	\begin{equation}
		H_{i\mu}^{(2)}(\sigma)
		=
		\frac{(\sigma/2)^{-i\mu}}{\sinh(\pi\mu)\Gamma(1-i\mu)}
		-\frac{e^{-\pi\mu}(\sigma/2)^{i\mu}}{\sinh(\pi\mu)\Gamma(1+i\mu)}
		+O(\sigma^2).
	\end{equation}
	Writing \(\theta_\mu(\sigma)=\arg\Gamma(1+i\mu)-\mu\ln(\sigma/2)\), the super-Hubble diffusion coefficients become
	\begin{align}
		D_{QQ}^{\rm heavy}
		&=
		\frac{H^2\sigma^3}{4\pi^2\mu\sinh(\pi\mu)}
		\left[
		\cosh(\pi\mu)-\cos(2\theta_\mu)
		\right]
		+O(H^2\sigma^5),
		\label{eq:DQQheavy}
		\\
		D_{QP}^{\rm heavy}
		&=
		-\frac{3}{2\mu\sinh(\pi\mu)}
		\left\{
		\frac32\left[\cosh(\pi\mu)-\cos(2\theta_\mu)\right]
		-\mu\sin(2\theta_\mu)
		\right\}
		+O(\sigma^2),
		\nonumber\\
		D_{PP}^{\rm heavy}
		&=
		\frac{9\pi^2}{H^2\sigma^3\mu\sinh(\pi\mu)}
		\left[
		\left(\frac94+\mu^2\right)\cosh(\pi\mu)
		-\left(\frac94-\mu^2\right)\cos(2\theta_\mu)
		-3\mu\sin(2\theta_\mu)
		\right]
		+O(H^{-2}\sigma^{-1}).
		\label{eq:HeavyDiffusionSuperHubbleMain}
	\end{align}
	The \(D_{QQ}\) entry therefore still vanishes as \(\sigma^3\), now with a bounded log-periodic prefactor. The mixed and momentum entries remain tied to it by the same one-channel Wronskian structure as in \cref{eq:CriticalDiffusionInvariantMain}; after the velocity factors entering \(J_Q\) are included, their field-current contributions are also \(O(H^2\sigma^3)\). The new heavy-field ingredient is the underdamped rotation of the retained \((Q,P)\) oscillator together with these log-periodic shell coefficients. Thus heavy-field \(D_{QQ}\) and its momentum feed-through terms do not support an unsuppressed field random walk, while the deterministic dynamics remain those of a damped oscillator.

	At fixed super-Hubble cutoff, the subsequent large-mass limit \(\mu\simeq m/H\to\infty\) gives
	\begin{equation}
		D_{QQ}^{\rm heavy}
		\sim
		\frac{H^2\sigma^3}{4\pi^2\mu},
		\qquad
		D_{QP}^{\rm heavy}
		\sim
		-\frac{9}{4\mu},
		\qquad
		D_{PP}^{\rm heavy}
		\sim
		\frac{9\pi^2\mu}{H^2\sigma^3}.
		\label{eq:HeavyDiffusionLargeMassMain}
	\end{equation}
	In this ordered limit the field noise and mixed noise disappear, while the momentum noise grows with the oscillator frequency. The stationary purity has the fixed-\(\mu\), super-Hubble expansion
	\begin{equation}
		\gamma_\infty^{\rm heavy}(\sigma)
		=
		\left[
		1+
		\frac{4\mu^2}
		{(9+4\mu^2)\sinh^2(\pi\mu)}
		\right]^{-1/2}
		+O(\sigma^2).
		\label{eq:HeavyPurityMain}
	\end{equation}
	The log-periodic terms in the individual diffusion coefficients cancel from this leading phase-space area. As \(\mu\to0^+\), \cref{eq:HeavyPurityMain} matches the critical value in \cref{eq:CriticalPurityLimitMain}; as \(\mu\simeq m/H\to\infty\),
	\begin{equation}
		\gamma_\infty^{\rm heavy}
		=
		1-2e^{-2\pi\mu}
		+O(e^{-4\pi\mu})
		+O(\sigma^2).
	\end{equation}
	The heavy stationary state therefore becomes exponentially close to a pure damped oscillator, rather than a classical field random walk.

	The super-Hubble behaviour of the \(D_{QQ}\) across the massless, light, critical and heavy regimes is summarised in Table~\ref{tab:diffusionScaling}. The massless and light-field rows refer back to \cref{sec:light_mass_limit}, while the remaining rows summarise the massive regimes of this section. The table isolates the field-diffusion entry relevant for comparison with a field-only random-walk limit; though the current analysis accounts for the \(D_{QP}\) and \(D_{PP}\) feed-through terms. The full fluctuation/dissipation structure remains given by a rank-one Kossakowski matrix.

	\begin{table}[t]
		\centering \small
		\begin{tabularx}{\textwidth}{>{\raggedright\arraybackslash}p{0.22\textwidth}>{\raggedright\arraybackslash}p{0.36\textwidth}>{\raggedright\arraybackslash}X}
			\hline Mass range & Small-\(\sigma\) field-diffusion scaling & Consequence \\ \hline \(0<m<3H/2\), \(\nu=\sqrt{9/4-m^2/H^2}\) & \(D_{QQ}^{\rm light}\sim\dfrac{H^2 2^{2\nu-3}\Gamma(\nu)^2}{\pi^3}\sigma^{3-2\nu}\) & \(D_{QQ}\) is suppressed at fixed non-zero light mass; the free covariance has a finite stationary state \\ \hline Massless limit & \(D_{QQ}^{m=0}=H^2/(4\pi^2)+O(\sigma^2)\) & Finite Wigner field diffusion; no finite stationary covariance, and \(\gamma_\infty^{m=0}=0\) \\ \hline \(m=3H/2\) & \(D_{QQ}^{\rm crit}=O(H^2\sigma^3\ln^2\sigma)\) & \(D_{QQ}\) is suppressed and \(\gamma_\infty^{\rm crit}>0\) \\ \hline \(m>3H/2\), \(\mu=\sqrt{m^2/H^2-9/4}\) & \(D_{QQ}^{\rm heavy}=\dfrac{H^2\sigma^3}{8\pi}e^{\pi\mu}|H_{i\mu}^{(2)}(\sigma)|^2\sim O(H^2\sigma^3)\) at fixed \(\mu\), with oscillations in \(\ln\sigma\) & Finite-purity quantum damped oscillator; no field random-walk limit \\ \hline
		\end{tabularx}
		\caption{Small-cutoff field-diffusion scaling for the direct comparison with Starobinsky diffusion across the regimes. The massless and light rows refer to \cref{sec:light_mass_limit}; the full fluctuation/dissipation structure is the rank-one \((Q,P)\) Kossakowski matrix.}
		\label{tab:diffusionScaling}
	\end{table}

	The Wigner and purity diagnostics now make the separation explicit. The massless and controlled light-field overdamped reductions give the classical stochastic-field marginal. Critical and heavy fields instead have no unsuppressed field random-walk channel after the momentum feed-through terms are included, no purity loss to \(\gamma=0\), and a retained quantum damped oscillator driven by the single boundary channel.

	\section{Summary}
	\label{sec:Summary}

	We have developed a canonical, quantum open-system derivation of stochastic inflation for a free spectator scalar in de Sitter space, using a sharp physical momentum cutoff. The retained system is the homogeneous canonical pair \((Q_N,P_N)\): the averaged field over the effective patch and the total momentum in that patch. The sharp \(k\)-space window defines an effective time-dependent comoving volume in real space, which is required to normalise the equal-time commutator.

	As the cutoff moves, Bunch-Davies modes enter the infrared sector. The homogeneous average couples only to one collective boundary mode, and the moving-basis redefinition of the bulk and boundary mode produces a redefined bulk entangled with a decoupled mode that is traced over. The resulting GKLS dynamics are generated by an effective Hamiltonian and a single non-Hermitian Lindblad operator demonstrating that for a free spectator scalar in de Sitter space diffusion and Hubble friction originate from the same quantum channel. We also provide several schemes under which one can unravel the GKLS dynamics into stochastic pure state dynamics as continuous measurement models of the decoupled mode, making contact with Langevin formulations of stochastic inflation.

	Taking the Wigner-Weyl transform of the GKLS dynamics provides a Fokker--Planck equation for the Wigner quasiprobabilty distribution. In the light field limit, one can perform an overdamped reduction by integrating out the momentum variable with the Wigner marginal distribution obeying the Starobinsky Fokker-Planck equation. Critical mass $m=3H/2$ and heavier fields are not subject to strong diffusive dynamics as \(\sigma\to0\); instead the retained \((Q,P)\) state is an underdamped oscillator with finite stationary purity. At criticality the purity approaches the high value of \(0.978\) with no linear \(\sigma\) correction. In the heavy regime the leading log-periodic shell factors cancel from the purity, which tends to one as \(m/H\) grows.

	For the free quadratic spectator theory considered here, the Wigner transform closes exactly into a second-order Fokker--Planck equation. A natural next step is to ask whether the same canonical coarse-graining framework can be extended to the inflaton and to genuinely interacting fields. In that setting, self-interactions and gravitational constraints are expected to generate higher Moyal derivatives in the Wigner-function dynamics that are not captured in the Fokker-Planck equation of stochastic inflation, together with non-Gaussian correlations and non-Markovian effects. These additional phenomena could modify the standard stochastic description of inflation, with possible consequences for the statistics of primordial fluctuations, rare-event tails, primordial black-hole production, and the size and shape of inflationary non-Gaussianities.

	\section{Code availability}

	The source files and supporting code for this work are hosted at \url{https://github.com/rchristie95/QuantumStochasticInflation}.

	\section{Acknowledgements}

	We thank Sebastian Cespedes, Thomas Colas and Yue-Zhou Li for useful discussions.	For open access, the authors have applied a Creative Commons Attribution (CC-BY) licence to any Author Accepted Manuscript version arising from this work.

	\bibliographystyle{JHEP} \bibliography{StochasticInflationBib} \newpage \appendix
	\section{Derivation of Stochastic-Inflation GKLS Dynamics}
	\label{app:bulk_boundary_derivation}

	This appendix is a technical and self-contained derivation of the effective one-mode Hamiltonian for the bulk part of the bulk-boundary dynamics and the one-channel dissipator used in Sections~\ref{sec:coarsegraining_hamiltonian} and \ref{sec:boundary_fluctuations}. The moving-cutoff step is formulated as a redefinition process over an infinitesimal interval \(\dd N\). At $e$-fold \(N\), the infrared field lives in the Fock space
	\begin{equation}
		\cF_N = \cF_{|\bm k|<k_\sigma(N)}
		=\cF_N^\parallel\otimes\cF_N^\perp .
	\end{equation}
	Here \(\cF_N^\parallel\) is the homogeneous one-mode Fock space on which \(\hat Q_N,\hat P_N\) act, while \(\cF_N^\perp\) collects the orthogonal infrared combinations. The density matrix \(\hat\rho_N\) used in the reduced dynamics is the restriction to \(\cF_N^\parallel\). After the cutoff moves from \(N\) to \(N+\dd N\), the enlarged infrared Fock space decomposes as
	\begin{equation}
		\cF_{N+\dd N} \simeq \cF_N\otimes\cF_\Delta,\qquad
		\cF_{N+\dd N}=\cF_{N+\dd N}^\parallel\otimes\cF_{N+\dd N}^\perp,\qquad
		\cF_\Delta=\cF_\Delta^\parallel\otimes\cF_\Delta^\perp ,
	\end{equation}
	where \(\cF_\Delta\) is the Fock space of the incoming boundary shell
	\begin{equation}
		k_\sigma(N)<|\bm k|<k_\sigma(N+\dd N).
	\end{equation}
	The Fock spaces \(\cF_N\otimes\cF_\Delta\) and \(\cF_{N+\dd N}\) are isomorphic, but the effective map involves only the incoming bulk and incoming boundary-shell modes. For this map we use the redefined-bulk/decoupled-mode decomposition
	\begin{equation}
		\cF_N^\parallel\otimes\cF_\Delta^\parallel
		\simeq \cF_{N+\dd N}^\parallel\otimes\cF^{\rm d,\,\perp}_{N+\dd N},
		\qquad \cF^{\rm d,\,\perp}_{N+\dd N}\subset\cF_{N+\dd N}^\perp .
		\label{eq:ParallelOutputDecompositionApp}
	\end{equation}
	The entangling unitary can correlate the redefined bulk and decoupled mode, and the reduced dynamics traces over \(\cF^{\rm d,\,\perp}_{N+\dd N}\). The notation used in this appendix is summarised in Table~\ref{tab:bulk-boundary-notation}, and the corresponding construction is shown schematically in Figure~\ref{fig:bulk_boundary}.

	\begin{table}[ht]
		\centering \small \renewcommand{\arraystretch}{1.25}
		\begin{tabularx}{\textwidth}{
				>{\raggedright\arraybackslash}p{0.27\textwidth} >{\raggedright\arraybackslash}X >{\raggedright\arraybackslash}p{0.30\textwidth}} \hline
			Symbol & Meaning & Acts on / lives on \\ \hline
			\(\cF_N^\parallel,\cF_N^\perp\) & homogeneous one-mode Fock space and its orthogonal complement before the cutoff moves & \(\cF_N=\cF_N^\parallel\otimes\cF_N^\perp\) \\ \hdashline
			\(\hat Q_N,\hat P_N,\hat\rho_N\) & coarse-grained canonical operators and reduced state & homogeneous mode \(\cF_N^\parallel\) \\ \hdashline
			\(d\hat b_\Delta,d\hat q_\Delta,d\hat p_\Delta\) & collective increments of the incoming boundary shell & incoming boundary-shell mode \(\cF_\Delta^\parallel\) \\ \hdashline
			\(\cF_N^\parallel\otimes\cF_\Delta^\parallel\) & product active input before the redefinition & pre-map homogeneous-plus-shell space \\ \hdashline
			\(\cF_{N+\dd N}^\parallel\) & redefined bulk mode & \(\cF_{N+\dd N}\) \\ \hdashline
			\(\cF^{\rm d,\,\perp}_{N+\dd N}\) & decoupled mode carrying any retained-mode correlations & \(\cF^{\rm d,\,\perp}_{N+\dd N}\subset\cF_{N+\dd N}^\perp\) \\ \hdashline
			\(\hat K_{\rm eff}\) & effective Hamiltonian generating deterministic homogeneous drift & \(\cF_N^\parallel\), lifted with \(\openone_{\cF_\Delta^\parallel}\) \\ \hdashline
			\(\hat U_{N+\dd N}=\exp(\hat S_{N+\dd N})\) & unitary implementing the bulk-boundary entangling in the redefined-bulk/decoupled-mode decomposition & \(\cF_N^\parallel\otimes\cF_\Delta^\parallel \to \cF_{N+\dd N}^\parallel\otimes\cF^{\rm d,\,\perp}_{N+\dd N}\) \\ \hdashline
			\(\hat\rho_{N+\dd N}^{\parallel\perp}\) & full active state before tracing the decoupled mode & \(\cF_{N+\dd N}^\parallel\otimes\cF^{\rm d,\,\perp}_{N+\dd N}\) \\ \hdashline
			\(\hat\rho_{N+\dd N}\) & reduced homogeneous state after the decoupled-mode trace & \(\cF_{N+\dd N}^\parallel\) \\ \hline
		\end{tabularx}
		\caption{\small Summary of the bulk-boundary notation used in the derivation.}
		\label{tab:bulk-boundary-notation}
	\end{table}

	We organise this appendix as follows:
	\begin{itemize}
		\item \ref{appA:bulkvars} defines the canonical variables \(\hat Q_N\) and \(\hat P_N\) on the bulk and shows their canonical commutation relation is preserved at all times, \([\hat Q_N,\hat P_N]=i\). \item \ref{appA:Bulk} derives the bulk contribution to the bulk-variable evolution arising from the Hamiltonian and shrinking comoving domain. \item \ref{appA:boundary} calculates the fluctuation relations between operators on the bulk and on the boundary. \item \ref{appA:Unitary} finds the unique unitary map acting on the bulk and boundary fields, under the physical conditions stated there, which is consistent with the dynamics derived in \ref{appA:Bulk} and \ref{appA:boundary}. \item \ref{appA:tracing} derives GKLS dynamics for the reduced bulk variables by continually partial tracing over the decoupled mode.
	\end{itemize}

	\subsection{Canonical Bulk Variables}
	\label{appA:bulkvars}

	The sharp infrared projector defining the bulk is
	\begin{equation}
		\Theta_\sigma(N,\bm k) = \Theta\!\left(k_\sigma(N)-|\bm k|\right), \qquad k_\sigma(N)=\sigma e^N H .
	\end{equation}
	In the one-step decomposition used below, this sector is the full bulk \(\cF_N=\cF_N^\parallel\otimes\cF_N^\perp\). The canonical averages below generate its homogeneous mode \(\cF_N^\parallel\). The next bulk-boundary system is obtained by adding the boundary shell \(\cF_\Delta=\cF_\Delta^\parallel\otimes\cF_\Delta^\perp\). The corresponding real-space window is
	\begin{equation}
		W(N,\bm x) = \int \frac{\dd^3k}{(2\pi)^3} \Theta_\sigma(N,\bm k)e^{i\bm k\cdot\bm x} = \frac{k_\sigma^3}{2\pi^2} \frac{j_1(k_\sigma|\bm x|)}{k_\sigma|\bm x|},
	\end{equation}
	where
	\begin{equation}
		j_1(z)=\frac{\sin z}{z^2}-\frac{\cos z}{z}.
	\end{equation}
	At the origin,
	\begin{equation}
		W(N,0)=\frac{k_\sigma^3(N)}{6\pi^2}.
	\end{equation}
	The effective comoving bulk volume is
	\begin{equation}
		\VF(N) = \frac{1}{W(N,0)} = \frac{6\pi^2}{k_\sigma^3(N)} .
	\end{equation}
	Since \(k_\sigma(N)=\sigma e^N H\), the corresponding physical bulk volume is time independent:
	\begin{equation}
		\Vphys = e^{3N}\VF(N) = \frac{6\pi^2}{\sigma^3H^3}.
		\label{eq:VphysApp}
	\end{equation}
	We have
	\begin{equation} \label{eq:QNPNmicroApp}
		\hat Q_N = \int \frac{\dd^3k}{(2\pi)^3} \Theta_\sigma(N,\bm k)\hvarphi_{\bm k}(N) \qquad \mathrm{and} \qquad \hat P_N = \VF(N) \int \frac{\dd^3k}{(2\pi)^3} \Theta_\sigma(N,\bm k)\hpi_{\bm k}(N)\ .
	\end{equation}
	The field variable \(\hat Q_N\) is an intensive bulk average. The conjugate momentum \(\hat P_N\) is extensive, and the factor \(\VF(N)\) makes it canonically conjugate to \(\hat Q_N\).

	Using
	\begin{equation}
		[\hvarphi_{\bm k}(N),\hpi_{\bm k'}(N)] = i(2\pi)^3\delta^{(3)}(\bm k+\bm k')
	\end{equation}
	and \(\Theta_\sigma^2=\Theta_\sigma\), one obtains
	\begin{align}
		[\hat Q_N,\hat P_N] &= \VF(N) \int\frac{\dd^3k}{(2\pi)^3} \int\frac{\dd^3k'}{(2\pi)^3} \Theta_\sigma(N,\bm k) \Theta_\sigma(N,\bm k') [\hvarphi_{\bm k},\hpi_{\bm k'}] \nonumber\\ &= i\VF(N) \int\frac{\dd^3k}{(2\pi)^3} \Theta_\sigma(N,\bm k) = i\VF(N)W(N,0) = i .
		\label{eq:Q_P_micro}
	\end{align}
	Thus the coarse-grained variables define a one-mode canonical algebra represented on the homogeneous one-mode Fock space \(\cF_N^\parallel\).

	\subsection{Bulk Part of the Bulk-Boundary Dynamics}
	\label{appA:Bulk}

	The one-step map from \(N\) to \(N+\dd N\) is obtained by evaluating the coarse-grained operators on the next bulk-boundary system, namely the bulk up to and including the incoming boundary shell at \(N+\dd N\). The resulting pair \((\hat Q_{N+\dd N},\hat P_{N+\dd N})\) generates \(\cF_{N+\dd N}^\parallel\):
	\begin{equation}
		\hat Q_{N+\dd N} = \int \frac{\dd^3k}{(2\pi)^3} \Theta_\sigma(N+\dd N,\bm k)\hvarphi_{\bm k}(N+\dd N),
	\end{equation}
	and
	\begin{equation}
		\hat P_{N+\dd N} = \VF(N+\dd N) \int \frac{\dd^3k}{(2\pi)^3} \Theta_\sigma(N+\dd N,\bm k)\hpi_{\bm k}(N+\dd N).
	\end{equation}
	The quadratic infrared Hamiltonian is
	\begin{equation}
		\hat H_{{\rm IR},N} = \int \frac{\dd^3k}{(2\pi)^3} \Theta_\sigma(N,\bm k) \left[ \frac{1}{2e^{3N}}\hpi_{\bm k}\hpi_{-\bm k} + \frac{e^{3N}}{2} \left( k^2e^{-2N}+m^2 \right) \hvarphi_{\bm k}\hvarphi_{-\bm k} \right].
	\end{equation}
	It gives, mode by mode,
	\begin{equation}
		\hvarphi_{\bm k}(N+\dd N) = \hvarphi_{\bm k}(N) + \frac{\dd N}{H e^{3N}}\hpi_{\bm k}(N) + O(\dd N^2),
	\end{equation}
	and
	\begin{equation}
		\hpi_{\bm k}(N+\dd N) = \hpi_{\bm k}(N) - \frac{e^{3N}\dd N}{H} \left( k^2e^{-2N}+m^2 \right) \hvarphi_{\bm k}(N) + O(\dd N^2).
	\end{equation}
	Split the updated projector as
	\begin{equation}
		\Theta_\sigma(N+\dd N,\bm k) = \Theta_\sigma(N,\bm k) + \Delta\Theta_N(\bm k), \qquad \Delta\Theta_N(\bm k) = \Theta_\sigma(N+\dd N,\bm k) - \Theta_\sigma(N,\bm k).
	\end{equation}
	We define the collective boundary increments on \(\cF_{\Delta}^\parallel\) as
	\begin{equation}
		d\hat q_{\Delta} = \int\frac{\dd^3k}{(2\pi)^3} \Delta\Theta_N(\bm k)\hvarphi_{\bm k}(N+\dd N),
		\label{eq:dQDeltaExactApp}
	\end{equation}
	and
	\begin{equation}
		d\hat p_{\Delta} = \VF(N+\dd N) \int\frac{\dd^3k}{(2\pi)^3} \Delta\Theta_N(\bm k)\hpi_{\bm k}(N+\dd N).
		\label{eq:dPDeltaExactApp}
	\end{equation}
	When these boundary increments appear in the active input space \(\cF_N^\parallel\otimes\cF_\Delta^\parallel\), we write the tensor product explicitly. We will revisit these boundary increments at the end of this section. For now we compute the homogeneous bulk part of the Heisenberg dynamics:
	\begin{align}
		\hat Q_{N+\dd N}-\openone_{\cF_N^\parallel}\otimes d\hat q_\Delta &= \int\frac{\dd^3k}{(2\pi)^3} \Theta_\sigma(N,\bm k)\hvarphi_{\bm k}(N+\dd N) \otimes \openone_{\cF_\Delta^\parallel} \nonumber\\ &= \left[ \hat Q_N + \frac{\dd N}{H e^{3N}} \int\frac{\dd^3k}{(2\pi)^3} \Theta_\sigma(N,\bm k)\hpi_{\bm k}(N) \right] \otimes \openone_{\cF_\Delta^\parallel} + O(\dd N^2) \nonumber\\ &= \left( \hat Q_N + \frac{\hat P_N}{H\Vphys}\,\dd N \right) \otimes \openone_{\cF_\Delta^\parallel} + O(\dd N^2).
		\label{eq:inputQWithIRApp}
	\end{align}
	For the momentum, using
	\begin{equation}
		\VF(N+\dd N) = \VF(N)e^{-3\dd N} = \VF(N)(1-3\dd N)+O(\dd N^2),
		\label{eq:VFexpandApp}
	\end{equation}
	one obtains
	\begin{align}
		&\hat P_{N+\dd N}-\openone_{\cF_N^\parallel}\otimes d\hat p_\Delta \nonumber\\ &= \VF(N+\dd N) \int\frac{\dd^3k}{(2\pi)^3} \Theta_\sigma(N,\bm k)\hpi_{\bm k}(N+\dd N) \otimes \openone_{\cF_\Delta^\parallel} \nonumber\\ &= \left( e^{-3\dd N}\hat P_N - \frac{\Vphys m^2}{H}\hat Q_N\,\dd N \right) \otimes \openone_{\cF_\Delta^\parallel} \nonumber\\ &\quad - \frac{\VF(N)e^N}{H} \int\frac{\dd^3k}{(2\pi)^3} \Theta_\sigma(N,\bm k)k^2\hvarphi_{\bm k}(N) \otimes \openone_{\cF_\Delta^\parallel}\,\dd N + O(\dd N^2).
		\label{eq:inputSectorWithGradientApp}
	\end{align}
	The last term is the inherited gradient contribution. It does not close on the one-mode algebra generated by \(\hat Q_N\) and \(\hat P_N\). At leading order in the super-Hubble gradient expansion, this term is \(O(\sigma^2)\) and will be discarded. Equivalently, since the infrared support obeys \(|\bm k|e^{-N}\leq \sigma H\), we neglect it consistently up to corrections \(O(\sigma^2 \dd N)\). Thus
	\begin{equation}
		\hat P_{N+\dd N}-\openone_{\cF_N^\parallel}\otimes d\hat p_\Delta = \left[ \hat P_N - \left( 3\hat P_N+\frac{m^2\Vphys}{H}\hat Q_N \right)\dd N \right] \otimes \openone_{\cF_\Delta^\parallel} + O(\dd N^2) + O(\sigma^2 \dd N).
		\label{eq:inputSectorWithIRApp}
	\end{equation}
	The commutator excluding the boundary increment is
	\begin{equation}
		\left[ \hat Q_{N+\dd N}-\openone_{\cF_N^\parallel}\otimes d\hat q_\Delta, \hat P_{N+\dd N}-\openone_{\cF_N^\parallel}\otimes d\hat p_\Delta \right] = i(1-3\dd N)+O(\dd N^2).
	\end{equation}
	As \(\cF_N^\parallel\) and \(\cF_\Delta^\parallel\) are disjoint incoming-bulk and incoming-boundary-shell modes, and as we have already demonstrated the preservation of the canonical commutator in \cref{eq:Q_P_micro}, we have
	\begin{equation}
		\left[ \openone_{\cF_N^\parallel}\otimes d\hat q_\Delta, \openone_{\cF_N^\parallel}\otimes d\hat p_\Delta \right] = 3i \dd N+O(\dd N^2).
	\end{equation}
	Combining \cref{eq:inputQWithIRApp,eq:inputSectorWithIRApp} and taking \(\dd N\to0\), we have the bulk contribution to the bulk-variable dynamics
	\begin{align}
		\left.\dv{\hat Q_{N}}{N}\right|_{\rm bulk} &= \frac{\hat P_N}{H\Vphys},
		\label{eq:BulkDrift1}
		\\ \left.\dv{\hat P_{N}}{N}\right|_{\rm bulk} &= -3\hat P_N - \frac{m^2\Vphys}{H}\hat Q_N,
		\label{eq:BulkDrift2}
	\end{align}
	where \(|_{\rm bulk}\) denotes the part that lives on the homogeneous bulk \(\cF_N^\parallel\), before the collective boundary contribution from \(\cF_\Delta^\parallel\) is added.

	\subsection{Bulk-Boundary Differential Relations}
	\label{appA:boundary}

	Expanding the field operators in terms of ladder operators,
	\begin{equation}
		\label{PhiPi_expa}
		\hvarphi_{{\bm k}}(N) = \phi_{k}(N) \hat a_{{\bm k}} + \phi^{\ast}_{k}(N) \hat a_{-{\bm k}}^{\dagger}, \qquad \hpi_{{\bm k}}(N) = \pi_{k}(N) \hat a_{{\bm k}} + \pi^{\ast}_{k}(N) \hat a_{-{\bm k}}^{\dagger},
	\end{equation}
	where
	\begin{equation}
		[\hat a_{{\bm k}}, \hat a_{{\bm k}'}^{\dagger}] = (2\pi)^3 \delta^{(3)}({\bm k} - {\bm k}'), \qquad [\hat a_{{\bm k}}, \hat a_{{\bm k}'}] = 0 .
	\end{equation}
	Define the boundary averaged annihilation increment on \(\cF_{\Delta}^\parallel\) as
	\begin{equation}
		\dd\hat b_\Delta(N) := \sqrt{\frac{\VF(N)}{3}} \int \frac{\dd^3k}{(2\pi)^3} \Delta\Theta_N(\bm k)\,\hat a_{\bm k},
		\label{eq:dbDeltaDefApp}
	\end{equation}
	for which it is straightforward to show that
	\begin{equation}
		[\dd\hat b_\Delta(N),\dd\hat b^{\dagger}_\Delta(N)] = \dd N+ O(\dd N^2)
	\end{equation}
	using the boundary phase-space volume
	\begin{equation}
		\int \frac{\dd^3k}{(2\pi)^3} \Delta\Theta_N(\bm k) = \frac{k_\sigma^3(N)}{2\pi^2}\dd N + O(\dd N^2) = \frac{3}{\VF(N)}\dd N + O(\dd N^2).
	\end{equation}
	We can write the boundary increments previously defined in \cref{eq:dQDeltaExactApp,eq:dPDeltaExactApp} in terms of the boundary creation and annihilation operators using \cref{PhiPi_expa}. To do so, start from the exact boundary increment \cref{eq:dQDeltaExactApp} and substitute the mode expansion \cref{PhiPi_expa} of \(\hvarphi_{\bm k}\) at $e$-fold \(N+\dd N\):
	\begin{equation}
		\dd\hat q_{\Delta} = \int\frac{\dd^3k}{(2\pi)^3} \Delta\Theta_N(\bm k) \left[ \phi_k(N+\dd N)\,\hat a_{\bm k} + \phi^*_k(N+\dd N)\,\hat a^\dagger_{-\bm k} \right].
		\label{eq:dQDeltaExpandedApp}
	\end{equation}
	The support of \(\Delta\Theta_N\) is the thin shell \(k_\sigma(N)<|\bm k|<k_\sigma(N+\dd N)\), on which \(|\bm k|=k_\sigma(N)+O(\dd N)\). Expanding the mode functions about \(k=k_\sigma(N)\) at time \(N\) produces corrections of order \(\dd N\) multiplying an integral whose phase-space volume is itself \(O(\dd N)\), so the resulting error in \cref{eq:dQDeltaExpandedApp} is \(O(\dd N^{3/2})\). To leading order we may therefore replace \(\phi_k(N+\dd N)\to\phi_{k_\sigma}\equiv\phi_{k_\sigma(N)}(N)\):
	\begin{equation}
		\dd\hat q_{\Delta} = \phi_{k_\sigma}\!\int\frac{\dd^3k}{(2\pi)^3} \Delta\Theta_N(\bm k)\,\hat a_{\bm k} + \phi^*_{k_\sigma}\!\int\frac{\dd^3k}{(2\pi)^3} \Delta\Theta_N(\bm k)\,\hat a^\dagger_{-\bm k} + O(\dd N^{3/2}).
		\label{eq:dQDeltaModeApp}
	\end{equation}
	The first integral is, by the definition \cref{eq:dbDeltaDefApp} of \(\dd\hat b_\Delta(N)\),
	\begin{equation}
		\int\frac{\dd^3k}{(2\pi)^3} \Delta\Theta_N(\bm k)\,\hat a_{\bm k} = \sqrt{\frac{3}{\VF(N)}}\,\dd\hat b_{\Delta}.
		\label{eq:bAnnIntegralApp}
	\end{equation}
	For the second integral, the projector \(\Delta\Theta_N(\bm k)\) depends only on \(|\bm k|\) and is therefore invariant under \(\bm k\to-\bm k\). Changing variables \(\bm k\to-\bm k\) gives
	\begin{equation}
		\int\frac{\dd^3k}{(2\pi)^3} \Delta\Theta_N(\bm k)\,\hat a^\dagger_{-\bm k} = \int\frac{\dd^3k}{(2\pi)^3} \Delta\Theta_N(\bm k)\,\hat a^\dagger_{\bm k} = \sqrt{\frac{3}{\VF(N)}}\,\dd\hat b^\dagger_{\Delta}.
		\label{eq:bCrIntegralApp}
	\end{equation}
	Substituting \cref{eq:bAnnIntegralApp} and \cref{eq:bCrIntegralApp} into \cref{eq:dQDeltaModeApp} yields
	\begin{equation}
		\dd\hat q_\Delta = \sqrt{\frac{3}{\VF(N)}} \left( \phi_{k_{\sigma}} \dd\hat b_\Delta + \phi_{k_{\sigma}}^* \dd\hat b_\Delta^\dagger \right) + O(\dd N^{3/2}).
		\label{eq:dQDeltaLeadingApp}
	\end{equation}
	The momentum case follows the same steps, starting from \cref{eq:dPDeltaExactApp}. Because the integral on the shell is \(O(\dd N^{1/2})\), the prefactor \(\VF(N+\dd N)=\VF(N)(1-3\dd N)\) may be replaced by \(\VF(N)\) up to an \(O(\dd N^{3/2})\) correction. Using the mode expansion of \(\hpi_{\bm k}\) from \cref{PhiPi_expa} and the integrals \cref{eq:bAnnIntegralApp,eq:bCrIntegralApp},
	\begin{align}
		\dd\hat p_\Delta &= \VF(N) \int\frac{\dd^3k}{(2\pi)^3} \Delta\Theta_N(\bm k) \left( \pi_{k_\sigma}\,\hat a_{\bm k} + \pi^*_{k_\sigma}\,\hat a^\dagger_{-\bm k} \right) + O(\dd N^{3/2}) \nonumber\\ &= \VF(N) \sqrt{\frac{3}{\VF(N)}} \left( \pi_{k_\sigma}\,\dd\hat b_\Delta + \pi^*_{k_\sigma}\,\dd\hat b_\Delta^\dagger \right) + O(\dd N^{3/2}) \nonumber\\ &= \sqrt{3\,\VF(N)} \left( \pi_{k_{\sigma}} \dd\hat b_\Delta + \pi_{k_{\sigma}}^* \dd\hat b_\Delta^\dagger \right) + O(\dd N^{3/2}).
		\label{eq:dPDeltaLeadingApp}
	\end{align}
	Using these relations, we can calculate the boundary increment of an arbitrary linear combination of \(\hat Q_{N}\) and \(\hat P_{N}\) on \(\cF_{N}^\parallel\). For complex coefficients \(u\) and \(v\), define
	\begin{equation}
		\hat X_{N} = u\hat Q_{N} + v\hat P_{N}.
	\end{equation}
	Its boundary contribution on the pre-trace bulk-boundary space is
	\begin{align}
		\left.\dd\hat X_N\right|_\Delta &= u\,\openone_{\cF_N^\parallel}\otimes \dd\hat q_\Delta + v\,\openone_{\cF_N^\parallel}\otimes \dd\hat p_\Delta \nonumber\\ &= \sqrt3 \left( \frac{u\phi_{k_{\sigma}}}{\sqrt{\VF}} + v\sqrt{\VF}\,\pi_{k_{\sigma}} \right)\openone_{\cF_N^\parallel}\otimes \dd\hat b_\Delta \nonumber\\ &\quad + \sqrt3 \left( \frac{u\phi_{k_{\sigma}}^*}{\sqrt{\VF}} + v\sqrt{\VF}\,\pi_{k_{\sigma}}^* \right)\openone_{\cF_N^\parallel}\otimes \dd\hat b_\Delta^\dagger + O(\dd N^{3/2}).
	\end{align}
	We can identify a boundary-adapted annihilation operator acting on \(\cF_N^\parallel\) whose boundary increment contains no \(\openone_{\cF_N^\parallel}\otimes \dd\hat b_\Delta^\dagger\) term. Such an operator satisfies the linear constraint
	\begin{equation}
		\frac{u\phi_{k_{\sigma}}^*}{\sqrt{\VF}} + v\sqrt{\VF}\,\pi_{k_{\sigma}}^* = 0.
		\label{eq:AnnihilationConstraintApp}
	\end{equation}
	A canonically normalised representative is
	\begin{equation}
		\hat A_{N} = i\left[ \frac{\phi_{k_{\sigma}}^*}{\sqrt{\VF(N)}}\hat P_{N} - \sqrt{\VF(N)}\,\pi_{k_{\sigma}}^*\hat Q_{N} \right].
		\label{eq:AdefApp}
	\end{equation}
	Using the Wronskian relation
	\begin{equation}
		\phi_{k_{\sigma}}\pi_{k_{\sigma}}^* - \phi_{k_{\sigma}}^*\pi_{k_{\sigma}} = i,
		\label{eq:WronskianApp}
	\end{equation}
	one obtains
	\begin{equation}
		[\hat A_{N},\hat A_{N}^\dagger]=1.
	\end{equation}
	The inverse relations are
	\begin{equation}
		\hat Q_{N} = \frac{\phi_{k_{\sigma}}}{\sqrt{\VF}}\hat A_{N} + \frac{\phi_{k_{\sigma}}^*}{\sqrt{\VF}}\hat A_{N}^\dagger, \qquad \hat P_{N} = \sqrt{\VF}\,\pi_{k_{\sigma}}\hat A_{N} + \sqrt{\VF}\,\pi_{k_{\sigma}}^*\hat A_{N}^\dagger .
		\label{eq:QPfromA}
	\end{equation}
	The boundary increment of \cref{eq:AdefApp} is given by
	\begin{align}
		\left.\dd\hat A_N\right|_\Delta &= i\left( \frac{\phi_{k_{\sigma}}^*}{\sqrt{\VF}} \openone_{\cF_N^\parallel}\otimes \dd\hat p_\Delta - \sqrt{\VF}\,\pi_{k_{\sigma}}^* \openone_{\cF_N^\parallel}\otimes \dd\hat q_\Delta \right) \nonumber\\ &= i\sqrt3 \left( \phi_{k_{\sigma}}^*\pi_{k_{\sigma}} - \pi_{k_{\sigma}}^*\phi_{k_{\sigma}} \right) \openone_{\cF_N^\parallel}\otimes \dd\hat b_\Delta + O(\dd N^{3/2}) \nonumber\\ &= \sqrt3\,\openone_{\cF_N^\parallel}\otimes \dd\hat b_\Delta + O(\dd N^{3/2}).
		\label{eq:dAboundary}
	\end{align}

	\subsection{Unitary Bulk-Boundary Entangling}
	\label{appA:Unitary}

	In order to eventually transition from the Heisenberg picture canonical variable evolution to the Schr\"odinger picture reduced bulk density operator dynamics, we now construct the joint unitary \(\hat V_{N+\dd N}\) on \(\cF_N^\parallel\otimes\cF_\Delta^\parallel\) whose Heisenberg action reproduces the one-step evolution derived in Sections~\ref{appA:Bulk} and \ref{appA:boundary}. The entangling unitary produces the decoupled mode. The relevant input boundary mode is the collective increment \(\dd\hat b_\Delta\) defined in \cref{eq:dbDeltaDefApp}. The boundary calculation in \cref{eq:dAboundary} requires that the adapted annihilation operator obey, to leading order,
	\begin{equation}
		\hat A_N\otimes\openone_{\cF_\Delta^\parallel} \longmapsto \hat A_N\otimes\openone_{\cF_\Delta^\parallel} + \sqrt3\,\openone_{\cF_N^\parallel}\otimes \dd\hat b_\Delta + O(\dd N), \qquad \hat A_N^\dagger\otimes\openone_{\cF_\Delta^\parallel} \longmapsto \hat A_N^\dagger\otimes\openone_{\cF_\Delta^\parallel} + \sqrt3\,\openone_{\cF_N^\parallel}\otimes \dd\hat b_\Delta^\dagger + O(\dd N),
		\label{eq:AtargetUnitaryApp}
	\end{equation}
	with no \(\openone_{\cF_N^\parallel}\otimes \dd\hat b_\Delta^\dagger\) term in the first equation and no \(\openone_{\cF_N^\parallel}\otimes \dd\hat b_\Delta\) term in the second. The same calculation gave the canonical variable Heisenberg dynamics:
	\begin{align}
		\hat Q_N\otimes\openone_{\cF_\Delta^\parallel} &\longmapsto \hat Q_N\otimes\openone_{\cF_\Delta^\parallel} + \frac{1}{H\Vphys}\hat P_N\otimes\openone_{\cF_\Delta^\parallel}\,\dd N + \openone_{\cF_N^\parallel}\otimes \dd\hat q_\Delta + O(\dd N^{3/2}), \nonumber\\ \hat P_N\otimes\openone_{\cF_\Delta^\parallel} &\longmapsto \hat P_N\otimes\openone_{\cF_\Delta^\parallel} - \left( 3\hat P_N + \frac{m^2\Vphys}{H}\hat Q_N \right)\otimes\openone_{\cF_\Delta^\parallel}\,\dd N + \openone_{\cF_N^\parallel}\otimes \dd\hat p_\Delta + O(\dd N^{3/2}).
		\label{eq:QPtargetUnitaryApp}
	\end{align}
	The unitary we seek must reproduce both \cref{eq:AtargetUnitaryApp} and \cref{eq:QPtargetUnitaryApp}. The thin-shell condition is essential here: the collective boundary increment is \(O(\dd N^{1/2})\), its commutator is \(O(\dd N)\), and all independent \(O(\dd N^{3/2})\) corrections are beyond the order needed to determine the GKLS generator. Therefore the leading bulk-boundary generator may be taken to be linear in \(\dd\hat b_\Delta\) and \(\dd\hat b_\Delta^\dagger\). Terms with more boundary increments start beyond the required order, while \(O(\dd N)\) bulk-boundary couplings only affect the Heisenberg equations at \(O(\dd N^{3/2})\).

	We start with the generic form
	\begin{equation}
		\hat V_{N+\dd N} = \exp\!\left[ \hat S_{N+\dd N} - i\bigl(\hat K_{\rm eff}\otimes\openone_{\cF_\Delta^\parallel}\bigr)\,\dd N \right] + O(\dd N^{3/2}),
		\label{eq:VNdefUnitaryApp}
	\end{equation}
	where \(\hat S_{N+\dd N}=O(\sqrt{\dd N})\) is anti-Hermitian and couples the bulk mode to the collective boundary increment, while \(\hat K_{\rm eff}=O(1)\) is a Hermitian bulk Hamiltonian. Since \([\hat S_{N+\dd N},(\hat K_{\rm eff}\otimes\openone_{\cF_\Delta^\parallel})\,\dd N]=O(\dd N^{3/2})\), the ordering of the two exponentials is immaterial here, and we may equivalently write
	\begin{equation}
		\hat V_{N+\dd N} = \hat U_{N+\dd N} \left( e^{-i\hat K_{\rm eff}\,\dd N}\otimes\openone_{\cF_\Delta^\parallel} \right) + O(\dd N^{3/2}), \qquad \hat U_{N+\dd N} = \exp(\hat S_{N+\dd N}).
	\end{equation}
	The most general anti-Hermitian bulk-boundary generator linear in \(\dd\hat b_\Delta\) and \(\dd\hat b_\Delta^\dagger\), and linear in the adapted bulk mode, is
	\begin{equation}
		\hat S_{N+\dd N} = c_1\,\hat A_N^\dagger\otimes \dd\hat b_\Delta - c_1^*\,\hat A_N\otimes \dd\hat b_\Delta^\dagger + c_2\,\hat A_N\otimes \dd\hat b_\Delta - c_2^*\,\hat A_N^\dagger\otimes \dd\hat b_\Delta^\dagger,
		\label{eq:SgeneralUnitaryApp}
	\end{equation}
	with \(c_1,c_2\in\mathbb C\). This is the only relevant \(O(\sqrt{\dd N})\) coupling under the stated physical conditions. A generator acting only on the bulk at \(O(\sqrt{\dd N})\) would produce deterministic \(O(\sqrt{\dd N})\) shifts of \(\hat Q_N\), \(\hat P_N\), or \(\hat A_N\), and such terms are absent from the Heisenberg equations \cref{eq:AtargetUnitaryApp} and \cref{eq:QPtargetUnitaryApp}. A boundary-only displacement commutes with the bulk operators and cannot reproduce the required stochastic increments. It is also physically excluded because it would displace the newly incoming boundary shell shell away from the Bunch-Davies vacuum. A boundary-only phase rotation merely changes the convention for the phase of \(\dd\hat b_\Delta\), and this freedom will reappear below as the physically irrelevant phase of the GKLS jump operator. Finally, a scalar phase of the full unitary is irrelevant, and a scalar addition to \(\hat K_{\rm eff}\) has no effect on the dynamics. Thus, modulo these physically irrelevant phase conventions, \cref{eq:SgeneralUnitaryApp} is the unique leading-order ansatz capable of satisfying all the Heisenberg equations derived in the previous two sections.

	Using the Baker-Campbell-Hausdorff expansion
	\begin{equation}
		\hat V_{N+\dd N}^\dagger \hat X \hat V_{N+\dd N} = \hat X + [\hat X,\hat S_{N+\dd N}] - i[\hat X,\hat K_{\rm eff}\otimes\openone_{\cF_\Delta^\parallel}]\,\dd N + \frac12[[\hat X,\hat S_{N+\dd N}],\hat S_{N+\dd N}] + O(\dd N^{3/2}),
		\label{eq:BCHUnitaryApp}
	\end{equation}
	together with \([\hat A_N,\hat A_N^\dagger]=1\) and \([\dd\hat b_\Delta,\dd\hat b_\Delta^\dagger]=\dd N+O(\dd N^2)\), one finds
	\begin{equation}
		[\hat A_N\otimes\openone_{\cF_\Delta^\parallel},\hat S_{N+\dd N}] = c_1\,\openone_{\cF_N^\parallel}\otimes \dd\hat b_\Delta - c_2^*\,\openone_{\cF_N^\parallel}\otimes \dd\hat b_\Delta^\dagger,
		\label{eq:firstCommAUnitaryApp}
	\end{equation}
	and
	\begin{equation}
		\frac12[[\hat A_N\otimes\openone_{\cF_\Delta^\parallel},\hat S_{N+\dd N}],\hat S_{N+\dd N}] = \frac12\left( |c_2|^2-|c_1|^2 \right)\hat A_N\otimes\openone_{\cF_\Delta^\parallel}\,\dd N + O(\dd N^{3/2}).
		\label{eq:secondCommAUnitaryApp}
	\end{equation}
	The conjugate relations for \(\hat A_N^\dagger\otimes\openone_{\cF_\Delta^\parallel}\) follow by Hermitian conjugation. The absence of an \(\openone_{\cF_N^\parallel}\otimes \dd\hat b_\Delta^\dagger\) term in the evolution of \(\hat A_N\otimes\openone_{\cF_\Delta^\parallel}\), together with the coefficient \(\sqrt3\) in \cref{eq:AtargetUnitaryApp}, fixes
	\begin{equation}
		c_1=\sqrt3, \qquad c_2=0,
		\label{eq:c1c2UnitaryApp}
	\end{equation}
	up to an irrelevant overall phase of the boundary mode. Therefore
	\begin{equation}
		\hat S_{N+\dd N} = \sqrt3 \left( \hat A_N^\dagger\otimes \dd\hat b_\Delta - \hat A_N\otimes \dd\hat b_\Delta^\dagger \right), \qquad \hat U_{N+\dd N} = \exp(\hat S_{N+\dd N})
		\label{eq:UNbeamsplitter}
	\end{equation}
	is the unique leading bulk-boundary entangling unitary at this order under the physical conditions just stated. Substituting \cref{eq:c1c2UnitaryApp} into \cref{eq:secondCommAUnitaryApp} reduces the Baker-Campbell-Hausdorff \(O(\dd N)\) term to \(-\tfrac32\hat A_N\otimes\openone_{\cF_\Delta^\parallel}\,\dd N\), so the Heisenberg action of \(\hat U_{N+\dd N}\) on the adapted ladder operator is
	\begin{equation}
		\hat U_{N+\dd N}^\dagger \left( \hat A_N\otimes\openone_{\cF_\Delta^\parallel} \right) \hat U_{N+\dd N} = \hat A_N\otimes\openone_{\cF_\Delta^\parallel} + \sqrt3\,\openone_{\cF_N^\parallel}\otimes \dd\hat b_\Delta - \frac32\,\hat A_N\otimes\openone_{\cF_\Delta^\parallel}\,\dd N + O(\dd N^{3/2}).
		\label{eq:UActionAUnitaryApp}
	\end{equation}
	Using the inverse relations \cref{eq:QPfromA}, equation \cref{eq:UActionAUnitaryApp} translates into an isotropic contraction of the canonical variables together with the boundary increments,
	\begin{align}
		\hat U_{N+\dd N}^\dagger \left( \hat Q_N\otimes\openone_{\cF_\Delta^\parallel} \right) \hat U_{N+\dd N} &= \hat Q_N\otimes\openone_{\cF_\Delta^\parallel} + \openone_{\cF_N^\parallel}\otimes \dd\hat q_\Delta - \frac32\,\hat Q_N\otimes\openone_{\cF_\Delta^\parallel}\,\dd N + O(\dd N^{3/2}), \nonumber\\ \hat U_{N+\dd N}^\dagger \left( \hat P_N\otimes\openone_{\cF_\Delta^\parallel} \right) \hat U_{N+\dd N} &= \hat P_N\otimes\openone_{\cF_\Delta^\parallel} + \openone_{\cF_N^\parallel}\otimes \dd\hat p_\Delta - \frac32\,\hat P_N\otimes\openone_{\cF_\Delta^\parallel}\,\dd N + O(\dd N^{3/2}).
		\label{eq:UActionQPUnitaryApp}
	\end{align}
	The bulk-boundary entangling therefore supplies the required stochastic increments and produces a deterministic contraction of both canonical variables. The remaining anisotropic drift must be generated by the bulk Hamiltonian \(\hat K_{\rm eff}\).

	Take the most general quadratic Hermitian bulk Hamiltonian on \(\cF_N^\parallel\) that can contribute to the required linear homogeneous Heisenberg drift,
	\begin{equation}
		\hat K_{\rm eff} = \frac{c_3}{2}\hat P_N^2 + \frac{c_4}{2}\hat Q_N^2 + \frac{c_5}{2}\{\hat Q_N,\hat P_N\}, \qquad c_3,c_4,c_5\in\mathbb R.
		\label{eq:KeffGeneralUnitaryApp}
	\end{equation}
	A completely general quadratic Hermitian Hamiltonian could also contain linear terms proportional to \(\hat Q_N\) and \(\hat P_N\), together with a constant. The linear terms would generate constant shifts in the Heisenberg equations for \(\hat P_N\) and \(\hat Q_N\), respectively. Since no such constant drifts appear in \cref{eq:QPtargetUnitaryApp}, their coefficients must vanish. The constant term is dynamically irrelevant and is omitted.

	Using \([\hat Q_N,\hat P_N]=i\),
	\begin{equation}
		-i[\hat Q_N,\hat K_{\rm eff}] = c_3\hat P_N+c_5\hat Q_N, \qquad -i[\hat P_N,\hat K_{\rm eff}] = -c_4\hat Q_N-c_5\hat P_N.
		\label{eq:KcommUnitaryApp}
	\end{equation}
	Embedding \(\hat K_{\rm eff}\) as \(\hat K_{\rm eff}\otimes\openone_{\cF_\Delta^\parallel}\) and combining \cref{eq:UActionQPUnitaryApp} with the Hamiltonian contribution from \cref{eq:KcommUnitaryApp} gives
	\begin{align}
		&\hat V_{N+\dd N}^\dagger \left( \hat Q_N\otimes\openone_{\cF_\Delta^\parallel} \right) \hat V_{N+\dd N} \nonumber\\ &\quad = \hat Q_N\otimes\openone_{\cF_\Delta^\parallel} + \openone_{\cF_N^\parallel}\otimes \dd\hat q_\Delta + \left[ c_3\hat P_N + \left( c_5-\frac32 \right)\hat Q_N \right] \otimes\openone_{\cF_\Delta^\parallel}\,\dd N + O(\dd N^{3/2}), \nonumber\\[2pt] &\hat V_{N+\dd N}^\dagger \left( \hat P_N\otimes\openone_{\cF_\Delta^\parallel} \right) \hat V_{N+\dd N} \nonumber\\ &\quad = \hat P_N\otimes\openone_{\cF_\Delta^\parallel} + \openone_{\cF_N^\parallel}\otimes \dd\hat p_\Delta - \left[ c_4\hat Q_N + \left( c_5+\frac32 \right)\hat P_N \right] \otimes\openone_{\cF_\Delta^\parallel}\,\dd N + O(\dd N^{3/2}).
		\label{eq:VActionQPUnitaryApp}
	\end{align}
	Matching \cref{eq:VActionQPUnitaryApp} with the target \cref{eq:QPtargetUnitaryApp} fixes
	\begin{equation}
		c_3=\frac{1}{H\Vphys}, \qquad c_4=\frac{m^2\Vphys}{H}, \qquad c_5=\frac32.
		\label{eq:coeffFinalUnitaryApp}
	\end{equation}
	The two equations in \cref{eq:VActionQPUnitaryApp} are consistent: both fix the same value \(c_5=\tfrac32\), reflecting the boundary \(O(\dd N)\) term strength \(\tfrac12|c_1|^2=\tfrac32\) inherited from \cref{eq:c1c2UnitaryApp}. Substituting \cref{eq:coeffFinalUnitaryApp} into \cref{eq:KeffGeneralUnitaryApp} yields
	\begin{equation}
		\hat K_{\rm eff} = \frac{\hat P_N^2}{2H\Vphys} + \frac{m^2\Vphys}{2H}\hat Q_N^2 + \frac{3}{4}\{\hat Q_N,\hat P_N\}.
		\label{eq:KeffFinal}
	\end{equation}

	\subsection{Tracing Over the Decoupled Mode}
	\label{appA:tracing}

	The reduced homogeneous state is obtained by applying the entangling joint unitary with the redefined-bulk/decoupled-mode decomposition
	\[
	\hat U_{N+\dd N}:\cF_N^\parallel\otimes\cF_\Delta^\parallel
	\longrightarrow \cF_{N+\dd N}^\parallel\otimes\cF^{\rm d,\,\perp}_{N+\dd N}
	\]
	and then tracing over \(\cF^{\rm d,\,\perp}_{N+\dd N}\), the decoupled mode. The incoming boundary shell mode in \(\cF_\Delta^\parallel\) is initially in the Bunch-Davies vacuum of the collective increment, denoted by \(\ket{0}_\Delta\). We denote the decoupled-mode Fock basis by \(\ket n_{\rm d}\). All modes orthogonal to this active two-mode construction remain in vacuum and decouple.

	Using
	\begin{equation}
		\hat V_{N+\dd N} = \hat U_{N+\dd N} \left( e^{-i\hat K_{\rm eff}\dd N}\otimes \openone_{\cF_\Delta^\parallel} \right) + O(\dd N^{3/2}),
	\end{equation}
	the one-step full state before the partial trace is
	\begin{equation}
		\hat\rho_{N+\dd N}^{\parallel\perp} = \hat U_{N+\dd N} \left( \tilde{\rho}_N \otimes \ket{0}_\Delta\!\bra{0}_\Delta \right) \hat U_{N+\dd N}^\dagger + O(\dd N^2),
	\end{equation}
	and the reduced one-step state is
	\begin{equation}
		\hat\rho_{N+\dd N} = \Tr_{\cF^{\rm d,\,\perp}_{N+\dd N}} \left( \hat\rho_{N+\dd N}^{\parallel\perp} \right) = \Tr_{\cF^{\rm d,\,\perp}_{N+\dd N}} \left[ \hat U_{N+\dd N} \left( \tilde{\rho}_N \otimes \ket{0}_\Delta\!\bra{0}_\Delta \right) \hat U_{N+\dd N}^\dagger \right] + O(\dd N^2),
		\label{eq:rhoTraceInputFull}
	\end{equation}
	where \(\tilde{\rho}_N\) is the Hamiltonian-evolved homogeneous state on \(\cF_N^\parallel\),
	\begin{equation}
		\tilde{\rho}_N := e^{-i\hat K_{\rm eff}\dd N} \hat\rho_N e^{i\hat K_{\rm eff}\dd N} = \hat\rho_N - i\,\dd N\,[\hat K_{\rm eff},\hat\rho_N] + O(\dd N^2).
		\label{eq:rhoTildeApp}
	\end{equation}
	The input collective-boundary number states used to evaluate the redefinition are defined by
	\begin{align}
		\ket{n}_\Delta &:= \frac{(\dd\hat b_\Delta^\dagger)^n} {\sqrt{n!\,(\dd N)^n}} \ket{0}_\Delta, \\ \dd\hat b_\Delta\ket{0}_\Delta &=0, \\ \dd\hat b_\Delta\ket{n}_\Delta &= \sqrt{n\,\dd N}\,\ket{n-1}_\Delta, \\ \dd\hat b_\Delta^\dagger\ket{n}_\Delta &= \sqrt{(n+1)\,\dd N}\,\ket{n+1}_\Delta,
		\label{eq:numberStatesTracingApp}
	\end{align}
	consistent with
	\begin{equation}
		[\dd\hat b_\Delta,\dd\hat b_\Delta^\dagger] = \dd N+O(\dd N^2).
	\end{equation}
	Inserting the collective-mode resolution of the identity,
	\begin{equation}
		\sum_{n=0}^{\infty} \ket{n} \!{}_{\rm d}\,{}_{\rm d}\!\bra{n} = \openone_{\cF^{\rm d,\,\perp}_{N+\dd N}},
	\end{equation}
	gives the Kraus representation
	\begin{equation}
		\hat\rho_{N+\dd N} = \sum_{n=0}^{\infty} \hat M_n^{(N)} \tilde{\rho}_N \hat M_n^{(N)\dagger} + O(\dd N^2),
		\label{eq:KrausResolutionApp}
	\end{equation}
	with Kraus operators on the bulk,
	\begin{equation}
		\hat M_n^{(N)} := \left( \openone_{\cF_{N+\dd N}^\parallel}\otimes{}_{\rm d}\!\bra{n} \right) \hat U_{N+\dd N} \left( \openone_{\cF_N^\parallel}\otimes\ket{0}_\Delta \right).
		\label{eq:KrausDefTracingApp}
	\end{equation}
	Each \(\hat M_n^{(N)}:\cF_N^\parallel\to\cF_{N+\dd N}^\parallel\) acts only between homogeneous one-mode spaces, which are identified in the stationary representation below. We denote the canonical identity under this one-mode identification by \(\openone^\parallel_{N\to N+\dd N}:\cF_N^\parallel\to\cF_{N+\dd N}^\parallel\). To evaluate these Kraus operators, expand
	\begin{equation}
		\hat U_{N+\dd N} = \openone_{\cF_N^\parallel\otimes\cF_\Delta^\parallel} + \hat S_{N+\dd N} + \frac12\hat S_{N+\dd N}^2 + \frac{1}{6}\hat S_{N+\dd N}^3 + \cdots,
		\label{eq:UNexpansionApp}
	\end{equation}
	where \(\hat S_{N+\dd N}\) was defined in \cref{eq:UNbeamsplitter}. Each factor of \(\hat S_{N+\dd N}\) contains one boundary ladder increment and is therefore \(O(\sqrt{\dd N})\). Acting on the incoming boundary vacuum, understood in the input representation as the map \(\openone_{\cF_N^\parallel}\otimes\ket{0}_\Delta:\cF_N^\parallel\to\cF_N^\parallel\otimes\cF_\Delta^\parallel\), one finds
	\begin{equation}
		\hat S_{N+\dd N} \left( \openone_{\cF_N^\parallel}\otimes\ket{0}_\Delta \right) = -\sqrt3\,\hat A_N\otimes \dd\hat b_\Delta^\dagger\ket{0}_\Delta = -\sqrt{3\,\dd N}\, \hat A_N\otimes\ket{1}_\Delta .
		\label{eq:Son0App}
	\end{equation}
	Acting once more gives
	\begin{align}
		\hat S_{N+\dd N}^2 \left( \openone_{\cF_N^\parallel}\otimes\ket{0}_\Delta \right) &= -\sqrt{3\,\dd N}\, \hat S_{N+\dd N} \left( \hat A_N\otimes\ket{1}_\Delta \right) \nonumber\\ &= -\sqrt{3\,\dd N}\,\sqrt3 \left( \hat A_N^\dagger\hat A_N \otimes \dd\hat b_\Delta\ket{1}_\Delta - \hat A_N^2 \otimes \dd\hat b_\Delta^\dagger\ket{1}_\Delta \right) \nonumber\\ &= -3\,\dd N\, \hat A_N^\dagger\hat A_N \otimes\ket{0}_\Delta + 3\sqrt2\,\dd N\, \hat A_N^2 \otimes\ket{2}_\Delta .
		\label{eq:S2on0App}
	\end{align}
	Here we used
	\begin{equation}
		\dd\hat b_\Delta\ket{1}_\Delta = \sqrt{\dd N}\,\ket{0}_\Delta, \qquad \dd\hat b_\Delta^\dagger\ket{1}_\Delta = \sqrt{2\,\dd N}\,\ket{2}_\Delta .
	\end{equation}
	Combining \cref{eq:Son0App} and \cref{eq:S2on0App} in \cref{eq:UNexpansionApp} gives
	\begin{align}
		\hat U_{N+\dd N} \left( \openone_{\cF_N^\parallel}\otimes\ket{0}_\Delta \right) &= \left( \openone^\parallel_{N\to N+\dd N} - \frac32\,\dd N\,\hat A_N^\dagger\hat A_N \right) \otimes\ket{0}_{\rm d} \nonumber\\ &\quad - \sqrt{3\,\dd N}\, \hat A_N\otimes\ket{1}_{\rm d} + \frac{3}{\sqrt2}\,\dd N\, \hat A_N^2\otimes\ket{2}_{\rm d} + O(\dd N^{3/2}).
		\label{eq:Uon0App}
	\end{align}
	Taking the boundary matrix elements in \cref{eq:KrausDefTracingApp} therefore yields
	\begin{align}
		\hat M_0^{(N)} &= \openone^\parallel_{N\to N+\dd N} - \frac32\,\dd N\,\hat A_N^\dagger\hat A_N + O(\dd N^2),
		\label{eq:KrausVacuumApp}
		\\[4pt] \hat M_1^{(N)} &= -\sqrt{3\,\dd N}\,\hat A_N + O(\dd N^{3/2}),
		\label{eq:KrausOneParticleApp}
		\\[4pt] \hat M_2^{(N)} &= \frac{3}{\sqrt2}\,\dd N\,\hat A_N^2 + O(\dd N^2),
		\label{eq:KrausTwoParticleApp}
		\\[4pt] \hat M_n^{(N)} &= O(\dd N^{n/2}) \qquad\text{for }n\ge 3.
		\label{eq:KrausHigherApp}
	\end{align}
	In the Kraus sum \cref{eq:KrausResolutionApp}, the contribution from \(\hat M_n^{(N)}\) is bilinear in \(\hat M_n^{(N)}\) and therefore scales as \(\dd N^n\) at leading order. Hence only \(n=0\) and \(n=1\) contribute at \(O(\dd N)\); the \(n\ge2\) terms are \(O(\dd N^2)\) or smaller in the reduced Kraus sum.

	It is useful, however, to display the pre-trace state before the decoupled-mode trace is performed. Projecting \(\cF^{\rm d,\,\perp}_{N+\dd N}\) onto the \(\{\ket{0}_{\rm d},\ket{1}_{\rm d}\}\) sector gives the \(2\times2\) decoupled-mode block
	\begin{equation}
		\left. \hat\rho_{N+\dd N}^{\parallel\perp} \right|_{0,1} =
		\begin{pmatrix}
			\hat M_0^{(N)}\tilde{\rho}_N\hat M_0^{(N)\dagger} & \hat M_0^{(N)}\tilde{\rho}_N\hat M_1^{(N)\dagger} \\[6pt] \hat M_1^{(N)}\tilde{\rho}_N\hat M_0^{(N)\dagger} & \hat M_1^{(N)}\tilde{\rho}_N\hat M_1^{(N)\dagger}
		\end{pmatrix}_{\{\ket{0}_{\rm d},\ket{1}_{\rm d}\}}
		+ O(\dd N^{3/2}) .
		\label{eq:FullStateBlock01App}
	\end{equation}
	Terms involving \(\ket{2}_{\rm d}\) lie outside this displayed block. They do not contribute to the diagonal reduced Kraus sum at \(O(\dd N)\). Off-diagonal coherences involving \(\ket{2}_{\rm d}\), such as \(\ket{0}\!{}_{\rm d}\,{}_{\rm d}\!\bra{2}\) and \(\ket{2}\!{}_{\rm d}\,{}_{\rm d}\!\bra{0}\), are also killed by the decoupled-mode trace and do not alter the \(O(\dd N)\) GKLS generator.

	Using \cref{eq:KrausVacuumApp} and \cref{eq:KrausOneParticleApp}, the displayed block is
	\begin{equation}
		\left. \hat\rho_{N+\dd N}^{\parallel\perp} \right|_{0,1} =
		\begin{pmatrix}
			\tilde{\rho}_N - \dfrac32 \dd N \bigl\{ \hat A_N^\dagger\hat A_N,\tilde{\rho}_N \bigr\} & -\sqrt{3\,\dd N}\, \tilde{\rho}_N\hat A_N^\dagger \\[8pt] -\sqrt{3\,\dd N}\, \hat A_N\tilde{\rho}_N & 3\dd N\,\hat A_N\tilde{\rho}_N\hat A_N^\dagger
		\end{pmatrix}_{\{\ket{0}_{\rm d},\ket{1}_{\rm d}\}}
		+ O(\dd N^{3/2}) .
		\label{eq:FullStateBlock01ExpandedApp}
	\end{equation}
	Here the matrix indices refer to the vacuum and one-particle sectors of the decoupled mode. Equivalently, after the entangling unitary has acted, the same two-dimensional Fock block is the vacuum-one-particle block of the decoupled mode. Each entry in the matrix is an operator between the homogeneous modes \(\cF_N^\parallel\to\cF_{N+\dd N}^\parallel\). The diagonal entries are the decoupled-mode number blocks, while the off-diagonal entries are the leading vacuum-one-particle coherences generated by the bulk-boundary entangling.

	The partial trace over the decoupled mode keeps only the diagonal decoupled-mode blocks,
	\begin{equation}
		\Tr_{\cF^{\rm d,\,\perp}_{N+\dd N}} \left( \left. \hat\rho_{N+\dd N}^{\parallel\perp} \right|_{0,1} \right) = \hat M_0^{(N)}\tilde{\rho}_N\hat M_0^{(N)\dagger} + \hat M_1^{(N)}\tilde{\rho}_N\hat M_1^{(N)\dagger},
	\end{equation}
	because
	\begin{equation}
		\Tr_{\cF^{\rm d,\,\perp}_{N+\dd N}} \left( \ket{0}{}\!_{\rm d}\,{}_{\rm d}\!\bra{1} \right) = \Tr_{\cF^{\rm d,\,\perp}_{N+\dd N}} \left( \ket{1}\!{}_{\rm d}\,{}_{\rm d}\!\bra{0} \right) = 0 .
	\end{equation}
	Thus the leading decoupled-mode coherence removed by the partial trace is
	\begin{equation}
		\hat\rho^{\rm off}_{N+\dd N} = -\sqrt{3\,\dd N} \left( \tilde{\rho}_N\hat A_N^\dagger \otimes \ket{0}\!{}_{\rm d}\,{}_{\rm d}\!\bra{1} + \hat A_N\tilde{\rho}_N \otimes \ket{1}\!{}_{\rm d}\,{}_{\rm d}\!\bra{0} \right) + O(\dd N^{3/2}) .
		\label{eq:DiscardedFockCoherenceApp}
	\end{equation}
	This term is not a decoupled-mode occupation probability. It is an off-diagonal coherence between the vacuum and one-particle sectors of the collective decoupled mode, correlated with the reduced bulk operators \(\tilde\rho_N\hat A_N^\dagger\) and \(\hat A_N\tilde\rho_N\).

	Before the decoupled-mode trace, this term lives in the operator subspace
	\begin{equation}
		\mathcal B(\cF_{N+\dd N}^\parallel) \otimes {\rm span} \left\{ \ket{0}\!{}_{\rm d}\,{}_{\rm d}\!\bra{1}, \ket{1}\!{}_{\rm d}\,{}_{\rm d}\!\bra{0} \right\} \subset \mathcal B\!\left(\cF_{N+\dd N}^\parallel\otimes\cF^{\rm d,\,\perp}_{N+\dd N}\right).
		\label{eq:DiscardedFockSubspaceApp}
	\end{equation}
	Equivalently, after the entangling redefinition, the relevant Hilbert space is the one-mode Fock space of the decoupled mode,
	\begin{equation}
		\cF^{\rm d,\,\perp}_{N+\dd N} \subset \cF_{N+\dd N}^\perp .
	\end{equation}
	The displayed coherence is supported on its vacuum-one-particle sector,
	\begin{equation}
		{\rm span} \left\{ \ket{0}_{\rm d},\ket{1}_{\rm d} \right\} \subset \cF^{\rm d,\,\perp}_{N+\dd N}.
		\label{eq:DecoupledModeFockSubspaceApp}
	\end{equation}
	The partial trace over the collective decoupled mode removes the off-diagonal operator components on this two-dimensional sector and retains only the diagonal contributions that enter the reduced homogeneous bulk state.

	Substituting \cref{eq:KrausVacuumApp} and \cref{eq:KrausOneParticleApp} gives
	\begin{align}
		\hat\rho_{N+\dd N}
		&= \hat M_0^{(N)} \tilde{\rho}_N \hat M_0^{(N)\dagger}
		+ \hat M_1^{(N)} \tilde{\rho}_N \hat M_1^{(N)\dagger}
		+ O(\dd N^2) \nonumber\\
		&= \left( \openone^\parallel_{N\to N+\dd N} - \frac32\,\dd N\,\hat A_N^\dagger\hat A_N \right)
		\tilde{\rho}_N
		\left( \openone^\parallel_{N\to N+\dd N} - \frac32\,\dd N\,\hat A_N^\dagger\hat A_N \right)^\dagger
		\nonumber\\
		&\hspace{2em} + 3\,\dd N\, \hat A_N\tilde{\rho}_N\hat A_N^\dagger
		+ O(\dd N^2) \nonumber\\
		&= \tilde{\rho}_N - \frac32\,\dd N\, \bigl\{ \hat A_N^\dagger\hat A_N,\tilde{\rho}_N \bigr\}
		+ 3\,\dd N\, \hat A_N\tilde{\rho}_N\hat A_N^\dagger
		+ O(\dd N^2).
		\label{eq:rhoExpandedApp}
	\end{align}
	The \(O(\dd N^2)\) term \[ \frac94(\dd N)^2 \hat A_N^\dagger\hat A_N\, \tilde{\rho}_N\, \hat A_N^\dagger\hat A_N \] generated by expanding \(\hat M_0^{(N)}\tilde{\rho}_N\hat M_0^{(N)\dagger}\) has been absorbed into the error term. Trace preservation is explicit to this order:
	\begin{align}
		\hat M_0^{(N)\dagger}\hat M_0^{(N)} + \hat M_1^{(N)\dagger}\hat M_1^{(N)} &= \openone_{\cF_N^\parallel} - 3\,\dd N\,\hat A_N^\dagger\hat A_N + 3\,\dd N\,\hat A_N^\dagger\hat A_N + O(\dd N^2) \nonumber\\ &= \openone_{\cF_N^\parallel} + O(\dd N^2).
		\label{eq:traceCheckApp}
	\end{align}
	Since the dissipative terms in \cref{eq:rhoExpandedApp} already carry an explicit factor of \(\dd N\), replacing \(\tilde{\rho}_N\) by \(\hat\rho_N\) inside those terms changes the state only by \(O(\dd N^2)\). Substituting \cref{eq:rhoTildeApp} in the leading term and writing the reduced state as the effective bulk state at the next step gives
	\begin{equation}
		\hat\rho_{N+\dd N} = \hat\rho_N - i\,\dd N\,[\hat K_{\rm eff},\hat\rho_N] + 3\,\dd N\,\hat A_N\hat\rho_N\hat A_N^\dagger - \frac32\,\dd N\, \bigl\{ \hat A_N^\dagger\hat A_N,\hat\rho_N \bigr\} + O(\dd N^2).
		\label{eq:rhoStepATracingApp}
	\end{equation}
	Define the jump operator on \(\cF_N^\parallel\) by
	\begin{equation}
		\hat L_N := \sqrt3\,\hat A_N, \qquad \hat L_N^\dagger\hat L_N = 3\,\hat A_N^\dagger\hat A_N, \qquad \hat L_N\hat\rho_N\hat L_N^\dagger = 3\,\hat A_N\hat\rho_N\hat A_N^\dagger .
		\label{eq:LdefTracingApp}
	\end{equation}
	Then \cref{eq:rhoStepATracingApp} becomes
	\begin{equation}
		\hat\rho_{N+\dd N} = \hat\rho_N - i\,\dd N\,[\hat K_{\rm eff},\hat\rho_N] + \left( \hat L_N\hat\rho_N\hat L_N^\dagger - \frac12 \bigl\{ \hat L_N^\dagger\hat L_N,\hat\rho_N \bigr\} \right)\dd N + O(\dd N^2).
		\label{eq:rhoStepWithHamiltonianApp}
	\end{equation}
	Taking the finite-difference quotient and then the continuum limit \(\dd N\to0\) gives
	\begin{equation}
		\frac{\dd\hat\rho_N}{\dd N} = -i[\hat K_{\rm eff},\hat\rho_N] + \hat L_N\hat\rho_N\hat L_N^\dagger - \frac12 \bigl\{ \hat L_N^\dagger\hat L_N,\hat\rho_N \bigr\}.
		\label{eq:GKLSFinalApp}
	\end{equation}
	This is the canonical GKLS equation with a single jump channel of strength \(\sqrt3\), inherited from the incoming boundary shell phase-space volume \(\VF\int\Delta\Theta_N=3\,\dd N\) per $e$-fold.

	The matching condition \cref{eq:c1c2UnitaryApp} fixes only \(|c_1|^2=3\); the overall phase of \(c_1\) is a convention. Equivalently, one may multiply the adapted ladder operator by a constant phase. Carrying this phase through the one-particle Kraus operator gives
	\begin{equation}
		\hat M_1^{(N)} = -\sqrt{3\,\dd N}\,e^{i\chi}\hat A_N + O(\dd N^{3/2}),
	\end{equation}
	and hence promotes the jump operator to
	\begin{equation}
		\hat L_N = \sqrt3\,e^{i\chi}\hat A_N, \qquad \chi\in\mathbb R.
	\end{equation}
	The GKLS generator is invariant under this phase choice, since
	\begin{equation}
		(e^{i\chi}\hat L_N)\hat\rho_N(e^{i\chi}\hat L_N)^\dagger = \hat L_N\hat\rho_N\hat L_N^\dagger, \qquad (e^{i\chi}\hat L_N)^\dagger(e^{i\chi}\hat L_N) = \hat L_N^\dagger\hat L_N .
	\end{equation}

	Although \(\hat A_N\) in \cref{eq:AdefApp} carries explicit \(N\)-dependence through the mode functions \(\phi_{k_\sigma},\pi_{k_\sigma}\) and through \(\VF(N)\), the coefficients of \(\hat Q_N\) and \(\hat P_N\) in the jump operator
	\begin{equation}
		\hat L_N = \sqrt3\,e^{i\chi}\hat A_N = i\sqrt3\,e^{i\chi} \left[ \frac{\phi_{k_\sigma}^*}{\sqrt{\VF(N)}}\,\hat P_N - \sqrt{\VF(N)}\,\pi_{k_\sigma}^*\,\hat Q_N \right]
		\label{eq:LGeneralBoundaryApp}
	\end{equation}
	are independent of \(N\), up to the arbitrary constant phase \(e^{i\chi}\). At the moving boundary,
	\begin{equation}
		k_\sigma(N)e^{-N} = \sigma H, \qquad \VF(N) = \frac{6\pi^2}{\sigma^3 e^{3N}H^3},
	\end{equation}
	and the explicit factors of \(e^N\) cancel between the mode functions evaluated at \(k=k_\sigma(N)\) and the powers of \(\VF(N)\). Using the Bunch-Davies mode functions, the jump operator may be written in closed form as
	\begin{equation}
		\hat L_\chi = e^{i\chi-\tfrac{i\pi}{4}(2\nu+1)} \left\{ \frac{3\pi^{3/2}}{\sqrt2\,\sigma^{3/2}H} \left[ \sigma H_{\nu-1}^{(2)}(\sigma) + \left( \frac32-\nu \right)H_\nu^{(2)}(\sigma) \right]\hat Q_N + \frac{H\sigma^{3/2}}{2\sqrt{2\pi}} H_\nu^{(2)}(\sigma)\,\hat P_N \right\}.
		\label{eq:LExplicitBoundaryApp}
	\end{equation}

	Since the coefficients in \cref{eq:KeffFinal} and \cref{eq:LExplicitBoundaryApp} are time independent, the same dynamics can be written as a stationary one-mode open-system evolution. We identify the instantaneous canonical pair with a fixed system pair,
	\begin{equation}
		\hat Q_N\to \hat Q, \qquad \hat P_N\to \hat P, \qquad [\hat Q,\hat P]=i, \qquad \hat\rho_N\to\hat\rho(N).
	\end{equation}
	The stationary effective Hamiltonian is
	\begin{equation}
		\hat K_{\rm eff} = \frac{\hat P^2}{2H\Vphys} + \frac{m^2\Vphys}{2H}\hat Q^2 + \frac{3}{4}\{\hat Q,\hat P\},
		\label{eq:KeffStationaryApp}
	\end{equation}
	and the stationary jump operator is
	\begin{equation}
		\hat L_\chi = e^{i\chi-\tfrac{i\pi}{4}(2\nu+1)} \left\{ \frac{3\pi^{3/2}}{\sqrt2\,\sigma^{3/2}H} \left[ \sigma H_{\nu-1}^{(2)}(\sigma) + \left( \frac32-\nu \right)H_\nu^{(2)}(\sigma) \right]\hat Q + \frac{H\sigma^{3/2}}{2\sqrt{2\pi}} H_\nu^{(2)}(\sigma)\,\hat P \right\}.
		\label{eq:LStationaryApp}
	\end{equation}
	The coarse-grained stationary GKLS equation is therefore
	\begin{equation}
		\frac{\dd\hat\rho}{\dd N} = -i[\hat K_{\rm eff},\hat\rho] + \hat L_\chi\hat\rho\hat L_\chi^\dagger - \frac12 \bigl\{ \hat L_\chi^\dagger\hat L_\chi,\hat\rho \bigr\}.
		\label{eq:GKLSStationaryApp}
	\end{equation}

	\section{POVMs and Stochastic Unravellings}
	\label{appA:povm_unravellings}

	In this appendix we use the stationary representation for the retained bulk system introduced above. Thus \(\hat\rho\), \(\hat L\), and \(\hat K_{\rm eff}\) act on the fixed one-mode algebra, while the labels \(N+\dd N\) are kept for the outgoing decoupled-mode Hilbert space and the pre-measurement joint state.

	Before tracing over the decoupled mode, the post-unitary state lives on \(\cF_{N+\dd N}^\parallel\otimes\cF^{\rm d,\,\perp}_{N+\dd N} \), and it is generically entangled. Its leading vacuum-one-particle block structure is displayed in \cref{eq:FullStateBlock01App} and \eqref{eq:FullStateBlock01ExpandedApp}. Because an entangled state cannot be written as a product of an independent pure bulk state and an independent pure decoupled-mode state, the redefined bulk alone does not generally have a pure state. There are then two options. If the decoupled-mode information is not observed, it is removed from the reduced description by taking the partial trace over \(\cF^{\rm d,\,\perp}_{N+\dd N}\). This gives the mixed reduced state on \(\cF_{N+\dd N}^\parallel\) and the non-selective GKLS evolution derived above. Alternatively, one may perform a perfectly efficient rank-one measurement of the decoupled mode. For a pure pre-measurement joint trajectory, a definite measurement outcome leaves the redefined bulk in the corresponding pure conditional state \cite{wiseman2009quantum}. This selective branch is illustrated in the main text in \cref{fig:PureStatePOVM}.

	A stochastic unravelling is the corresponding selective description. One conditions the bulk state on a chosen continuous measurement record of the decoupled mode. Different choices of record give different conditioned trajectories, but their ensemble average reproduces the same reduced GKLS equation. The appropriate language for this choice of boundary record is a positive-operator-valued measure, or POVM. For the present purpose a POVM is simply a set of positive operators \(\{\hat\Pi_r\}\) acting on the decoupled mode, labelled by possible outcomes \(r\), and satisfying
	\begin{equation}
		\sum_r \hat\Pi_r=\openone_{\cF^{\rm d,\,\perp}_{N+\dd N}}, \qquad \hat\Pi_r\ge 0 .
	\end{equation}
	For a continuous outcome, the sum is replaced by an integral. If \(\hat\rho_{N+\dd N}^{\parallel\perp}\) is the full active state before the decoupled mode is measured, the probability for outcome \(r\) is
	\begin{equation}
		p(r) = \Tr_{\cF_{N+\dd N}^\parallel\otimes\cF^{\rm d,\,\perp}_{N+\dd N}} \left[ \left( \openone_{\cF_{N+\dd N}^\parallel}\otimes \hat\Pi_r \right) \hat\rho_{N+\dd N}^{\parallel\perp} \right].
	\end{equation}
	The POVM fixes these probabilities but does not by itself fix a unique state update. In the efficient rank-one measurements of the decoupled mode used below, we take the associated rank-one instrument. The corresponding unnormalised conditional homogeneous state is
	\begin{equation}
		\mathcal I_r(\hat\rho) = \Tr_{\cF^{\rm d,\,\perp}_{N+\dd N}} \left[ \left( \openone_{\cF_{N+\dd N}^\parallel}\otimes \hat\Pi_r \right) \hat\rho_{N+\dd N}^{\parallel\perp} \right],
		\label{eq:InstrumentIntroApp}
	\end{equation}
	and the normalised conditional state is
	\begin{equation}
		\hat\rho_{c,r} = \frac{\mathcal I_r(\hat\rho)} {\Tr_{\cF_{N+\dd N}^\parallel}\mathcal I_r(\hat\rho)} .
	\end{equation}
	The map \(\mathcal I_r\) is called the instrument associated with the outcome \(r\). The POVM specifies the probabilities, while the instrument specifies both the probabilities and the corresponding conditioned state update \cite{wiseman2009quantum}.

	In the present problem there is no external detector added by hand. The ``meter'' is the decoupled mode generated by the entangling unitary. The \(2\times2\) Fock block in \cref{eq:FullStateBlock01App} makes this explicit. The diagonal blocks describe the vacuum and one-particle decoupled-mode sectors. The off-diagonal blocks, displayed in \cref{eq:DiscardedFockCoherenceApp}, are the leading \(\ket{0}\!{}_{\rm d}\,{}_{\rm d}\!\bra{1}\) and \(\ket{1}\!{}_{\rm d}\,{}_{\rm d}\!\bra{0}\) coherences between the homogeneous bulk algebra and the decoupled mode. The unconditional trace removes these coherences. A continuous measurement of the decoupled mode instead turns them into a classical stochastic record.

	Different choices of \(\hat\Pi_r\) give different unravellings. A Fock-basis POVM gives the jump, or counting, unravelling. A quadrature POVM gives a single-real-noise homodyne unravelling. A coherent-state POVM gives the two-real-noise heterodyne, or quantum-state-diffusion, unravelling. These are different measurements of the same decoupled mode, not different reduced generators.

	This distinction will be important below. The usual Starobinsky Langevin equation arises only after choosing a decoupled-mode readout and then taking the appropriate super-Hubble and overdamped limits. In particular, the homodyne readout with the field-adapted phase gives a one-real-noise trajectory whose field-centre equation reduces to the standard Starobinsky noise amplitude. The heterodyne readout keeps both real components of the same decoupled-mode coherence and provides a phase-neutral comparison.

	We now apply this general measurement language to the specific decoupled-mode block constructed in Appendix~\ref{appA:tracing}. The goal is to show explicitly how different decoupled-mode POVMs contract the same pre-trace state.

	It is convenient to combine the Hamiltonian and no-jump factor into
	\begin{equation}
		\hat R := \openone - \left( i\hat K_{\rm eff} + \frac12\hat L^\dagger\hat L \right)\dd N, \qquad \hat J := -\sqrt{\dd N}\,\hat L,
		\label{eq:RJDefPOVMApp}
	\end{equation}
	where \(\hat L\) denotes the stationary jump operator with a fixed arbitrary phase convention. To \(O(\dd N)\), the pre-trace state in the \(\{\ket{0}_{\rm d},\ket{1}_{\rm d}\}\) decoupled-mode sector is
	\begin{equation}
		\left. \hat\rho_{N+\dd N}^{\parallel\perp} \right|_{0,1} =
		\begin{pmatrix}
			\hat R\hat\rho\hat R^\dagger & \hat R\hat\rho\hat J^\dagger \\[6pt] \hat J\hat\rho\hat R^\dagger & \hat J\hat\rho\hat J^\dagger
		\end{pmatrix}_{\{\ket{0}_{\rm d},\ket{1}_{\rm d}\}}
		+ O(\dd N^{3/2}) .
		\label{eq:FullBlockRJPOVMApp}
	\end{equation}
	This is the same information as \cref{eq:FullStateBlock01ExpandedApp}, but with the Hamiltonian evolution included in the vacuum block. The off-diagonal entries are precisely the leading decoupled-mode coherences identified in \cref{eq:DiscardedFockCoherenceApp}. Tracing over the decoupled mode keeps only the diagonal entries of \cref{eq:FullBlockRJPOVMApp}.

	Let a decoupled-mode POVM outcome \(r\) have, in the same Fock sector, the matrix
	\begin{equation}
		\hat\Pi_r = \sum_{m,n=0}^{1} \pi_{mn}^{(r)} \ket{m}\!{}_{\rm d}\,{}_{\rm d}\!\bra{n} + \cdots ,
		\label{eq:GeneralPOVMBlockApp}
	\end{equation}
	where the ellipsis denotes higher-Fock components not needed at It\^o order. The corresponding instrument on the homogeneous mode is
	\begin{align}
		\mathcal I_r(\hat\rho) &:= \Tr_{\cF^{\rm d,\,\perp}_{N+\dd N}} \left[ \left( \openone_{\cF_{N+\dd N}^\parallel}\otimes\hat\Pi_r \right) \hat\rho_{N+\dd N}^{\parallel\perp} \right] \nonumber\\ &= \sum_{m,n=0}^{1} \pi_{nm}^{(r)} \hat K_m\hat\rho\hat K_n^\dagger + O(\dd N^{3/2}), \qquad \hat K_0=\hat R,\quad \hat K_1=\hat J .
		\label{eq:GeneralInstrumentPOVMApp}
	\end{align}
	Thus the same pre-trace block produces different stochastic equations depending on the POVM placed on the decoupled mode.

	\subsection{Jump unravelling}
	\label{appA:jump_unravelling}

	The counting, or jump, instrument uses the diagonal Fock POVM \(\hat\Pi_n=\ket{n}\!{}_{\rm d}\,{}_{\rm d}\!\bra{n}\). Retaining the outcomes that contribute through \(O(\dd N)\) gives
	\begin{equation}
		\hat\Pi_0=\ket{0}\!{}_{\rm d}\,{}_{\rm d}\!\bra{0}, \qquad \hat\Pi_1=\ket{1}\!{}_{\rm d}\,{}_{\rm d}\!\bra{1} .
		\label{eq:CountingPOVMApp}
	\end{equation}
	Equation~\eqref{eq:GeneralInstrumentPOVMApp} gives
	\begin{align}
		\mathcal I_0(\hat\rho) &= \hat R\hat\rho\hat R^\dagger,
		\label{eq:JumpNoCountInstrumentApp}
		\\ \mathcal I_1(\hat\rho) &= \hat J\hat\rho\hat J^\dagger = \dd N\,\hat L\hat\rho\hat L^\dagger .
		\label{eq:JumpCountInstrumentApp}
	\end{align}
	The probability for a count in the interval is
	\begin{equation}
		\mathbb E[\dd J] = \Tr\!\left( \hat L^\dagger\hat L\hat\rho_c \right)\dd N, \qquad \dd J^2=\dd J .
		\label{eq:JumpRateApp}
	\end{equation}
	The normalised conditional state obeys
	\begin{align}
		\dd\hat\rho_c ={}& -i[\hat K_{\rm eff},\hat\rho_c]\dd N - \frac12 \bigl\{ \hat L^\dagger\hat L,\hat\rho_c \bigr\}\dd N + \Tr\!\left( \hat L^\dagger\hat L\hat\rho_c \right)\hat\rho_c\,\dd N \nonumber\\ &+ \left[ \frac{ \hat L\hat\rho_c\hat L^\dagger }{ \Tr(\hat L^\dagger\hat L\hat\rho_c) } - \hat\rho_c \right]\dd J .
		\label{eq:JumpSMEApp}
	\end{align}
	This unravelling uses only the diagonal decoupled-mode Fock blocks. It does not read the off-diagonal coherences in \cref{eq:DiscardedFockCoherenceApp}.

	\subsection{Homodyne unravelling}
	\label{app:single_comp}

	A homodyne readout measures one quadrature of the decoupled mode. For a real phase \(\chi\), define the phase-rotated jump operator
	\begin{equation}
		\hat L_\chi := e^{i\chi}\hat L .
		\label{eq:LchiPOVMApp}
	\end{equation}
	For one crossing interval we write the normalised decoupled-mode oscillator as
	\begin{equation}
		\hat b_{\rm d} = \frac{\dd\hat b_{\rm d}}{\sqrt{\dd N}}, \qquad [\hat b_{\rm d},\hat b_{\rm d}^\dagger]=1 .
		\label{eq:NormalizedOutgoingShellOscillatorApp}
	\end{equation}
	The homodyne POVM is the spectral measure of one signed decoupled-mode quadrature,
	\begin{equation}
		\hat X_{{\rm d},\chi} = -\left( e^{i\chi}\hat b_{\rm d} + e^{-i\chi}\hat b_{\rm d}^\dagger \right), \qquad \hat X_{{\rm d},\chi}\ket{\dd Y_\chi;\chi}_{\rm d} = \frac{\dd Y_\chi}{\sqrt{\dd N}}\ket{\dd Y_\chi;\chi}_{\rm d}.
	\end{equation}
	The minus sign is only the convention induced by \(\hat J=-\sqrt{\dd N}\,\hat L\). To It\^o order only the vacuum-one-particle components of the generalised quadrature eigenstate are needed,
	\begin{equation}
		\ket{\dd Y_\chi;\chi}_{\rm d} = p_0(\dd Y_\chi)^{1/2} \left[ \ket{0}_{\rm d} - e^{-i\chi}\frac{\dd Y_\chi}{\sqrt{\dd N}}\ket{1}_{\rm d} +\cdots \right].
		\label{eq:HomodyneQuadratureEigenketApp}
	\end{equation}
	Thus the infinitesimal decoupled-mode quadrature POVM effect \(\hat\Pi_\chi(\dd Y_\chi)=\ket{\dd Y_\chi;\chi}\!{}_{\rm d}\,{}_{\rm d}\!\bra{\dd Y_\chi;\chi}\,\dd Y_\chi\) has the \(\{\ket{0}_{\rm d},\ket{1}_{\rm d}\}\) block
	\begin{equation}
		\hat\Pi_\chi(\dd Y_\chi) = p_0(\dd Y_\chi)
		\begin{pmatrix}
			1 & -e^{i\chi}\dfrac{\dd Y_\chi}{\sqrt{\dd N}} \\[8pt] -e^{-i\chi}\dfrac{\dd Y_\chi}{\sqrt{\dd N}} & \dfrac{\dd Y_\chi^2}{\dd N}
		\end{pmatrix}_{\{\ket{0}_{\rm d},\ket{1}_{\rm d}\}}
		\dd Y_\chi + O(\dd N^{1/2}),
		\label{eq:HomodynePOVMBlockApp}
	\end{equation}
	where the vacuum density is
	\begin{equation}
		p_0(\dd Y_\chi) = \frac{1}{\sqrt{2\pi \dd N}} \exp\!\left( -\frac{\dd Y_\chi^2}{2\dd N} \right).
		\label{eq:HomodyneOstensibleApp}
	\end{equation}
	The signs in \cref{eq:HomodynePOVMBlockApp} compensate the convention \(\hat J=-\sqrt{\dd N}\,\hat L\). Contracting this POVM with \cref{eq:FullBlockRJPOVMApp} gives the efficient homodyne instrument
	\begin{equation}
		\mathcal I_\chi(\dd Y_\chi)(\hat\rho) = \hat M_\chi(\dd Y_\chi)\hat\rho \hat M_\chi^\dagger(\dd Y_\chi)\,\dd Y_\chi,
		\label{eq:HomodyneInstrumentApp}
	\end{equation}
	with
	\begin{equation}
		\hat M_\chi(\dd Y_\chi) = p_0(\dd Y_\chi)^{1/2} \left[ \openone - \left( i\hat K_{\rm eff} + \frac12\hat L^\dagger\hat L \right)\dd N + \hat L_\chi\,\dd Y_\chi \right] + O(\dd N^{3/2}).
		\label{eq:HomodyneMeasurementOperatorApp}
	\end{equation}
	Here \(\hat\Pi_\chi(\dd Y_\chi)\) is the positive decoupled-mode operator that defines the probability of the homodyne outcome, whereas \(\hat M_\chi(\dd Y_\chi)\) is the induced Kraus operator acting on the bulk state after the decoupled-mode quadrature eigenstate has been projected out. The physical distribution of \(\dd Y_\chi\) is shifted relative to \(p_0(\dd Y_\chi)\):
	\begin{equation}
		\dd Y_\chi = \ev{\hat L_\chi+\hat L_\chi^\dagger}_c \dd N + \dd W_\chi, \qquad \dd W_\chi^2=\dd N .
		\label{eq:HomodyneRecordPOVMApp}
	\end{equation}
	The normalised conditional density matrix obeys
	\begin{equation}
		\dd\hat\rho_c = -i[\hat K_{\rm eff},\hat\rho_c]\dd N + \mathcal D[\hat L]\hat\rho_c\,\dd N + \mathcal H[\hat L_\chi]\hat\rho_c\,\dd W_\chi ,
		\label{eq:HomodyneSMEPOVMApp}
	\end{equation}
	where
	\begin{equation}
		\mathcal D[\hat C]\hat\rho = \hat C\hat\rho\hat C^\dagger - \frac12 \bigl\{ \hat C^\dagger\hat C,\hat\rho \bigr\}, \qquad \mathcal H[\hat C]\hat\rho = \hat C\hat\rho+\hat\rho\hat C^\dagger - \Tr\!\left[ (\hat C+\hat C^\dagger)\hat\rho \right]\hat\rho .
		\label{eq:DHDefinitionsPOVMApp}
	\end{equation}
	Unlike the jump readout, the homodyne instrument contracts the off-diagonal decoupled-mode coherences \(\ket{0}\!{}_{\rm d}\,{}_{\rm d}\!\bra{1}\) and \(\ket{1}\!{}_{\rm d}\,{}_{\rm d}\!\bra{0}\). The phase \(\chi\) selects which real linear combination of those coherences is read.

	For pure states, the corresponding normalised stochastic Schr\"odinger equation is
	\begin{align}
		d\ket{\psi_c} &= \left[ -i\hat K_{\rm eff} -\frac12\left( \hat L_\chi^\dagger\hat L_\chi -\ev*{\hat X_\chi}_c\hat L_\chi +\frac14\ev*{\hat X_\chi}_c^2 \right) \right]\ket{\psi_c}\dd N \nonumber\\ &\quad + \left( \hat L_\chi -\frac12\ev*{\hat X_\chi}_c \right)\ket{\psi_c}\dd W_\chi ,
		\label{eq:homodyneSSEshort}
	\end{align}
	where
	\begin{equation}
		\hat X_\chi = \hat L_\chi + \hat L_\chi^\dagger .
	\end{equation}

	\subsubsection{Field-adapted homodyne phase and the Starobinsky limit}

	We now specialise the homodyne phase using the explicit stationary jump operator obtained in \cref{eq:LStationaryApp}. Separating the arbitrary homodyne phase from the fixed Bunch-Davies phase, write
	\begin{equation}
		\hat L_\chi = e^{i\chi}\hat L_0 ,
	\end{equation}
	where \(\hat L_0\) denotes \cref{eq:LStationaryApp} with \(\chi=0\). Thus
	\begin{equation}
		\hat L_0 = e^{-\tfrac{i\pi}{4}(2\nu+1)} \bigg[ \frac{3\pi^{3/2}}{\sqrt2\,\sigma^{3/2}H} T_Q(\nu,\sigma)\,\hat Q + \frac{H\sigma^{3/2}}{2\sqrt{2\pi}} H_\nu^{(2)}(\sigma)\,\hat P \bigg],
		\label{eq:L0FieldAdaptedApp}
	\end{equation}
	with
	\begin{equation}
		T_Q(\nu,\sigma) := \sigma H_{\nu-1}^{(2)}(\sigma) + \left( \frac32-\nu \right)H_\nu^{(2)}(\sigma).
		\label{eq:TQdefAppendix}
	\end{equation}
	The field-adapted phase is chosen so that the coefficient of \(\hat Q\) in \(\hat L_{\chi_Q}\) is purely imaginary. Equivalently,
	\begin{equation}
		e^{i\chi_Q} = -i\, \frac{ \left[ e^{-\tfrac{i\pi}{4}(2\nu+1)} T_Q(\nu,\sigma) \right]^* }{ \left| e^{-\tfrac{i\pi}{4}(2\nu+1)} T_Q(\nu,\sigma) \right| } .
		\label{eq:chiQDefNoAlphaBetaApp}
	\end{equation}
	With this choice, the phase-rotated jump operator takes the form
	\begin{equation}
		\hat L_{\chi_Q} = -i\Lambda_Q\,\hat Q + \mathcal P_Q\,\hat P,
		\label{eq:LchiQFormNoAlphaBetaApp}
	\end{equation}
	where
	\begin{equation}
		\Lambda_Q = \frac{3\pi^{3/2}}{\sqrt2\,\sigma^{3/2}H} \left|T_Q(\nu,\sigma)\right|, \qquad \mathcal P_Q = e^{i\chi_Q-\tfrac{i\pi}{4}(2\nu+1)} \frac{H\sigma^{3/2}}{2\sqrt{2\pi}} H_\nu^{(2)}(\sigma).
		\label{eq:PQLambdaQNoAlphaBetaApp}
	\end{equation}
	The homodyne signal is the Hermitian component
	\begin{equation}
		\hat X_{\chi_Q} = \hat L_{\chi_Q} + \hat L_{\chi_Q}^\dagger = 2\mathrm{Re}(\mathcal P_Q)\hat P .
		\label{eq:XchiQNoAlphaBetaApp}
	\end{equation}
	Thus the field-adapted homodyne readout does not directly monitor the field amplitude \(\hat Q\). It monitors the decoupled-mode momentum component, while the field displacement appears through the corresponding measurement back-action.

	For real \(\nu\), using the Hankel Wronskian
	\begin{equation}
		H_\nu^{(1)}(\sigma)H_{\nu-1}^{(2)}(\sigma) - H_{\nu-1}^{(1)}(\sigma)H_\nu^{(2)}(\sigma) = -\frac{4i}{\pi\sigma},
	\end{equation}
	one obtains
	\begin{equation}
		2\mathrm{Re}(\mathcal P_Q) = \Gamma_Q(\nu,\sigma), \qquad \Gamma_Q(\nu,\sigma) = \frac{2H\sigma^{3/2}} {\pi\sqrt{2\pi}\,|T_Q(\nu,\sigma)|}.
		\label{eq:GammaQNoAlphaBetaApp}
	\end{equation}
	For complex \(\nu\), the same definition is used with the adjoint order conjugated.

	For a Gaussian conditional state, let
	\begin{equation}
		Z_c=
		\begin{pmatrix}
			\ev*{\hat Q}\\ \ev*{\hat P}
		\end{pmatrix},
		\qquad \mathbf{\Sigma}_c=
		\begin{pmatrix}
			\Delta_{QQ} & \Delta_{QP}\\ \Delta_{QP} & \Delta_{PP}
		\end{pmatrix},
		\qquad \Omega=
		\begin{pmatrix}
			0&1\\ -1&0
		\end{pmatrix}.
	\end{equation}
	The single-component innovation gives
	\begin{equation}
		dZ_{c,{\rm noise}} = \left( 2\mathbf{\Sigma}_c\,\mathrm{Re}\nabla L_\chi - \Omega\,\mathrm{Im}\nabla L_\chi \right)\dd W_\chi .
	\end{equation}
	For the field-adapted phase this becomes
	\begin{align}
		\dd\ev*{\hat Q}\big|_{\rm noise} &= \left[ \Gamma_Q\Delta_{QP} - \mathrm{Im}(\mathcal P_Q) \right]\dd W_{\chi_Q}, \\ \dd\ev*{\hat P}\big|_{\rm noise} &= \left( \Gamma_Q\Delta_{PP} - \Lambda_Q \right)\dd W_{\chi_Q}.
	\end{align}

	In the massless case, \(\nu=3/2\),
	\begin{equation}
		T_Q(3/2,\sigma) = \sigma H_{1/2}^{(2)}(\sigma) = i\sqrt{\frac{2\sigma}{\pi}}e^{-i\sigma}.
	\end{equation}
	Equation~\eqref{eq:chiQDefNoAlphaBetaApp} then gives
	\begin{equation}
		e^{i\chi_Q}=e^{i\sigma}.
	\end{equation}
	The phase-rotated jump operator is
	\begin{equation}
		\hat L_\sigma = -i\frac{3\pi}{\sigma H}\hat Q + \frac{H}{2\pi}(\sigma-i)\hat P,
	\end{equation}
	so that
	\begin{equation}
		\Lambda_Q=\frac{3\pi}{\sigma H}, \qquad \mathcal P_Q=\frac{H}{2\pi}(\sigma-i), \qquad \Gamma_Q=\frac{H\sigma}{\pi}.
	\end{equation}
	The monitored component is
	\begin{equation}
		\hat X_\sigma = \hat L_\sigma+\hat L_\sigma^\dagger = \frac{H\sigma}{\pi}\hat P .
	\end{equation}
	The field- and momentum-centre equations become
	\begin{align}
		\dd\ev*{\hat Q} &= \frac{\ev*{\hat P}}{H\Vphys}\dd N + \frac{H}{2\pi} \left( 1+2\sigma\Delta_{QP} \right)\dd W_\sigma, \\ \dd\ev*{\hat P} &= \left( -3\ev*{\hat P} \right)\dd N + \left( -\frac{3\pi}{H\sigma} + \frac{H\sigma}{\pi}\Delta_{PP} \right)\dd W_\sigma .
	\end{align}
	Thus
	\begin{equation}
		\dd\ev*{\hat Q}\big|_{\rm noise} = \frac{H}{2\pi}\dd W_\sigma + O(\sigma)\dd W_\sigma .
	\end{equation}
	After adiabatic elimination of the fast momentum in the free massless limit,
	\begin{equation}
		\dd\ev*{\hat Q} = \frac{H}{2\pi}\dd W_\sigma .
	\end{equation}
	This is the one-real-noise Starobinsky diffusion limit.

	\subsection{Heterodyne, or two-component, unravelling}
	\label{app:two_comp}

	The heterodyne readout does not choose a single real quadrature of the decoupled mode. Instead it uses the coherent-state POVM of the decoupled-mode oscillator. This records the full complex displacement of the decoupled mode, rather than one chosen real projection of it.

	Using the normalised decoupled-mode oscillator of \cref{eq:NormalizedOutgoingShellOscillatorApp}, the coherent states satisfy
	\begin{equation}
		\hat b_{\rm d}\ket{z}_{\rm d} = z\ket{z}_{\rm d}, \qquad \ket{z}_{\rm d} = e^{-|z|^2/2} \sum_{n=0}^{\infty} \frac{z^n}{\sqrt{n!}}\ket{n}_{\rm d} .
	\end{equation}
	They form the overcomplete POVM
	\begin{equation}
		\hat\Pi_{\rm coh}(\dd^2 z) = \frac{\dd^2 z}{\pi} \ket{z}\!{}_{\rm d}\,{}_{\rm d}\!\bra{z}, \qquad \int \hat\Pi_{\rm coh}(\dd^2 z)=\openone_{\cF^{\rm d,\,\perp}_{N+\dd N}}.
		\label{eq:CoherentPOVMFullApp}
	\end{equation}
	This is not a pair of sharp quadrature projectors. It is the simultaneous unsharp measurement of the two non-commuting decoupled-mode quadratures, with the extra vacuum noise required by such a simultaneous readout.

	For consistency with the sign convention \(\hat J=-\sqrt{\dd N}\,\hat L\), introduce the complex infinitesimal outcome \(\dd Y\) by
	\begin{equation}
		z=-\frac{\dd Y}{\sqrt{\dd N}} .
		\label{eq:CoherentOutcomeConventionApp}
	\end{equation}
	Expanding \(\ket{z}\!{}_{\rm d}\,{}_{\rm d}\!\bra{z}\) in the \(\{\ket{0}_{\rm d},\ket{1}_{\rm d}\}\) sector gives
	\begin{equation}
		\hat\Pi_{\rm het}(\dd^2 Y) = p_0(\dd^2 Y)
		\begin{pmatrix}
			1 & -\dfrac{\dd Y^*}{\sqrt{\dd N}} \\[8pt] -\dfrac{\dd Y}{\sqrt{\dd N}} & \dfrac{|\dd Y|^2}{\dd N}
		\end{pmatrix}_{\{\ket{0}_{\rm d},\ket{1}_{\rm d}\}}
		\dd^2 Y + O(\dd N^{1/2}),
		\label{eq:HeterodynePOVMBlockApp}
	\end{equation}
	where
	\begin{equation}
		p_0(\dd^2 Y) = \frac{1}{\pi \dd N} \exp\!\left( -\frac{|\dd Y|^2}{\dd N} \right).
		\label{eq:HeterodyneOstensibleApp}
	\end{equation}
	Equivalently, writing the complex record in terms of two real records,
	\begin{equation}
		\dd Y=\frac{\dd Y_1+i\,\dd Y_2}{\sqrt2}, \qquad \dd^2 Y=\frac12\,\dd Y_1\,\dd Y_2,
	\end{equation}
	the ostensible vacuum distribution is
	\begin{equation}
		p_0(\dd Y_1,\dd Y_2) = \frac{1}{2\pi \dd N} \exp\!\left( -\frac{\dd Y_1^2+\dd Y_2^2}{2\dd N} \right).
	\end{equation}
	Thus the coherent-state POVM produces two real Gaussian readouts of equal variance \(\dd N\). These two readouts are the two noisy components of a single complex decoupled-mode record.

	Contracting \cref{eq:HeterodynePOVMBlockApp} with the pre-trace block \cref{eq:FullBlockRJPOVMApp} gives the heterodyne instrument
	\begin{equation}
		\mathcal I_{\rm het}(\dd^2 Y)(\hat\rho) = \hat M_{\rm het}(\dd Y)\hat\rho \hat M_{\rm het}^\dagger(\dd Y)\,\dd^2 Y,
		\label{eq:HeterodyneInstrumentApp}
	\end{equation}
	where
	\begin{equation}
		\hat M_{\rm het}(\dd Y) = p_0(\dd^2 Y)^{1/2} \left[ \openone - \left( i\hat K_{\rm eff} + \frac12\hat L^\dagger\hat L \right)\dd N + \hat L\,\dd Y^* \right] + O(\dd N^{3/2}).
		\label{eq:HeterodyneMeasurementOperatorApp}
	\end{equation}
	The physical complex record is
	\begin{equation}
		\dd Y = \ev{\hat L}_c \dd N + \dd \zeta, \qquad \dd \zeta=\frac{\dd W_1+i\,\dd W_2}{\sqrt2},
		\label{eq:HeterodyneRecordPOVMApp}
	\end{equation}
	with It\^o products
	\begin{equation}
		\dd \zeta^2=0, \qquad \dd \zeta\,\dd \zeta^*=\dd N, \qquad \dd W_i\,\dd W_j=\delta_{ij}\,\dd N .
	\end{equation}
	In real form this is
	\begin{align}
		\dd Y_1 &= \sqrt2\,\mathrm{Re}\ev{\hat L}_c\,\dd N + \dd W_1,
		\label{eq:HeterodyneRealRecord1App}
		\\ \dd Y_2 &= \sqrt2\,\mathrm{Im}\ev{\hat L}_c\,\dd N + \dd W_2 .
		\label{eq:HeterodyneRealRecord2App}
	\end{align}
	The two real records therefore monitor the two real components of the same complex decoupled-mode displacement. In contrast with homodyne detection, there is no selected phase \(\chi\); a phase rotation of \(\hat L\) simply rotates the pair \((\dd Y_1,\dd Y_2)\).

	The normalised conditional state obeys
	\begin{align}
		\dd\hat\rho_c ={}& -i[\hat K_{\rm eff},\hat\rho_c]\dd N + \mathcal D[\hat L]\hat\rho_c\,\dd N \nonumber\\ &+ \left( \hat L\hat\rho_c - \ev{\hat L}_c\hat\rho_c \right)\dd \zeta^* + \left( \hat\rho_c\hat L^\dagger - \ev{\hat L^\dagger}_c\hat\rho_c \right)\dd \zeta .
		\label{eq:HeterodyneSMEPOVMApp}
	\end{align}
	Equivalently, for pure states,
	\begin{equation}
		d\ket{\psi_c} = \left( -i\hat K_{\rm eff} -\frac12\hat L^\dagger\hat L +\ev*{\hat L^\dagger}_c\hat L -\frac12\ev*{\hat L^\dagger}_c\ev*{\hat L}_c \right)\ket{\psi_c}\dd N + \left( \hat L-\ev*{\hat L}_c \right)\ket{\psi_c}\dd \zeta^* .
		\label{eq:heterodyneSSEshort}
	\end{equation}
	This instrument reads both real components of the same off-diagonal decoupled-mode coherence. In terms of the explicit stationary jump operator \cref{eq:LStationaryApp}, the complex record is
	\begin{equation}
		\dd Y = \ev{\hat L}_c \dd N+\dd \zeta .
		\label{eq:HeterodyneComplexRecordCompactApp}
	\end{equation}
	Thus the coherent-state POVM records the full complex displacement of the decoupled mode. The real and imaginary parts of this displacement are the two classical readout channels.

	\subsubsection{Gaussian dynamics for the heterodyne readout}
	\label{app:Gaussian}

	For a quadratic Hamiltonian and a jump operator linear in \((\hat Q,\hat P)\), Gaussian conditional states remain Gaussian. Let
	\begin{equation}
		W(z,N)=\frac{1}{2\pi\sqrt{\det\mathbf{\Sigma}_c(N)}} \exp\left\{ -\frac12 \left[z-Z_c(N)\right]^\top \mathbf{\Sigma}_c^{-1}(N) \left[z-Z_c(N)\right] \right\},
	\end{equation}
	where
	\begin{equation}
		z=(Q,P)^\top, \qquad Z_c=\begin{pmatrix}\ev*{\hat Q}\\\ev*{\hat P}\end{pmatrix}, \qquad \mathbf{\Sigma}_c=
		\begin{pmatrix}
			\Delta_{QQ} & \Delta_{QP}\\ \Delta_{QP} & \Delta_{PP}
		\end{pmatrix}.
	\end{equation}
	Write the linear jump operator as
	\begin{equation}
		\hat L = \ell^\top
		\begin{pmatrix}\hat Q\\ \hat P\end{pmatrix},
		\qquad \ell=\begin{pmatrix}\ell_Q\\ \ell_P\end{pmatrix}, \qquad \Omega=\begin{pmatrix}0&1\\-1&0\end{pmatrix}.
	\end{equation}
	For the two-component readout,
	\begin{equation}
		dZ_c = A Z_c\,\dd N+n_1\dd W_1+n_2\dd W_2,
		\label{eq:appQSDcenter}
	\end{equation}
	where
	\begin{equation}
		n_1 = \frac{1}{\sqrt2} \left( 2\mathbf{\Sigma}_c\,\mathrm{Re}\ell - \Omega\,\mathrm{Im}\ell \right), \qquad n_2 = \frac{1}{\sqrt2} \left( 2\mathbf{\Sigma}_c\,\mathrm{Im}\ell + \Omega\,\mathrm{Re}\ell \right).
		\label{eq:appQSDnoiseVectors}
	\end{equation}
	The deterministic drift matrix is
	\begin{equation}
		A =
		\begin{pmatrix}
			0 & \dfrac{1}{H\Vphys}\\[6pt] -\dfrac{m^2\Vphys}{H} & -3
		\end{pmatrix}
		=
		\begin{pmatrix}
			0 & \dfrac{H^2\sigma^3}{6\pi^2}\\[6pt] -\dfrac{6\pi^2m^2}{H^4\sigma^3} & -3
		\end{pmatrix}.
	\end{equation}
	The conditional covariance obeys the Riccati equation
	\begin{equation}
		\frac{d\mathbf{\Sigma}_c}{\dd N} = A\mathbf{\Sigma}_c+\mathbf{\Sigma}_cA^\top+D-n_1n_1^\top-n_2n_2^\top,
		\label{eq:appQSDRiccati}
	\end{equation}
	with
	\begin{equation}
		D=\Omega\,\mathrm{Re}(\ell\ell^\dagger)\,\Omega^\top .
	\end{equation}
	For the massless operator \cref{eq:LStationaryApp},
	\begin{equation}
		D =
		\begin{pmatrix}
			\dfrac{H^2(1+\sigma^2)}{4\pi^2} & -\dfrac{3}{2\sigma} \\[8pt] -\dfrac{3}{2\sigma} & \dfrac{9\pi^2}{H^2\sigma^2}
		\end{pmatrix}.
		\label{eq:appUnconditionalD}
	\end{equation}
	The corresponding two real noise amplitudes are
	\begin{align}
		n_Q^{(1)} &= \frac{H}{2\sqrt2\,\pi} \left( 1+2\sigma\Delta_{QP} \right), & n_P^{(1)} &= \frac{1}{\sqrt2} \left( -\frac{3\pi}{H\sigma} + \frac{H\sigma}{\pi}\Delta_{PP} \right), \nn\\ n_Q^{(2)} &= \frac{1}{\sqrt2} \left( \frac{H\sigma}{2\pi} - \frac{H}{\pi}\Delta_{QP} - \frac{6\pi}{H\sigma}\Delta_{QQ} \right), & n_P^{(2)} &= \frac{1}{\sqrt2} \left( -\frac{H}{\pi}\Delta_{PP} - \frac{6\pi}{H\sigma}\Delta_{QP} \right).
		\label{eq:appQSDMasslessNoises}
	\end{align}
	A phase rotation of the massless jump operator rotates the pair \((\dd W_1,\dd W_2)\) and leaves the field-centre diffusion invariant:
	\begin{equation}
		D_{QQ}^{\rm centres} = (n_1)_Q^2+(n_2)_Q^2 .
		\label{eq:appDQQcenters}
	\end{equation}
	On the stabilising branch of \cref{eq:appQSDRiccati},
	\begin{equation}
		D_{QQ}^{\rm centres} = \frac{H^2}{4\pi^2} + O(\sigma),
		\label{eq:appDQQStarobinskyQSD}
	\end{equation}	which is the Starobinsky amplitude squared.

	\section{Rank-One Boundary Noise and the Singular \texorpdfstring{$Q,P$}{Q,P} Kossakowski Matrix}
	\label{app:diagonal}

	The single boundary channel may be written after choosing the irrelevant overall phase of the jump operator so that \(\hat L=\ell_Q\hat Q+\ell_P\hat P\),
	\begin{equation}
		\hat L=\ell_Q\hat Q+\ell_P\hat P, \qquad \ell_Q=-\sqrt{3\VF}\,\pi_{k_\sigma}^*, \qquad \ell_P=\sqrt{\frac{3}{\VF}}\,\phi_{k_\sigma}^* .
		\label{eq:ellQPdef}
	\end{equation}
	In the operator basis \(\hat F_1=\hat Q\), \(\hat F_2=\hat P\), the GKLS equation \cref{eq:Master1D} can be written as
	\begin{equation}
		\dv{}{N}\hat\rho =-i [\hat K_{\rm eff},\hat\rho] +\sum_{i,j=1}^2 \bm\Xi_{ij} \left( \hat F_i\hat\rho\hat F_j -\frac12\{\hat F_j\hat F_i,\hat\rho\} \right),
	\end{equation}
	where
	\begin{equation}
		\bm\Xi=\ell\ell^\dagger, \qquad \ell=\begin{pmatrix}\ell_Q\\ \ell_P\end{pmatrix}.
	\end{equation}
	Thus
	\begin{equation}
		\bm\Xi = 3
		\begin{pmatrix}
			\VF|\pi_{k_\sigma}|^2 & -\pi_{k_\sigma}^*\phi_{k_\sigma} \\[4pt] -\phi_{k_\sigma}^*\pi_{k_\sigma} & |\phi_{k_\sigma}|^2/\VF
		\end{pmatrix}.
		\label{eq:RankOneQPKossakowski}
	\end{equation}
	This is the Kossakowski matrix in the operator basis \((\hat Q,\hat P)\), and it should not be identified directly with the diffusion matrix appearing in the Wigner equation. The latter is obtained from the real symmetric part of \(\bm\Xi\) with the canonical symplectic rotation
	\begin{equation}
		\bm D = \Omega\,{\rm Re}\,\bm\Xi\,\Omega^\top, \qquad \Omega=
		\begin{pmatrix}
			0 & 1\\ -1 & 0
		\end{pmatrix}.
		\label{eq:DiffusionFromKossakowski}
	\end{equation}
	Equivalently, in the \((Q,P)\) ordering,
	\begin{equation}
		\bm D = 3
		\begin{pmatrix}
			|\phi_{k_\sigma}|^2/\VF & \Re(\phi_{k_\sigma}^*\pi_{k_\sigma}) \\[4pt] \Re(\phi_{k_\sigma}^*\pi_{k_\sigma}) & \VF|\pi_{k_\sigma}|^2
		\end{pmatrix}.
	\end{equation}
	Thus the rank-one statement belongs to the Hermitian operator-space Kossakowski matrix, while the Fokker-Planck diffusion matrix is the real canonical phase-space diffusion tensor obtained from it. The Kossakowski matrix in \cref{eq:RankOneQPKossakowski} is positive and rank one. Its determinant vanishes identically,
	\begin{equation}
		\det\bm\Xi=0,
	\end{equation}
	and its non-zero eigenvalue is
	\begin{equation}
		\lambda=\Tr\bm\Xi =3\left(\VF|\pi_{k_\sigma}|^2+\frac{|\phi_{k_\sigma}|^2}{\VF}\right).
	\end{equation}
	Diagonalising \(\bm\Xi\) therefore recovers the same single Lindblad operator. The second eigenvalue vanishes and gives no independent channel. In the massless limit the Kossakowski matrix is
	\begin{equation}
		\bm\Xi=
		\begin{pmatrix}
			\dfrac{9\pi^2}{\sigma^2H^2} & \dfrac{3}{2\sigma}(1-i\sigma) \\[8pt] \dfrac{3}{2\sigma}(1+i\sigma) & \dfrac{H^2}{4\pi^2}(1+\sigma^2)
		\end{pmatrix},
		\qquad \det\bm\Xi=0.
		\label{eq:MasslessQPKossakowski}
	\end{equation}
	The entries in \cref{eq:MasslessQPKossakowski} are therefore perfectly correlated at the operator-space level. Applying \cref{eq:DiffusionFromKossakowski} gives
	\begin{equation}
		\bm D^{m=0} =
		\begin{pmatrix}
			\dfrac{H^2}{4\pi^2}(1+\sigma^2) & -\dfrac{3}{2\sigma} \\[8pt] -\dfrac{3}{2\sigma} & \dfrac{9\pi^2}{\sigma^2H^2}
		\end{pmatrix},
	\end{equation}
	which is the diffusion matrix in \cref{eq:appUnconditionalD}. The swap of the diagonal entries and the sign of the mixed entry are consequences of the symplectic rotation between the operator Kossakowski form and the Wigner diffusion tensor.

\end{document}

%% file: figures/GKLS_tikz.tex
\colorlet{bulkcol}{blue!14}
\colorlet{boundarycol}{orange!25}
\colorlet{reducedcol}{bulkcol!50!boundarycol}
\begin{tikzpicture}[
  scale=0.90,
  transform shape,
  x=1cm,
  y=1cm,
  >=Latex,
  flow/.style={->, line width=0.8pt, draw=teal!60!black},
  callout/.style={->, line width=0.45pt, draw=black!60},
  entangle/.style={<->, line width=0.55pt, draw=black!60, bend left=28},
  small/.style={font=\scriptsize, align=center},
  label/.style={font=\footnotesize, align=center},
  inner/.style={draw=black, line width=0.7pt, fill=bulkcol},
  ring/.style={draw=black, line width=0.7pt, fill=boundarycol},
  retainedmix/.style={draw=black, line width=0.7pt,
    top color=bulkcol, bottom color=boundarycol},
  complementmix/.style={draw=black, line width=0.7pt,
    top color=boundarycol, bottom color=bulkcol},
  reducedmix/.style={draw=black, line width=0.7pt,
    fill=reducedcol},
  jointbox/.style={draw=black!45, dashed, rounded corners=2pt, inner sep=7pt}
]
  \coordinate (input) at (1.45,0);
  \filldraw[ring] (input) circle[radius=1.28];
  \filldraw[inner] (input) circle[radius=0.83];
  \node[align=center] at (input)
    {$\hat\rho_N$\\[-2pt]{\scriptsize bulk $\parallel$}};
  \node[small] at ($(input)+(0,1.05)$) {$\ket{0}_\Delta$};
  \node[small, text width=2.2cm] at ($(input)+(0,1.82)$)
    {entering vacuum\\$\mathcal F_\Delta^\parallel$};
  \draw[callout] ($(input)+(0.50,1.48)$) -- ($(input)+(0.42,1.10)$);
  \node[label, text width=3.1cm] at ($(input)+(0,-1.68)$)
    {$ \hat \rho_{N} \otimes \ket{0}_{\Delta} \in \mathcal F_N^\parallel	\otimes\mathcal F_\Delta^\parallel$};

  \node[retainedmix, circle, minimum size=2.56cm, align=center] (postbulk) at (5.55,0)
    {{\scriptsize retained}\\[-2pt]{\scriptsize bulk $\parallel$}};
  \node[complementmix, circle, minimum size=2.56cm, align=center, small] (postde) at (8.65,0)
    {$\perp$\\[-2pt]{\scriptsize decoupled mode}};
  \node[jointbox, fit=(postbulk) (postde)] (joint) {};
  \draw[entangle] (postbulk.north east) to
    node[small, above] {entangled} (postde.north west);
  \node[small, text width=2.8cm] at ($(postde)+(0,2.12)$)
    {decoupled from $\parallel$\\$\mathcal F^{\rm d,\,\perp}_{N+dN}$};
  \draw[callout] ($(postde)+(0.25,1.82)$) -- ($(postde)+(0.15,1.29)$);
  \node[label, text width=5.2cm] at ($(joint)+(0,-2.02)$)
    {$\hat{\rho}_{N+\mathrm dN}^{\parallel\perp} \in \mathcal F_{N+\mathrm dN}^\parallel
      \otimes \mathcal F^{\rm d,\,\perp}_{N+\mathrm dN}$};

  \coordinate (reduced) at (12.75,0);
  \filldraw[reducedmix] (reduced) circle[radius=1.28];
  \node[align=center] at (reduced)
    {$\hat\rho_{N+\mathrm dN}$\\[-2pt]{\scriptsize bulk $\parallel$}};
  \node[label, text width=2.7cm] at ($(reduced)+(0,-1.68)$)
    {$\hat{\rho}_{N+\mathrm dN} \in \mathcal F_{N+\mathrm dN}^\parallel$};

  \draw[flow] ($(input)+(1.32,0)$) --
    node[small, above, text width=2.55cm]
      {entangling redefinition\\$\hat V_{N+\mathrm dN}$}
    (joint.west);
  \draw[flow] (postde.east) --
    node[small, above, text width=2.35cm]
      {trace over\\$\mathcal F^{\rm d,\,\perp}_{N+\rm d N}$}
    ($(reduced)+(-1.32,0)$);
  \draw[flow] ($(reduced)+(1.32,0)$) -- ++(0.36,0)
    node[right, font=\large] {$\cdots$};
\end{tikzpicture}

%% file: figures/pure_state_povm_tikz.tex
\colorlet{bulkcol}{blue!14}
\colorlet{boundarycol}{orange!25}
\colorlet{reducedcol}{bulkcol!50!boundarycol}
\begin{tikzpicture}[
  scale=0.90,
  transform shape,
  x=1cm,
  y=1cm,
  >=Latex,
  flow/.style={->, line width=0.8pt, draw=teal!60!black},
  callout/.style={->, line width=0.45pt, draw=black!60},
  entangle/.style={<->, line width=0.55pt, draw=black!60, bend left=28},
  small/.style={font=\scriptsize, align=center},
  label/.style={font=\footnotesize, align=center},
  inner/.style={draw=black, line width=0.7pt, fill=bulkcol},
  ring/.style={draw=black, line width=0.7pt, fill=boundarycol},
  retainedmix/.style={draw=black, line width=0.7pt,
    top color=bulkcol, bottom color=boundarycol},
  complementmix/.style={draw=black, line width=0.7pt,
    top color=boundarycol, bottom color=bulkcol},
  reducedmix/.style={draw=black, line width=0.7pt,
    fill=reducedcol},
  jointbox/.style={draw=black!45, dashed, rounded corners=2pt, inner sep=7pt}
]
  \coordinate (input) at (1.45,0);
  \filldraw[ring] (input) circle[radius=1.28];
  \filldraw[inner] (input) circle[radius=0.83];
  \node[align=center] at (input)
    {$\ket{\psi_{c,N}}$\\[-2pt]{\scriptsize bulk $\parallel$}};
  \node[small] at ($(input)+(0,1.05)$) {$\ket{0}_\Delta$};
  \node[small, text width=2.2cm] at ($(input)+(0,1.82)$)
    {entering vacuum\\$\mathcal F_\Delta^\parallel$};
  \draw[callout] ($(input)+(0.50,1.48)$) -- ($(input)+(0.42,1.10)$);
  \node[label, text width=3.3cm] at ($(input)+(0,-1.68)$)
    {$\ket{\Psi}_{N}^{\parallel} \otimes \ket{0}_{\Delta} \in \mathcal F_N^\parallel
      \otimes\mathcal F_\Delta^\parallel$};

  \node[retainedmix, circle, minimum size=2.56cm, align=center] (postbulk) at (5.55,0)
    {{\scriptsize retained}\\[-2pt]{\scriptsize bulk $\parallel$}};
  \node[complementmix, circle, minimum size=2.56cm, align=center, small] (postde) at (8.65,0)
    {$\perp$\\[-2pt]{\scriptsize decoupled mode}};
  \node[jointbox, fit=(postbulk) (postde)] (joint) {};
  \draw[entangle] (postbulk.north east) to
    node[small, above] {entangled} (postde.north west);
  \node[small, text width=2.8cm] at ($(postde)+(0,2.12)$)
    {decoupled from $\parallel$\\$\mathcal F^{\rm d,\,\perp}_{N+\mathrm dN}$};
  \draw[callout] ($(postde)+(0.25,1.82)$) -- ($(postde)+(0.15,1.29)$);
  \node[small, text width=5.2cm] at ($(joint)+(0,-2.02)$)
    {$\ket{\Psi}_{N+\mathrm dN}^{\parallel\perp}
      \in \mathcal F_{N+\mathrm dN}^\parallel
      \otimes \mathcal F^{\rm d,\,\perp}_{N+\mathrm dN}$};

  \coordinate (conditioned) at (12.75,0);
  \filldraw[reducedmix] (conditioned) circle[radius=1.28];
  \node[] at (conditioned)
    {$\ket{\psi_{c,r}}^\parallel$};
  \node[label, text width=2.9cm] at ($(conditioned)+(0,-1.68)$)
    {pure conditioned\\bulk state $\in \mathcal F_{N+\mathrm dN}^\parallel$};
  \draw[flow] ($(input)+(1.32,0)$) --
    node[small, above, text width=2.55cm]
      {entangling redefinition\\$\hat V_{N+\mathrm dN}$}
    (joint.west);
  \draw[flow] (postde.east) --
    node[small, above, text width=1.65cm]
      {rank-one\\readout on\\$\mathcal{F}^{\rm d,\,\perp}_{N+\mathrm dN}$}
    ($(conditioned)+(-1.32,0)$);
  \draw[flow] ($(conditioned)+(1.32,0)$) -- ++(0.36,0)
    node[right, font=\large] {$\cdots$};
\end{tikzpicture}